\documentclass[aps,prd,superscriptaddress,nofootinbib,secnumarabic]{revtex4-2}[10pt]
\pdfoutput=1
\usepackage{latexsym}
\usepackage{mathrsfs}
\usepackage{slashed}
\usepackage{amsmath}
\usepackage{amssymb}
\usepackage{unicode}
\usepackage{diagbox}
\usepackage{makecell}
\usepackage{multirow}
\usepackage{hyperref}
\usepackage{url}
\usepackage{graphicx}
\usepackage{appendix}
\usepackage{indentfirst}
\usepackage{layout}
\usepackage{geometry}

\begin{document}
\title{Di-Higgs production as a probe of flavor changing neutral Yukawa couplings}
\date{\today}
\author{Shi-Ping He}
\email{sphe@ihep.ac.cn}
\affiliation{Center for Future High Energy Physics and Theoretical Physics Division, Institute of High Energy Physics, Chinese Academy of Sciences, Beijing 100049, China}

\begin{abstract}
Top partners are well motivated in many new physics models. Usually, vector like quarks $T_{L,R}$ are introduced to avoid the quantum anomaly. It is crucial to probe their interactions with the standard model particles. However, flavor changing neutral couplings are always difficult to detect directly in the current and future experiments. In this paper, we will show how to constrain the flavor changing neutral Yukawa coupling $Tth$ through the di-Higgs production indirectly. We consider the simplified model including a pair of gauge singlet $T_{L,R}$. Under the perturbative unitarity and experimental constraints, we choose $m_T=400~\mathrm{GeV},s_L=0.2$ and $m_T=800~\mathrm{GeV},s_L=0.1$ as benchmark points. After the analysis of amplitude and evaluation of the numerical cross sections, we find that the present constraints from di-Higgs production have already surpassed the unitarity bound because of the $(y_{L,R}^{tT})^4$ behavior. For the case of $m_T=400~\mathrm{GeV}$ and $s_L=0.2$, $\mathrm{Re}y_{L,R}^{tT}$ and $\mathrm{Im}y_{L,R}^{tT}$ can be bounded optimally in the range $(-0.4, 0.4)$ at HL-LHC with $2\sigma$ CL. For the case of $m_T=800~\mathrm{GeV}$ and $s_L=0.1$, $\mathrm{Re}y_{L,R}^{tT}$ and $\mathrm{Im}y_{L,R}^{tT}$ can be bounded optimally in the range $(-0.5, 0.5)$ at HL-LHC with $2\sigma$ CL. The anomalous triple Higgs coupling $\delta_{hhh}$ can also affect the constraints on $y_{L,R}^{tT}$. Finally, we find that the top quark electric dipole moment can give stronger bounds of $y_{L,R}^{tT}$ in the off-axis regions for some scenarios.
\end{abstract}
\maketitle
\tableofcontents
\section{Introduction}
The standard model (SM) of elementary particle physics has been proposed for more than fifty years \cite{Glashow:1961tr, *Weinberg:1967tq, *Salam:1968rm}, and it is proved to be a very effective description of this field \cite{Tanabashi:2018oca}. The electro-weak symmetry breaking (EWSB) \cite{Englert:1964et, *Higgs:1964ia, *Guralnik:1964eu, *Kibble:1967sv} mechanism predicts the existence of a physical Higgs boson, which is observed at the Large Hadron Collider (LHC) in 2012 \cite{Aad:2012tfa, *Chatrchyan:2012xdj}. Although the SM goes such strong, there are still some clouds in the sky of particle physics. The typical problems are Higgs mass naturalness, gauge coupling unification, fermion mass hierarchy, electro-weak vacuum stability, dark matter, matter anti-matter asymmetry, and so on. Many new physics beyond the SM (BSM) are aimed at solving or partially solving these problems. In some of these BSM models, vector-like quarks (VLQs)\cite{AguilarSaavedra:2009es, Aguilar-Saavedra:2013qpa} are introduced to avoid the quantum anomaly.

For the up-type VLQ, the $T$ quark can interact with the SM particles via the $TbW,TtZ$, and $Tth$ interactions. Constraints on these couplings are crucial, because they may help us unveil the nature of the EWSB. For the strong interaction mediated pair production of VLQs, we can only constrain the partial decay branching ratios of the $T$ quark. To bound these couplings, single production of VLQ needs to be considered. Unfortunately, it is difficult to detect the flavor changing neutral (FCN) interactions $TtZ,Tth$ directly because of the suppression of the single $T$ production from $tZ,th$ fusion. Here, we will focus on the FCN Yukawa (FCNY) interaction $Tth$. After the Higgs boson discovery, we can get more and more information on new physics from the Higgs precision measurements \cite{Dittmaier:2011ti, *Dittmaier:2012vm, *Heinemeyer:2013tqa, *deFlorian:2016spz}. The FCNY interaction can show up in loop induced processes, for example $h\rightarrow\gamma Z$ and $gg\rightarrow hh$. In our previous work \cite{He:2020suf}, we show that it is possible to constrain the FCNY coupling through the $h\rightarrow\gamma Z$ decay mode indirectly. The double Higgs production is also an appealing channel for unravelling the FCNY interaction, which is free of electro-weak gauge interactions. The constraints from $h\rightarrow\gamma Z$ and $gg\rightarrow hh$ are independent of exotic decay modes and total width of the $T$ quark.

In this paper, we build the framework of FCN couplings in Sec.~\ref{sec:framework} first. Sec.~\ref{sec:constraints} is devoted to the theoretical and experimental constraints on the simplified model. In Sec.~\ref{sec:analysis}, we compute the new physics contributions to the parton level cross section of $gg\rightarrow hh$. Then we perform the numerical constraints on the FCNY interactions in Sec.~\ref{sec:numerical}. Finally, we give the summary and conclusions in Sec.~\ref{sec:summary}.
\section{Framework of flavor changing neutral couplings}\label{sec:framework}
\subsection{Minimal singlet vector-like quark model}
The SM gauge group is $SU_C(3)\otimes SU_L(2)\otimes U_Y(1)$, under which the singlet up-type VLQs have the representation (3, 0, 2/3). Let us start with the minimal extension of SM by adding a pair of singlet $T_L,T_R$ \cite{AguilarSaavedra:2009es, Aguilar-Saavedra:2013qpa}, which is named as the VLQT model. Note that the mass mixing term $\bar{T}_Lt_R$ can be eliminated with field redefinition \cite{Dawson:2012di, DeSimone:2012fs}. Then, the Lagrangian can be written as \cite{Aguilar-Saavedra:2013qpa}
\begin{align}
&\mathcal{L}=\mathcal{L}_{SM}+\mathcal{L}_T^{Yukawa}+\mathcal{L}_T^{gauge},\nonumber\\
&\mathcal{L}_T^{Yukawa}=-\Gamma_T^i\bar{Q}_L^i\widetilde{\Phi}T_R-M_T\bar{T}_LT_R+\mathrm{h.c.},\quad\mathcal{L}_T^{gauge}=\bar{T}_Li\slash\!\!\!\!DT_L+\bar{T}_Ri\slash\!\!\!\!DT_R,
\end{align}
where $\widetilde{\Phi}=i\sigma_2\Phi^{\ast}$ and the covariant derivative is defined as $D_{\mu}=\partial_{\mu}-ig^{\prime}Y_TB_{\mu}$. $Y_T$ and $Q_T$ are the $U_Y(1)$ and electric charge of the $T$ quark, respectively. The Higgs doublet is parametrized as $\Phi^T=[\phi^+,~\frac{v+h+i\chi}{\sqrt{2}}]$. It is reasonable to neglect the mixings between heavy particles and the first two generations because of mass hierarchy and the bounds from flavor physics \cite{delAguila:2000aa, delAguila:2000rc, AguilarSaavedra:2002kr}. Here, we only consider the mixings between the third generation and heavy quarks for simplicity.

To diagonalize the $t$ and $T$ quark mass terms, we can perform the transformations
\begin{align}\label{eqn:quark:rotation}
\left[\begin{array}{c}t_L\\T_L\end{array}\right]\rightarrow
	\left[\begin{array}{cc}\cos\theta_L&\sin\theta_L\\-\sin\theta_L&\cos\theta_L\end{array}\right]
	\left[\begin{array}{c}t_L\\T_L\end{array}\right],\quad
\left[\begin{array}{c}t_R\\T_R\end{array}\right]\rightarrow
	\left[\begin{array}{cc}\cos\theta_R&\sin\theta_R\\-\sin\theta_R&\cos\theta_R\end{array}\right]
	\left[\begin{array}{c}t_R\\T_R\end{array}\right].
\end{align}
In fact, we have the relation $m_T\tan\theta_R=m_t\tan\theta_L$. In the following, we will take $s_L,c_L,s_R,c_R$ as shorthands for $\sin\theta_L,\cos\theta_L,\sin\theta_R,\cos\theta_R$, respectively. Then, we can obtain the mass eigenstate Yukawa interactions
\begin{align}
&\mathcal{L}_{Yukawa}\supset-m_t\bar{t}t-m_T\bar{T}T-\frac{m_t}{v}c_L^2h\bar{t}t-\frac{m_T}{v}s_L^2h\bar{T}T-\frac{m_T}{v}s_Lc_Lh(\bar{t}_LT_R+\bar{T}_Rt_L)-\frac{m_t}{v}s_Lc_Lh(\bar{T}_Lt_R+\bar{t}_RT_L),
\end{align}
and gauge interactions
\begin{align}\label{eqn:quark:gauge0}
&\mathcal{L}_{gauge}\supset\frac{g}{c_W}Z_{\mu}[(\frac{1}{2}c_L^2-\frac{2}{3}s_W^2)\bar{t}_L\gamma^{\mu}t_L+(\frac{1}{2}s_L^2-\frac{2}{3}s_W^2)\bar{T}_L\gamma^{\mu}T_L+\frac{1}{2}s_Lc_L(\bar{t}_L\gamma^{\mu}T_L+\bar{T}_L\gamma^{\mu}t_L)\nonumber\\
&-\frac{2}{3}s_W^2\bar{t}_R\gamma^{\mu}t_R-\frac{2}{3}s_W^2\bar{T}_R\gamma^{\mu}T_R]+\frac{gc_L}{\sqrt{2}}(W_{\mu}^+\bar{t}_L\gamma^{\mu}b_L+W_{\mu}^-\bar{b}_L\gamma^{\mu}t_L)+\frac{gs_L}{\sqrt{2}}(W_{\mu}^+\bar{T}_L\gamma^{\mu}b_L+W_{\mu}^-\bar{b}_L\gamma^{\mu}T_L).
\end{align}
Here, we have two independent extra parameters $m_T$ and $\theta_L$. For more details, please refer to our previous work \cite{He:2020suf}.
\subsection{Simplified model}
Besides the singlet VLQs, the scalar sector can also be enlarged in the non-minimally extended models. For example, we can also introduce a real gauge singlet scalar \cite{Dolan:2016eki, Kim:2018mks}, a Higgs doublet \cite{Aguilar-Saavedra:2017giu}, and even both the singlet-doublet scalars at the same time \cite{Muhlleitner:2016mzt, Aguilar-Saavedra:2017giu}. In these models, the $T$ quark can exist other decay channels \cite{Cheung:2018ljx, Cacciapaglia:2019zmj}. Here, we will adopt a general framework \cite{Buchkremer:2013bha}. Then, the simplified related mass eigenstate interactions can be read as \cite{He:2020suf}
\begin{align}\label{eqn:frame:simplify}
&\mathcal{L}\supset-m_t\bar{t}t-m_T\bar{T}T-eA_{\mu}\sum_{f=t,T}Q_f\bar{f}\gamma^{\mu}f+eZ_{\mu}[\bar{t}\gamma^{\mu}(g_L^t\omega_-+g_R^t\omega_+)t+\bar{T}\gamma^{\mu}(g_L^T\omega_-+g_R^T\omega_+)T\nonumber\\
&+\bar{t}\gamma^{\mu}(g_L^{tT}\omega_-+g_R^{tT}\omega_+)T+\bar{T}\gamma^{\mu}(g_L^{tT}\omega_-+g_R^{tT}\omega_+)t]-\frac{m_t}{v}h\bar{t}(\kappa_t+i\gamma^5\widetilde{\kappa}_t)t+h\bar{T}(y_T+i\gamma^5\widetilde{y}_T)T\nonumber\\
&+h\bar{t}(y_L^{tT}\omega_-+y_R^{tT}\omega_+)T+h\bar{T}((y_L^{tT})^{*}\omega_++(y_R^{tT})^{*}\omega_-)t+\frac{gc_L}{\sqrt{2}}(W_{\mu}^+\bar{t}_L\gamma^{\mu}b_L+W_{\mu}^-\bar{b}_L\gamma^{\mu}t_L)\nonumber\\
&+\frac{gs_L}{\sqrt{2}}(W_{\mu}^+\bar{T}_L\gamma^{\mu}b_L+W_{\mu}^-\bar{b}_L\gamma^{\mu}T_L)-\lambda_{hhh}h^3,
\end{align}
where $\omega_{\pm}$ are the chirality projection operators $(1\pm\gamma^5)/2$ and the gauge couplings are listed as
\begin{align}
&g_L^t=\frac{1}{s_Wc_W}(\frac{1}{2}c_L^2-\frac{2}{3}s_W^2),~g_L^T=\frac{1}{s_Wc_W}(\frac{1}{2}s_L^2-\frac{2}{3}s_W^2),~g_L^{tT}=\frac{s_Lc_L}{2s_Wc_W},\nonumber\\
&\qquad\qquad~~~g_R^t=-\frac{2s_W}{c_W},~g_R^T=-\frac{2s_W}{c_W},~g_R^{tT}=0.
\end{align}
The triple Higgs coupling $\lambda_{hhh}$ can deviate from the SM value $\lambda_{hhh}^{SM}=\frac{m_h^2}{2v}$ in many new physics models \cite{Kanemura:2004mg, He:2016sqr, Kanemura:2016lkz, Arhrib:2015hoa, Kanemura:2017gbi, Chiang:2018xpl, Braathen:2019pxr, Englert:2019eyl, Kanemura:2019slf, Englert:2020gcp}. Here, $m_T,\theta_L,\kappa_t,\widetilde{\kappa}_t,y_T,\widetilde{y}_T$ are all real parameters, while $y_L^{tT},y_R^{tT}$ can be complex. From now on, we will turn off the parameters $\widetilde{\kappa}_t$ and $\widetilde{y}_T$ for simplicity.

In the following context, we will show how to constrain the FCNY couplings through the $gg\rightarrow hh$ channel. Although the FCNY couplings $y_{L,R}^{tT}$ are not free parameters in the above VLQT model, they can be free in more complex models. If we can extend the SM by the singlet $T_L,T_R$ and many new scalars, there can be enough degrees of freedom. Thus, we can take them as free parameters to make a general analysis.
\section{Constraints on the simplified model}\label{sec:constraints}
\begin{figure}[!h]
\begin{center}
\includegraphics[scale=0.3]{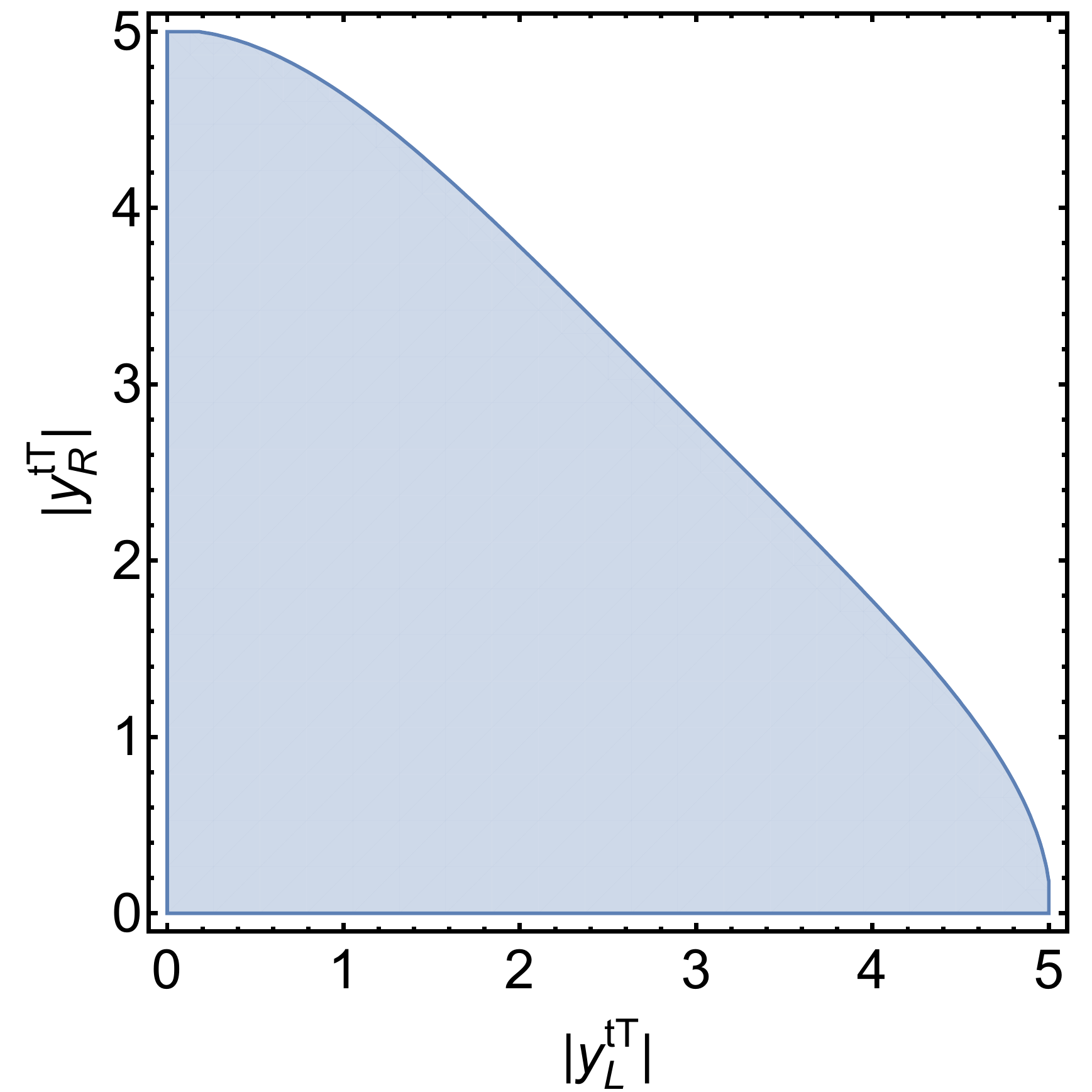}
\caption{The region allowed by the perturbative unitarity.}\label{fig:unitarity}
\end{center}
\end{figure}

\begin{figure}[!h]
\begin{center}
\includegraphics[scale=0.35]{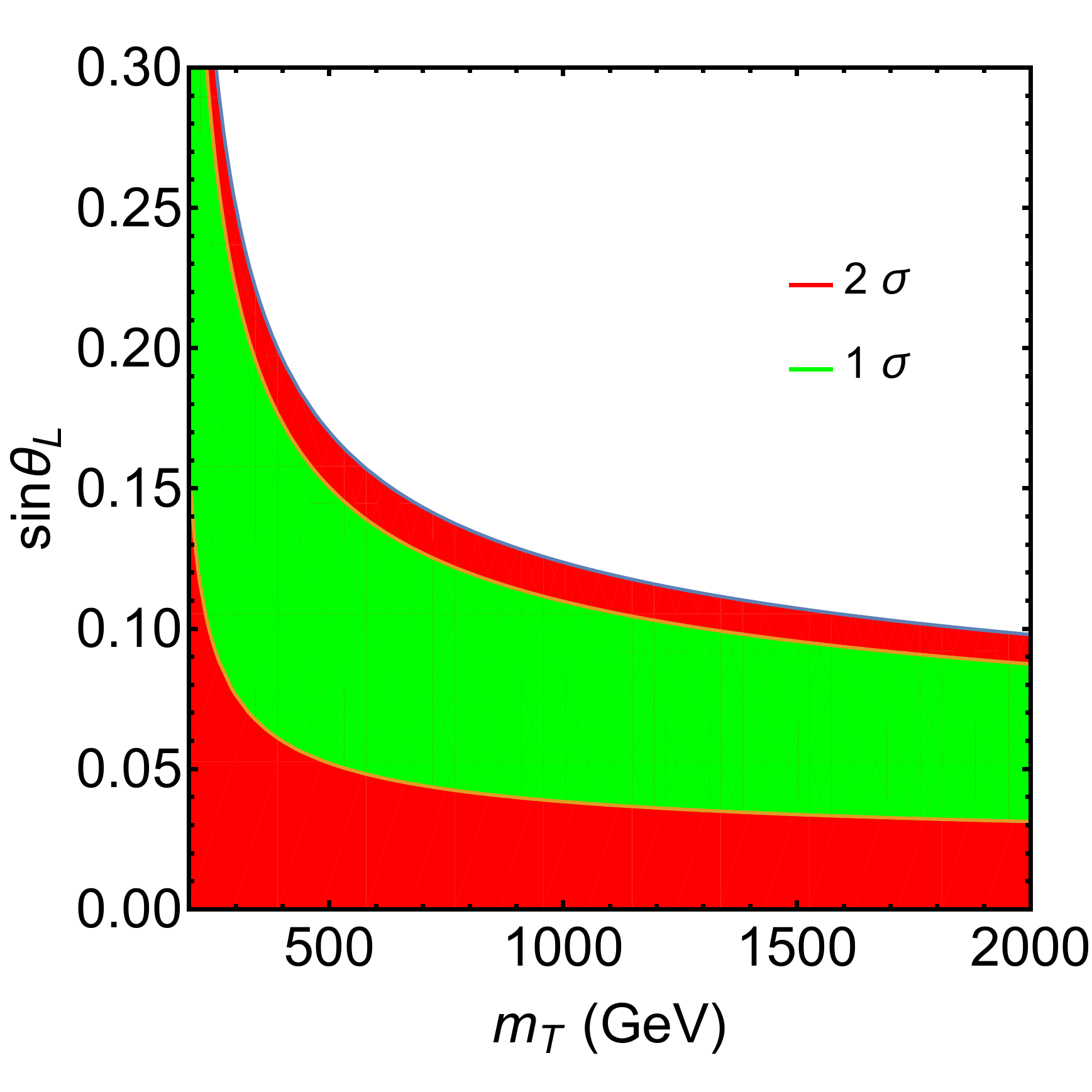}
\caption{The constraints on $m_T,s_L$ from the $S$ and $T$ parameters. Here, the green and red label the allowed regions at $1\sigma$ and $2\sigma$ CL, respectively.}\label{fig:constraints:EWPO}
\end{center}
\end{figure}

In this section, we will review the theoretical and experimental constraints on the simplified model. Specific details are already given in our previous study \cite{He:2020suf}. $S$-wave unitarity will lead to the bound \cite{He:2020suf}
\begin{align}\label{eqn:unitarity}
\sqrt{(|y_L^{tT}|^2+|y_R^{tT}|^2)^2+12|y_L^{tT}|^2|y_R^{tT}|^2}+|y_L^{tT}|^2+|y_R^{tT}|^2\leq16\pi.
\end{align}
In Fig.~\ref{fig:unitarity}, we show the unitarity allowed region in the plane of $|y_L^{tT}|-|y_R^{tT}|$.
Higgs signal strength and top quark physics give quite loose constraints. Direct search can bound the VLQ mass as light as 400 GeV without specific assumptions \cite{Cacciapaglia:2019zmj}. The strongest constraints on $m_T$ and $s_L$ come from the electro-weak precision measurements. Here, we consider the $S$ and $T$ parameters \cite{Peskin:1990zt, Peskin:1991sw, Lavoura:1992np, AguilarSaavedra:2002kr, Chen:2017hak}. In Fig.~\ref{fig:constraints:EWPO}, we show the allowed parameter space regions from the global fits at $1\sigma$ and $2\sigma$ confidence level (CL).
In this paper, the input parameters are chosen as $m_Z=91.1876~\mathrm{GeV}$, $m_W=80.387~\mathrm{GeV}$, $m_h=125.09~\mathrm{GeV}$, $
m_t=172.74~\mathrm{GeV}$, $G_F=1.1664\times10^{-5}~\mathrm{GeV}^{-2}$, and $c_W=m_W/m_Z$ \cite{Tanabashi:2018oca}.

Then, we turn to the constraints from the top quark electric dipole moment (EDM) \cite{Baron:2013eja, Andreev:2018ayy, Kamenik:2011dk, Cirigliano:2016njn, Cirigliano:2016nyn}. If there exists CP violation in the FCNY interactions, it will contribute to the EDM type interaction $-\frac{i}{2}d_t^{EDM}\bar{t}\sigma^{\mu\nu}\gamma^5tF_{\mu\nu}$. The $d_t^{EDM}$ is computed as
\begin{align}
&d_t^{EDM}=\frac{eQ_Tm_T[y_R^{tT}(y_L^{tT})^*-y_L^{tT}(y_R^{tT})^*]}{16\pi^2}C_1,
\end{align}
with $C_1$ defined as
\begin{align*}
&C_1=\frac{1}{4m_t^2}[B_0(m_t^2,m_T^2,m_h^2)-B_0(0,m_T^2,m_T^2)+(m_T^2-m_t^2-m_h^2)C_0(m_t^2,0,m_t^2,m_h^2,m_T^2,m_T^2)].
\end{align*}
As can be seen from the identity $[y_R^{tT}(y_L^{tT})^*-y_L^{tT}(y_R^{tT})^*]=2i(\mathrm{Re}y_L^{tT}\mathrm{Im}y_R^{tT}-\mathrm{Re}y_R^{tT}\mathrm{Im}y_L^{tT})$, $d_t^{EDM}$ will vanish if the imaginary parts of $y_{L,R}^{tT}$ are both turned off. If we take $m_T=400~\mathrm{GeV}$, top EDM sets the upper limit of $|y_R^{tT}(y_L^{tT})^*-y_L^{tT}(y_R^{tT})^*|$ to be 0.12 at $90\%$ CL. If we take $m_T=800~\mathrm{GeV}$, the corresponding upper limit of $|y_R^{tT}(y_L^{tT})^*-y_L^{tT}(y_R^{tT})^*|$ is 0.24 at $90\%$ CL. The larger $m_T$ is, the looser will the constraints be.
\section{The analysis of double Higgs production}\label{sec:analysis}
\subsection{New physics results of the amplitude}\label{sec:gg2hh:NP}
The double Higgs production is a hot topic in the field of Higgs physics. The di-Higgs production cross section has been calculated in SM for many years \cite{Glover:1987nx, Djouadi:2005gi}. The new physics effects have also drawn much attention of this community \cite{Asakawa:2010xj, Dolan:2012ac, Dawson:2015oha, He:2015spf, DiMicco:2019ngk}. Some works on di-Higgs production are based on the SM effective field theory (EFT) \cite{Goertz:2014qta, Azatov:2015oxa, Lu:2015jza, Cao:2015oaa, Cao:2016zob, Li:2019uyy} and non-linearly realized EFT \cite{Contino:2010mh, Contino:2012xk, Grober:2017gut, Buchalla:2018yce}. There are also many studies considered in specific models, for example, Higgs singlet model \cite{Chen:2014ask, Dawson:2015haa, Lewis:2017dme}, two Higgs doublet model \cite{Lu:2015qqa, DeCurtis:2017gzi, Kon:2018vmv, Ren:2017jbg}, VLQ models \cite{Dawson:2012mk, Cacciapaglia:2017gzh, Cheung:2020xij}, composite Higgs models \cite{Gillioz:2012se, Grober:2016wmf}, minimal supersymmetric standard model (MSSM) \cite{Plehn:1996wb, Dawson:1998py, Djouadi:2005gj}, next-to-MSSM \cite{Cao:2013si, Basler:2018dac}, and many other new physics models.

\begin{figure}[!h]
\includegraphics[scale=0.45]{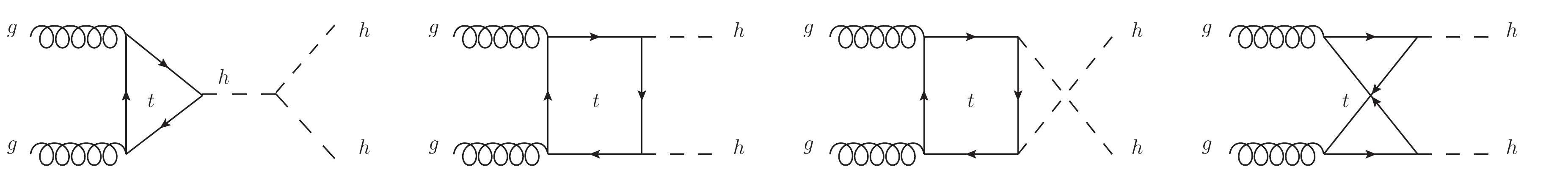}
\caption{Typical Feynman diagrams contributing to the $g(k_1,\mu)g(k_2,\nu)\rightarrow h(p_1)h(p_2)$ production with pure top (also $T$) quarks running in the loops, where the counter-clockwise diagrams should be included.}\label{fig:Feyn:pure}
\end{figure}

\begin{figure}[!h]
\includegraphics[scale=0.45]{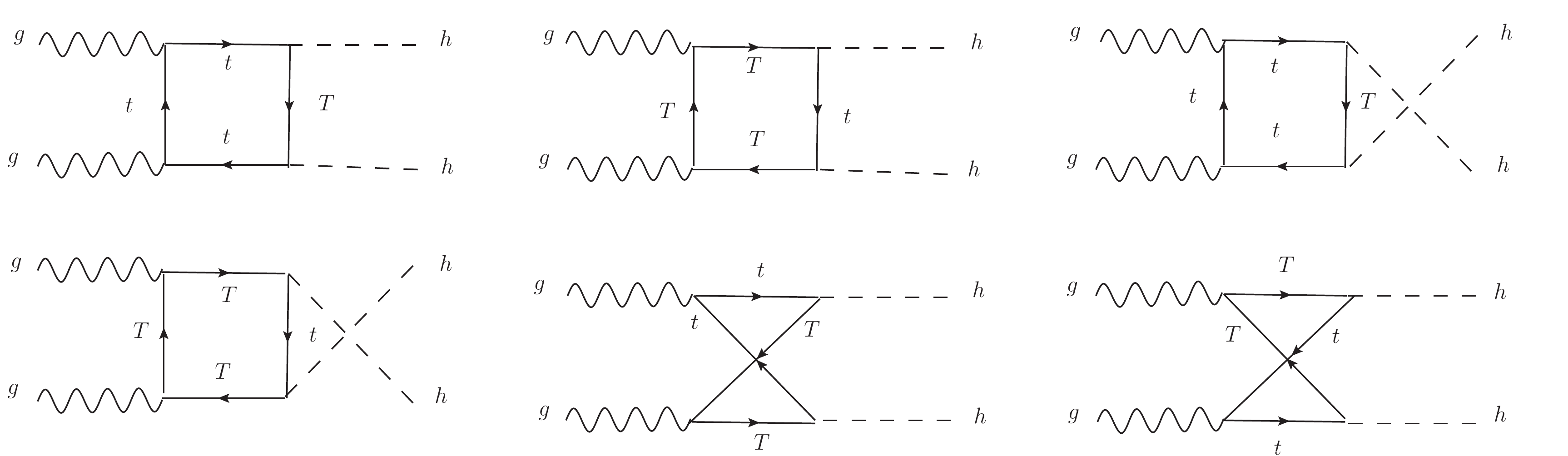}
\caption{Typical Feynman diagrams contributing to the $g(k_1,\mu)g(k_2,\nu)\rightarrow h(p_1)h(p_2)$ production with both top and $T$ quarks running in the loops, where the counter-clockwise diagrams should be included.}\label{fig:Feyn:mixed}
\end{figure}

For VLQ models, there are additional fermion contributions: the pure new quark loops and the loops with both SM and new quarks. In Fig.~\ref{fig:Feyn:pure} and Fig.~\ref{fig:Feyn:mixed}, we show the Feynman diagrams from the pure quark loops and mixed quark loops, respectively \footnote{The diagrams are drawn by JaxoDraw \cite{Binosi:2008ig}.}. The latter will induced by the FCNY interactions. The FCNY contributions are less considered in most of the studies, because they are small compared to the same flavor terms. As a second thought, this channel can be sensitive to large FCNY couplings.

Starting from the Lagrangian in Eq.~\eqref{eqn:frame:simplify}, the amplitude of $gg\rightarrow hh$ can be parametrized as
\begin{align}
i\mathcal{M}(\hat{s})=-i\frac{g_s^2\hat{s}}{16\pi^2v^2}\epsilon_{\mu}^{a,r_1}(k_1)\epsilon_{\nu}^{a,r_2}(k_2)(A^{\mu\nu}f_A+B^{\mu\nu}f_B+C^{\mu\nu}f_C),
\end{align}
where $a$ and $r_{1,2}$ are the color and spin indices, and the tensor structures $A^{\mu\nu},~B^{\mu\nu},~C^{\mu\nu}$ are given by
\begin{align}
&A^{\mu\nu}\equiv g^{\mu\nu}-\frac{k_2^{\mu}k_1^{\nu}}{k_1\cdot k_2},\qquad\qquad\qquad C^{\mu\nu}\equiv \frac{k_{1\rho}k_{2\sigma}\epsilon^{\mu\nu\rho\sigma}}{k_1\cdot k_2},\nonumber\\
&B^{\mu\nu}\equiv g^{\mu\nu}+\frac{m_h^2k_2^{\mu}k_1^{\nu}}{p_T^2(k_1\cdot k_2)}-\frac{2(k_1\cdot p_1)k_2^{\mu}p_1^{\nu}}{p_T^2(k_1\cdot k_2)}-\frac{2(k_2\cdot p_1)p_1^{\mu}k_1^{\nu}}{p_T^2(k_1\cdot k_2)}+\frac{2p_1^{\mu}p_1^{\nu}}{p_T^2}\quad(p_T^2\equiv\frac{\hat{t}\hat{u}-m_h^4}{\hat{s}}).
\end{align}
They have the orthonormal relations $A^{\mu\nu}A_{\mu\nu}=B^{\mu\nu}B_{\mu\nu}=C^{\mu\nu}C_{\mu\nu}=2$ and $A^{\mu\nu}B_{\mu\nu}=A^{\mu\nu}C_{\mu\nu}=B^{\mu\nu}C_{\mu\nu}=0$. The coefficients $f_{A,B,C}$ receive contributions from the $t$ and $T$ quark loops. They can be written as $f_{A,B,C}\equiv f_{A,B,C}^t+f_{A,B,C}^T+f_{A,B,C}^{tT}$. Here $f_{A,B,C}^{t(T)}$ mark the contributions from pure $t(T)$ quark loops, while $f_{A,B,C}^{tT}$ labels the contributions from mixed $t$ and $T$ quark loops. When we set $\kappa_t^2=1$ and turn off the couplings $y_{L,R}^{tT}$, they will go to the SM result. After some lengthy calculations, we can obtain their explicit expressions.

The pure top quark contribution to $f_A$ is given by\footnote{During the calculations, we have used the FeynCalc to simplify the results \cite{Mertig:1990an, Shtabovenko:2016sxi}.} $f_A^t=\kappa_tf_A^{t,\triangle}+\kappa_t^2f_A^{t,\Box1}$. Here, $f_A^{t,\triangle}$ and $f_A^{t,\Box1}$ are defined as
\begin{align}
&f_A^{t,\triangle}=\frac{6m_h^2m_t^2}{\hat{s}(\hat{s}-m_h^2)}[2+(4m_t^2-\hat{s})C_0^t(\hat{s})],\nonumber\\
&f_A^{t,\Box1}=\frac{2m_t^2}{\hat{s}}\Big\{4m_t^2C_0^t(\hat{s})+\frac{2(m_h^2-4m_t^2)}{\hat{s}}[(\hat{t}-m_h^2)C_0^t(\hat{t})+(\hat{u}-m_h^2)C_0^t(\hat{u})]\nonumber\\
	&+m_t^2(8m_t^2-\hat{s}-2m_h^2)[D_0^t(\hat{t},\hat{s})+D_0^t(\hat{u},\hat{s})+D_0^t(\hat{t},\hat{u})]+2+\frac{\hat{t}\hat{u}-m_h^4}{\hat{s}}(4m_t^2-m_h^2)D_0^t(\hat{t},\hat{u})\Big\}.
\end{align}

The pure $T$ quark contribution to $f_A$ is given by $f_A^T=(-\frac{vy_T}{m_T})f_A^{T,\triangle}+(\frac{vy_T}{m_T})^2f_A^{T,\Box1}$. Here, $f_A^{T,\triangle}$ and $f_A^{T,\Box1}$ are defined as
\begin{align}
&f_A^{T,\triangle}=\frac{6m_h^2m_T^2}{\hat{s}(\hat{s}-m_h^2)}[2+(4m_T^2-\hat{s})C_0^T(\hat{s})],\nonumber\\
&f_A^{T,\Box1}=\frac{2m_T^2}{\hat{s}}\Big\{4m_T^2C_0^T(\hat{s})+\frac{2(m_h^2-4m_T^2)}{\hat{s}}[(\hat{t}-m_h^2)C_0^T(\hat{t})+(\hat{u}-m_h^2)C_0^T(\hat{u})]\nonumber\\
	&+m_T^2(8m_T^2-\hat{s}-2m_h^2)[D_0^T(\hat{t},\hat{s})+D_0^T(\hat{u},\hat{s})+D_0^T(\hat{t},\hat{u})]+2+\frac{\hat{t}\hat{u}-m_h^4}{\hat{s}}(4m_T^2-m_h^2)D_0^T(\hat{t},\hat{u})\Big\}.
\end{align}
The top and $T$ quark mixed contribution to $f_A$ is given by $f_A^{tT}=(|y_L^{tT}|^2+|y_R^{tT}|^2)f_A^{tT,\Box1}+[y_L^{tT}(y_R^{tT})^*+y_R^{tT}(y_L^{tT})^*]f_A^{tT,\Box2}$. Here, $f_A^{tT,\Box1}$ and $f_A^{tT,\Box2}$ are defined as
\begin{align}
&f_A^{tT,\Box1}\nonumber\\
	=&\frac{2v^2}{\hat{s}}\Big\{2m_t^2C_0^t(\hat{s})+2m_T^2C_0^T(\hat{s})+\frac{m_h^2-m_t^2-m_T^2}{\hat{s}}[(\hat{t}-m_h^2)(C_0^{tT}(\hat{t})+C_0^{Tt}(\hat{t}))+(\hat{u}-m_h^2)(C_0^{tT}(\hat{u})+C_0^{Tt}(\hat{u}))]\nonumber\\
	&+(m_t^2+m_T^2-m_h^2)[m_t^2(D_0^{tT}(\hat{t},\hat{s})+D_0^{tT}(\hat{u},\hat{s})+D_0^{tT}(\hat{t},\hat{u}))+m_T^2(D_0^{Tt}(\hat{t},\hat{s})+D_0^{Tt}(\hat{u},\hat{s})+D_0^{Tt}(\hat{t},\hat{u}))]\nonumber\\
	&+2+\frac{\hat{t}\hat{u}-m_h^4}{\hat{s}}(m_t^2+m_T^2-m_h^2)D_0^{tT}(\hat{t},\hat{u})\Big\},\nonumber\\
&f_A^{tT,\Box2}=\frac{m_tm_Tv^2}{\hat{s}^2}\Big\{4(m_h^2-\hat{t})[C_0^{tT}(\hat{t})+C_0^{Tt}(\hat{t}))]+4(m_h^2-\hat{u})[C_0^{tT}(\hat{u})+C_0^{Tt}(\hat{u})]\nonumber\\
	&+\hat{s}(4m_t^2-\hat{s})[D_0^{tT}(\hat{t},\hat{s})+D_0^{tT}(\hat{u},\hat{s})+D_0^{tT}(\hat{t},\hat{u})]+\hat{s}(4m_T^2-\hat{s})[D_0^{Tt}(\hat{t},\hat{s})+D_0^{Tt}(\hat{u},\hat{s})+D_0^{Tt}(\hat{t},\hat{u})]\nonumber\\
	&+4(\hat{t}\hat{u}-m_h^4)D_0^{tT}(\hat{t},\hat{u})\Big\}.
\end{align}
The pure top quark contribution to $f_B$ is given by $f_B^t=\kappa_t^2f_B^{t,\Box1}$. Here, $f_B^{t,\Box1}$ is defined as
\begin{align}
&f_B^{t,\Box1}=\frac{m_t^2}{\hat{s}}\Big\{-2\hat{s}C_0^t(\hat{s})+2(m_h^2-\hat{t})C_0^t(\hat{t})+2(m_h^2-\hat{u})C_0^t(\hat{u})-2(8m_t^2+\hat{s}-2m_h^2)C_0^t(m_h^2)\nonumber\\
	&+2m_t^2(8m_t^2+\hat{s}-2m_h^2)[D_0^t(\hat{t},\hat{s})+D_0^t(\hat{u},\hat{s})+D_0^t(\hat{t},\hat{u})]\nonumber\\
	&+\frac{1}{\hat{t}\hat{u}-m_h^4}\Big[\hat{s}\hat{t}(8m_t^2\hat{t}-\hat{t}^2-m_h^4)D_0^t(\hat{t},\hat{s})+\hat{s}\hat{u}(8m_t^2\hat{u}-\hat{u}^2-m_h^4)D_0^t(\hat{u},\hat{s})\nonumber\\
	&+(8m_t^2+\hat{s}-2m_h^2)\Big(\hat{s}(\hat{s}-2m_h^2)C_0^t(\hat{s})+\hat{s}(\hat{s}-4m_h^2)C_0^t(m_h^2)+2\hat{t}(m_h^2-\hat{t})C_0^t(\hat{t})+2\hat{u}(m_h^2-\hat{u})C_0^t(\hat{u})\Big)\Big]\Big\}.
\end{align}
The pure $T$ quark contribution to $f_B$ is given by $f_B^T=(\frac{vy_T}{m_T})^2f_B^{T,\Box1}$. Here, $f_B^{T,\Box1}$ is defined as
\begin{align}
&f_B^{T,\Box1}=\frac{m_T^2}{\hat{s}}\Big\{-2\hat{s}C_0^T(\hat{s})+2(m_h^2-\hat{t})C_0^T(\hat{t})+2(m_h^2-\hat{u})C_0^T(\hat{u})-2(8m_T^2+\hat{s}-2m_h^2)C_0^T(m_h^2)\nonumber\\
	&+2m_T^2(8m_T^2+\hat{s}-2m_h^2)[D_0^T(\hat{t},\hat{s})+D_0^T(\hat{u},\hat{s})+D_0^T(\hat{t},\hat{u})]\nonumber\\
	&+\frac{1}{\hat{t}\hat{u}-m_h^4}\Big[\hat{s}\hat{t}(8m_T^2\hat{t}-\hat{t}^2-m_h^4)D_0^T(\hat{t},\hat{s})+\hat{s}\hat{u}(8m_T^2\hat{u}-\hat{u}^2-m_h^4)D_0^T(\hat{u},\hat{s})\nonumber\\
	&+(8m_T^2+\hat{s}-2m_h^2)\Big(\hat{s}(\hat{s}-2m_h^2)C_0^T(\hat{s})+\hat{s}(\hat{s}-4m_h^2)C_0^T(m_h^2)+2\hat{t}(m_h^2-\hat{t})C_0^T(\hat{t})+2\hat{u}(m_h^2-\hat{u})C_0^T(\hat{u})\Big)\Big]\Big\}.
\end{align}
The top and $T$ quark mixed contribution to $f_B$ is given by $f_B^{tT}=(|y_L^{tT}|^2+|y_R^{tT}|^2)f_B^{tT,\Box1}+[y_L^{tT}(y_R^{tT})^*+y_R^{tT}(y_L^{tT})^*]f_B^{tT,\Box2}$. Here, $f_B^{tT,\Box1}$ is defined as
\begin{small}
\begin{align}
&f_B^{tT,\Box1}\nonumber\\
	&=\frac{2v^2}{\hat{s}}\Big\{-\frac{\hat{s}}{2}[C_0^t(\hat{s})+C_0^T(\hat{s})]-\frac{1}{2}(\hat{s}-2m_h^2+2m_t^2+2m_T^2)[C_0^{tT}(m_h^2)+C_0^{Tt}(m_h^2)]+\frac{m_h^2-\hat{t}}{2}[C_0^{tT}(t)+C_0^{Tt}(t)]\nonumber\\
	&+\frac{m_h^2-\hat{u}}{2}[C_0^{tT}(u)+C_0^{Tt}(u)]+\frac{1}{2}(m_t^2-m_T^2)(\hat{s}+m_t^2+m_T^2-m_h^2)[D_0^{tT}(\hat{t},\hat{s})+D_0^{tT}(\hat{u},\hat{s})-D_0^{Tt}(\hat{t},\hat{s})-D_0^{Tt}(\hat{u},\hat{s})]\nonumber\\
	&+\frac{(m_t^2+m_T^2)}{4}(\hat{s}-2m_h^2+2m_t^2+2m_T^2)[D_0^{tT}(\hat{t},\hat{s})+D_0^{tT}(\hat{u},\hat{s})+D_0^{tT}(\hat{t},\hat{u})+D_0^{Tt}(\hat{t},\hat{s})+D_0^{Tt}(\hat{u},\hat{s})+D_0^{Tt}(\hat{t},\hat{u})]\nonumber\\
	&+\frac{1}{4(\hat{t}\hat{u}-m_h^4)}\Big[\hat{s}(\hat{s}-2m_h^2)(\hat{s}-2m_h^2+2m_t^2+2m_T^2)\Big(C_0^t(\hat{s})+C_0^T(\hat{s})\Big)\nonumber\\
	&+2\hat{s}(m_T^2-m_t^2)(\hat{s}-2m_h^2+2m_T^2+2m_t^2)\Big(C_0^t(\hat{s})-C_0^T(\hat{s})\Big)+\hat{s}(\hat{s}-4m_h^2)(\hat{s}-2m_h^2+2m_t^2+2m_T^2)\Big(C_0^{tT}(m_h^2)+C_0^{Tt}(m_h^2)\Big)\nonumber\\
	&+2\hat{t}(m_h^2-\hat{t})(\hat{s}-2m_h^2+2m_t^2+2m_T^2)\Big(C_0^{tT}(\hat{t})+C_0^{Tt}(\hat{t})\Big)+2\hat{u}(m_h^2-\hat{u})(\hat{s}-2m_h^2+2m_t^2+2m_T^2)\Big(C_0^{tT}(\hat{u})+C_0^{Tt}(\hat{u})\Big)\nonumber\\
	&+\hat{s}(m_t^2-m_T^2)^2(\hat{s}-2m_h^2+2m_t^2+2m_T^2)\Big(D_0^{tT}(\hat{t},\hat{s})+D_0^{tT}(\hat{u},\hat{s})+D_0^{tT}(\hat{t},\hat{u})+D_0^{Tt}(\hat{t},\hat{s})+D_0^{Tt}(\hat{u},\hat{s})+D_0^{Tt}(\hat{t},\hat{u})\Big)\nonumber\\
	&+2\hat{s}\hat{t}(m_t^2-m_T^2)(\hat{s}-2m_h^2+2m_t^2+2m_T^2)\Big(D_0^{tT}(\hat{t},\hat{s})-D_0^{Tt}(\hat{t},\hat{s})\Big)\nonumber\\
	&+2\hat{s}\hat{u}(m_t^2-m_T^2)(\hat{s}-2m_h^2+2m_t^2+2m_T^2)\Big(D_0^{tT}(\hat{u},\hat{s})-D_0^{Tt}(\hat{u},\hat{s})\Big)\nonumber\\
	&-\hat{s}\hat{t}\Big(\hat{t}^2+m_h^4-2\hat{t}(m_t^2+m_T^2)\Big)\Big(D_0^{tT}(\hat{t},\hat{s})+D_0^{tT}(\hat{u},\hat{s})\Big)-\hat{s}\hat{u}\Big(\hat{u}^2+m_h^4-2\hat{u}(m_t^2+m_T^2)\Big)\Big(D_0^{Tt}(\hat{t},\hat{s})+D_0^{Tt}(\hat{u},\hat{s})\Big)\Big]\Big\},\nonumber\\
\end{align}
and $f_B^{tT,\Box2}$ is defined as
\begin{align}
&f_B^{tT,\Box2}=\frac{2m_tm_Tv^2}{\hat{s}}\Big\{-2[C_0^{tT}(m_h^2)+C_0^{Tt}(m_h^2)]+2m_t^2[D_0^{tT}(\hat{t},\hat{s})+D_0^{tT}(\hat{u},\hat{s})+D_0^{tT}(\hat{t},\hat{u})]\nonumber\\
	&+2m_T^2[D_0^{Tt}(\hat{t},\hat{s})+D_0^{Tt}(\hat{u},\hat{s})+D_0^{Tt}(\hat{t},\hat{u})]+\frac{1}{\hat{t}\hat{u}-m_h^4}\Big[\hat{s}(\hat{s}-2m_h^2)\Big(C_0^t(\hat{s})+C_0^T(\hat{s})\Big)-2\hat{s}(m_t^2-m_T^2)\Big(C_0^t(\hat{s})-C_0^T(\hat{s})\Big)\nonumber\\
	&+\hat{s}(\hat{s}-4m_h^2)\Big(C_0^{tT}(m_h^2)+C_0^{Tt}(m_h^2)\Big)+
2\hat{t}(m_h^2-\hat{t})\Big(C_0^{tT}(\hat{t})+C_0^{Tt}(\hat{t})\Big)+2\hat{u}(m_h^2-\hat{u})\Big(C_0^{tT}(\hat{u})+C_0^{Tt}(\hat{u})\Big)\nonumber\\
	&+\hat{s}\Big(\hat{t}^2+(m_t^2-m_T^2)^2\Big)\Big(D_0^{tT}(\hat{t},\hat{s})+D_0^{Tt}(\hat{t},\hat{s})\Big)+\hat{s}\Big(\hat{u}^2+(m_t^2-m_T^2)^2\Big)\Big(D_0^{tT}(\hat{u},\hat{s})+D_0^{Tt}(\hat{u},\hat{s})\Big)\nonumber\\
	&+2\hat{s}\hat{t}(m_t^2-m_T^2)\Big(D_0^{tT}(\hat{t},\hat{s})-D_0^{Tt}(\hat{t},\hat{s})\Big)+2\hat{s}\hat{u}(m_t^2-m_T^2)\Big(D_0^{tT}(\hat{u},\hat{s})-D_0^{Tt}(\hat{u},\hat{s})\Big)+2\hat{s}(m_t^2-m_T^2)^2D_0^{tT}(\hat{t},\hat{u})\Big]\Big\}.
\end{align}
\end{small}
The pure top quark contribution to $f_C$ is proportional to $\widetilde{\kappa}_t$, so we set $f_C^t$ to be zero. Similarly, the pure $T$ quark contribution to $f_C$ is also turned off.
The top and $T$ quark mixed contribution to $f_C$ is given by $f_C^{tT}=-i[y_L^{tT}(y_R^{tT})^*-y_R^{tT}(y_L^{tT})^*]f_C^{tT,\Box}$. Here, $f_C^{tT,\Box}$ is defined as
\begin{align}
&f_C^{tT,\Box}=m_tm_Tv^2[D_0^{tT}(\hat{t},\hat{s})+D_0^{tT}(\hat{u},\hat{s})+D_0^{tT}(\hat{t},\hat{u})+D_0^{Tt}(\hat{t},\hat{s})+D_0^{Tt}(\hat{u},\hat{s})+D_0^{Tt}(\hat{t},\hat{u})].
\end{align}
\subsection{Heavy quark expansion}
In the limit of $\frac{m_h^2,\hat{s},\hat{t},\hat{u}}{m_{t}^2}\ll1$, the coefficients of pure top quark loops $f_A^t,~f_B^t,~f_C^t$ can be expanded as
\begin{align}
&f_A^{t,\triangle}=\frac{2m_h^2}{\hat{s}-m_h^2}\left[1+\frac{7\hat{s}}{120m_t^2}+\mathcal{O}(\frac{1}{m_t^4})\right],\quad f_A^{t,\Box1}=-\frac{2}{3}\left[1+\frac{7m_h^2}{20m_t^2}+\mathcal{O}(\frac{1}{m_t^4})\right],\nonumber\\
&f_B^{t,\Box1}=\frac{11(m_h^4-\hat{t}\hat{u})}{90m_t^2\hat{s}}+\mathcal{O}(\frac{1}{m_t^4}),\quad f_B^{t,\Box2}=\frac{m_h^4-\hat{t}\hat{u}}{6m_t^2\hat{s}}+\mathcal{O}(\frac{1}{m_t^4}),\nonumber\\
&f_C^{t,\triangle}=\frac{3m_h^2}{\hat{s}-m_h^2}\left[1+\frac{\hat{s}}{12m_t^2}+\mathcal{O}(\frac{1}{m_t^4})\right],\quad f_C^{t,\Box}=-2\left[1+\frac{m_h^2}{3m_t^2}+\mathcal{O}(\frac{1}{m_t^4})\right].
\end{align}
For the coefficients of pure $T$ quark loops, they are just the ones with $m_t$ replaced by $m_T$.

For the case of $t$ and $T$ quark mixed loops, things are more complicated. In the limit of $\frac{m_h^2,\hat{s},\hat{t},\hat{u}}{m_{t,T}^2}\ll1$, the coefficient $f_A^{tT},~f_B^{tT},~f_C^{tT}$ can be expanded as
\begin{align}
&f_A^{tT,\Box1}=\mathcal{O}(\frac{1}{m_{t,T}^4}),\nonumber\\
&f_B^{tT,\Box1}=\frac{v^2(\hat{t}^2-\hat{u}^2)}{m_T^2(\hat{t}\hat{u}-m_h^4)}\cdot\frac{(1+r_{tT}^2)(1+2r_{tT}^2\log r_{tT}^2-r_{tT}^4)}{2r_{tT}^2(1-r_{tT}^2)^2}+\mathcal{O}(\frac{1}{m_{t,T}^4}),\nonumber\\
&f_C^{tT,\Box}=\frac{v^2}{m_tm_T}+\mathcal{O}(\frac{1}{m_{t,T}^4}).
\end{align}
\subsection{The cross section analysis}
When averaging the initial spin and color degrees of freedom, we can get the partonic cross section of $gg\rightarrow hh$ at leading order (LO) as follows
\begin{align}\label{eqn:xs:partonic}
&\hat{\sigma}_{LO}(gg\rightarrow hh;\hat{s})=\frac{\alpha_S^2G_F^2\sqrt{\hat{s}(\hat{s}-4m_h^2)}}{128(4\pi)^3}\int_{-1}^1d\cos\theta~(|f_A|^2+|f_B|^2+|f_C|^2)\nonumber\\
&=\frac{\alpha_S^2G_F^2}{64(4\pi)^3}\int_{\hat{t}_{min}}^{\hat{t}_{max}}d\hat{t}~(|f_A|^2+|f_B|^2+|f_C|^2)(\hat{t}_{min}=-\frac{1}{4}(\sqrt{\hat{s}}+\sqrt{\hat{s}-4m_h^2})^2,~\hat{t}_{max}=-\frac{1}{4}(\sqrt{\hat{s}}-\sqrt{\hat{s}-4m_h^2})^2),
\end{align}
where $f_A,f_B,f_C$ are calculated as
\begin{align}\label{eqn:xs:ff}
&f_A=f_A^t+f_A^T+f_A^{tT}=\kappa_tf_A^{t,\triangle}+\kappa_t^2f_A^{t,\Box1}+(-\frac{vy_T}{m_T})f_A^{T,\triangle}+(\frac{vy_T}{m_T})^2f_A^{T,\Box1}\nonumber\\
&+(|y_L^{tT}|^2+|y_R^{tT}|^2)f_A^{tT,\Box1}+[y_L^{tT}(y_R^{tT})^*+y_R^{tT}(y_L^{tT})^*]f_A^{tT,\Box2},\nonumber\\
&f_B=f_B^t+f_B^T+f_B^{tT}=\kappa_t^2f_B^{t,\Box1}+(\frac{vy_T}{m_T})^2f_B^{T,\Box1}+(|y_L^{tT}|^2+|y_R^{tT}|^2)f_B^{tT,\Box1}+[y_L^{tT}(y_R^{tT})^*+y_R^{tT}(y_L^{tT})^*]f_B^{tT,\Box2},\nonumber\\
&f_C=f_C^t+f_C^T+f_C^{tT}=-i[y_L^{tT}(y_R^{tT})^*-y_R^{tT}(y_L^{tT})^*]f_C^{tT,\Box}.
\end{align}
Note that there is a $\frac{1}{2}$ factor in the partonic cross section because of the identical final states. In general, the anomalous triple Higgs coupling $\lambda_{hhh}$ will also alter the di-Higgs production cross section. Its effects can be captured with $f_A^{f,\triangle},f_C^{f,\triangle}(f=t,T)$ multiplied by the factor $1+\delta_{hhh}\equiv\lambda_{hhh}/\lambda_{hhh}^{SM}$.

After folding the partonic cross section with the gluon luminosity, we can get the hadron level cross section
\begin{align}\label{eqn:xs:hadron}
\sigma_{LO}(pp\rightarrow hh)=\int_\frac{4m_h^2}{s}^1d\tau\int_\tau^1\frac{dx}{x}f(x,\mu_F^2)f(\frac{\tau}{x},\mu_F^2)\hat{\sigma}_{LO}(gg\rightarrow hh;\hat{s}=\tau s),
\end{align}
where $f$ represents the gluon parton distribution function (PDF) and $\mu_F$ is the factorization scale.
\section{The numerical results and constraint prospects}\label{sec:numerical}
Just similar to the VLQT model, we take $\kappa_t=c_L^2,y_T=-\frac{m_T}{v}s_L^2$ for simplicity, but let $\mathrm{Re}(y_L^{tT}),\mathrm{Re}(y_R^{tT}),\mathrm{Im}(y_L^{tT}),\mathrm{Im}(y_R^{tT})$ to be free. Then we can choose several benchmark scenarios and estimate the constraints on the magnitude and sign of the FCNY couplings. Now, we need to normalize the cross section to the SM ones numerically for fixed $m_T$ and $s_L$, which is defined as 
\begin{align}
\mu_{hh}\equiv\frac{\sigma_{LO}(pp\rightarrow hh)}{\sigma_{LO}^{SM}(pp\rightarrow hh)}.
\end{align}
Up to LO level, $\mu_{hh}$ can be parametrized as
\begin{align}\label{eqn:num:hhss}
&\mu_{hh}=1+A_1+A_0^{hhh}\delta_{hhh}+A_1^{hhh}\delta_{hhh}^2+(A_2+A_2^{hhh}\delta_{hhh})(|y_L^{tT}|^2+|y_R^{tT}|^2)\nonumber\\
&+(A_3+A_3^{hhh}\delta_{hhh})[y_L^{tT}(y_R^{tT})^*+y_R^{tT}(y_L^{tT})^*]+A_4(|y_L^{tT}|^2+|y_R^{tT}|^2)^2+A_5[y_L^{tT}(y_R^{tT})^*+y_R^{tT}(y_L^{tT})^*]^2\nonumber\\
&+A_6(|y_L^{tT}|^2+|y_R^{tT}|^2)[y_L^{tT}(y_R^{tT})^*+y_R^{tT}(y_L^{tT})^*]-A_7[y_L^{tT}(y_R^{tT})^*-y_R^{tT}(y_L^{tT})^*]^2,
\end{align}
From the observation of Eq.~\eqref{eqn:xs:partonic} and Eq.~\eqref{eqn:xs:ff}, we can find that $A_1,~A_2,~A_3$, $A_0^{hhh},~A_1^{hhh},~A_2^{hhh},~A_3^{hhh}$ depend on the choices of both $m_T$ and $s_L$, while $A_4,~A_5,~A_6,~A_7$ only depend on $m_T$. Moreover, $A_1^{hhh}$, $A_4,~A_5,~A_7$ are always non-negative and $A_1$ vanishes as $s_L$ goes to zero.

Although $\sigma^{SM}(pp\rightarrow hh)$ has been calculated with high precision \cite{Shao:2013bz, deFlorian:2013jea, deFlorian:2015moa, Degrassi:2016vss, Borowka:2016ehy, Grazzini:2018bsd, Baglio:2018lrj, Chen:2019lzz, Chen:2019fhs}, we will not do that hard work here. We only keep the LO results because a large part of the QCD corrections can be cancelled in the ratio \cite{Azatov:2015oxa, Carvalho:2015ttv, Carvalho:2016rys, Buchalla:2018yce}. To get the numerical results of cross sections, we write a model file through FeynRules \cite{Alloul:2013bka, Degrande:2011ua}, FeynArts \cite{Hahn:2000kx} and NLOCT \cite {Degrande:2014vpa}. Then it is linked to MadGraph \cite{Alwall:2014hca}. Before the numerical calculations, we take the following default settings:\\
$\bullet$ Proton contains $b,\bar{b}$, that is, we use the 5FS (5 flavor scheme).\\
$\bullet$ We adopt the PDF choice of "MSTW2008lo68cl" (LHAPDF ID 21000).\\
$\bullet$ The default $dynamical_{\_}scale_{\_}choice$ is set to be 3 (see \cite{Hirschi:2015iia}). \\
$\bullet$ The input parameters are choose as $m_h=125.09~\mathrm{GeV},~G_F=1.1664\times
10^{-5}~\mathrm{GeV}^{-2}$, $m_t=172.74~\mathrm{GeV}$, and $\alpha_s(m_Z)=0.1184$. Thus we have $v=246.221~\mathrm{GeV}$.

Currently, the Higgs pair production is bounded to be $|\mu_{hh}|\leq6.9$ at $95\%$ confidence level (CL) \cite{Sirunyan:2018ayu, Aad:2019uzh}. At the high luminosity LHC (HL-LHC), di-Higgs production measurement is accessible. The expected signal strength is $\mu_{hh}=1.00_{-0.39}^{+0.41}$ with $1\sigma$ uncertainty \cite{Cepeda:2019klc}. We take the benchmark points as $m_T=400~\mathrm{GeV},~s_L=0.2$ and $m_T=800~\mathrm{GeV},~s_L=0.1$, and all the following discussions are based on the two benchmark points. Now, we should determine the specific values of $A_1,~A_2,~A_3,~A_4,~A_5,~A_6,~A_7$ and $A_0^{hhh},~A_1^{hhh},~A_2^{hhh},~A_3^{hhh}$. First of all, we have $\sigma_{LO}^{SM}(pp\rightarrow hh)=$ 24.7 fb. When setting different values of $\delta_{hhh},~y_L^{tT},~y_R^{tT}$, we can obtain different normalized cross sections (see Tab.~\ref{tab:ss:solve1} and Tab.~\ref{tab:ss:solve2}). Then the numerical values of $A_1,...,A_7$ and $A_0^{hhh},~A_1^{hhh},~A_2^{hhh},~A_3^{hhh}$ can be solved from the first seven and last four equations individually. Their results are given in Tab.~\ref{tab:Ai:value}.

\begin{table}[!h]
\begin{tabular}{c|c|c|c}
\hline
$\delta_{hhh}$ & $(~y_L^{tT},~y_R^{tT})$ & expressions of $\mu_{hh}$ & numerical values of $\mu_{hh}$\\
\hline
\multirow{7}{*}{0}&$(0,~0)$ & $1+A_1$ & 0.8081\\
\cline{2-4}
&$(0,~1)$ & $1+A_1+A_2+A_4$ & 1.254\\
\cline{2-4}
&$(0,~\frac{1}{2})$ & $1+A_1+\frac{1}{4}A_2+\frac{1}{16}A_4$ & 0.9057\\
\cline{2-4}
&$(1,~1)$ & $1+A_1+2A_2+2A_3+4A_4+4A_5+4A_6$ & 10.92\\
\cline{2-4}
&$(1,~-1)$ & $1+A_1+2A_2-2A_3+4A_4+4A_5-4A_6$ & 1.695\\
\cline{2-4}
&$(1,~i)$ & $1+A_1+2A_2+4A_4+4A_7$ & 14.13\\
\cline{2-4}
&$(\frac{1}{2},~\frac{1}{2})$ & $1+A_1+\frac{1}{2}A_2+\frac{1}{2}A_3+\frac{1}{4}A_4+\frac{1}{4}A_5+\frac{1}{4}A_6$ & 2.206\\
\hline
\multirow{3}{*}{1}&$(0,~0)$ & \makecell*[c]{$1+A_1+A_0^{hhh}+A_1^{hhh}$} & 0.3877\\
\cline{2-4}
&$(0,~1)$ & $1+A_1+A_0^{hhh}+A_1^{hhh}+A_2+A_2^{hhh}+A_4$ & 0.6996\\
\cline{2-4}
&$(1,~1)$ & \makecell*[c]{$1+A_1+A_0^{hhh}+A_1^{hhh}+2(A_2+A_2^{hhh})+2(A_3+A_3^{hhh})$\\$+4A_4+4A_5+4A_6$} & 8.235\\
\hline
$-1$&$(0,~0)$ & $1+A_1-A_0^{hhh}+A_1^{hhh}$ & 1.779\\
\hline
\end{tabular}
\caption{The normalized cross sections for different $\delta_{hhh},~y_L^{tT},~y_R^{tT}$ values with $m_T=$ 400 GeV and $s_L=0.2$ at $\sqrt{s}$=14TeV.}\label{tab:ss:solve1}
\end{table}

\begin{table}[!h]
\begin{tabular}{c|c|c|c}
\hline
$\delta_{hhh}$ & $(~y_L^{tT},~y_R^{tT})$ & expressions of $\mu_{hh}$ & numerical values of $\mu_{hh}$\\
\hline
\multirow{7}{*}{0}&$(0,~0)$ & $1+A_1$ & 0.9506\\
\cline{2-4}
&$(0,~1)$ & $1+A_1+A_2+A_4$ & 1.098\\
\cline{2-4}
&$(0,~\frac{1}{2})$ & $1+A_1+\frac{1}{4}A_2+\frac{1}{16}A_4$ & 0.9838\\
\cline{2-4}
&$(1,~1)$ & $1+A_1+2A_2+2A_3+4A_4+4A_5+4A_6$ & 5.255\\
\cline{2-4}
&$(1,~-1)$ & $1+A_1+2A_2-2A_3+4A_4+4A_5-4A_6$ & 0.3376\\
\cline{2-4}
&$(1,~i)$ & $1+A_1+2A_2+4A_4+4A_7$ & 5.247\\
\cline{2-4}
&$(\frac{1}{2},~\frac{1}{2})$ & $1+A_1+\frac{1}{2}A_2+\frac{1}{2}A_3+\frac{1}{4}A_4+\frac{1}{4}A_5+\frac{1}{4}A_6$ & 1.675\\
\hline
\multirow{3}{*}{1}&$(0,~0)$ & \makecell*[c]{$1+A_1+A_0^{hhh}+A_1^{hhh}$} & 0.4502\\
\cline{2-4}
&$(0,~1)$ & $1+A_1+A_0^{hhh}+A_1^{hhh}+A_2+A_2^{hhh}+A_4$ & 0.5526\\
\cline{2-4}
&$(1,~1)$ & \makecell*[c]{$1+A_1+A_0^{hhh}+A_1^{hhh}+2(A_2+A_2^{hhh})+2(A_3+A_3^{hhh})$\\$+4A_4+4A_5+4A_6$} & 3.519\\
\hline
$-1$&$(0,~0)$ & $1+A_1-A_0^{hhh}+A_1^{hhh}$ & 2.013\\
\hline
\end{tabular}
\caption{The normalized cross sections for different $\delta_{hhh},~y_L^{tT},~y_R^{tT}$ values with $m_T=$ 800 GeV and $s_L=0.1$ at $\sqrt{s}$=14TeV.}\label{tab:ss:solve2}
\end{table}

\begin{table}[!h]
\begin{tabular}{c|c|c|c|c|c|c|c|c}
\hline
$\sqrt{s}$ (TeV) & ($m_T$/GeV, $s_L$) &$A_1$ &$A_2$ &$A_3$ &$A_4$ &$A_5$ &$A_6$ &$A_7$ \\ \hline
\multirow{2}{*}{14} & (400, 0.2) & $-0.1919$ & 0.3717 & 1.672 & 0.07449 & 1.114 & 0.3166 & 3.071 \\ \cline{2-9}
 & (800, 0.1) & $-0.04939$ & 0.1279 & 1.087 & 0.01943 & 0.378 & 0.0711 & 0.9907 \\ \hline
$\sqrt{s}$ (TeV) & ($m_T$/GeV, $s_L$) &$A_0^{hhh}$ &$A_1^{hhh}$ &$A_2^{hhh}$ &$A_3^{hhh}$ & & &  \\ \hline
\multirow{2}{*}{14} & (400, 0.2) & $-0.6958$ & 0.2754 & $-0.1343$ & $-0.9956$ & & \\ \cline{2-9}
 & (800, 0.1) & $-0.7814$ & 0.281 & $-0.04494$ & $-0.5731$ &  & \\ \hline
\end{tabular}
\caption{The coefficients in Eq.~\eqref{eqn:num:hhss} solved through the signal strength values in Tab.~\ref{tab:ss:solve1} and Tab.~\ref{tab:ss:solve2}.}\label{tab:Ai:value}
\end{table}
\subsection{The benchmark point $m_T=$ 400 GeV and $s_L=0.2$}
For the case of $m_T=$ 400 GeV and $s_L=0.2$, the numerical results of $\mu_{hh}$ are evaluated as
\begin{align}
&\mu_{hh}=1-0.1919-0.6958~\delta_{hhh}+0.2754~\delta_{hhh}^2+(0.3717-0.1343~\delta_{hhh})(|y_L^{tT}|^2+|y_R^{tT}|^2)\nonumber\\
&+(1.672-0.9956~\delta_{hhh})[y_L^{tT}(y_R^{tT})^*+y_R^{tT}(y_L^{tT})^*]+0.07449(|y_L^{tT}|^2+|y_R^{tT}|^2)^2+1.114[y_L^{tT}(y_R^{tT})^*+y_R^{tT}(y_L^{tT})^*]^2\nonumber\\
&+0.3166(|y_L^{tT}|^2+|y_R^{tT}|^2)[y_L^{tT}(y_R^{tT})^*+y_R^{tT}(y_L^{tT})^*]-3.071[y_L^{tT}(y_R^{tT})^*-y_R^{tT}(y_L^{tT})^*]^2.
\end{align}
In this case, the present di-Higgs production experiments give the constraints $\delta_{hhh}\in(-3.61, 6.13)$ and $\mathrm{Re}y_L^{tT},\mathrm{Im}y_L^{tT},\mathrm{Re}y_R^{tT},\mathrm{Im}y_R^{tT}\in(-2.62, 2.62)$ at $95\%$ CL by setting one parameter at a time. In Tab.~\ref{tab:mT400:bound}, we give the expected constraints on the parameters $\delta_{hhh},\mathrm{Re}y_L^{tT},\mathrm{Im}y_L^{tT},\mathrm{Re}y_R^{tT},\mathrm{Im}y_R^{tT}$ at HL-LHC. It can be seen that both of the current and expected constraints at HL-LHC are stronger than the unitarity bound, the reason is that the highest power in di-Higgs production cross section is proportional to $(y_{L,R}^{tT})^4$.

\begin{table}[!h]
\begin{center}
\begin{tabular}{c|c|c|c|c|c|c}
\hline
\multicolumn{2}{c|}{\diagbox[width=3.2cm,trim=lr]{method}{parameters}} &$\delta_{hhh}$ & $\mathrm{Re}y_L^{tT}$ & $\mathrm{Im}y_L^{tT}$ & $\mathrm{Re}y_R^{tT}$ & $\mathrm{Im}y_R^{tT}$\\
\hline
\multirow{2}{*}{individual}& $1~\sigma$ & (-0.681, 0.327)$\cup$(2.20, 3.21) & (-1.13, 1.13) & (-1.13, 1.13) & (-1.13, 1.13) & (-1.13, 1.13)\\
\cline{2-7}
 & $2~\sigma$ & (-1.03, 3.56) & (-1.40, 1.40) & (-1.40, 1.40) & (-1.40, 1.40) & (-1.40, 1.40)\\
\hline
\multirow{2}{*}{marginalized}& $1~\sigma$ & (-3.76, 3.99) & (-1.91, 1.91) & (-1.91, 1.91) & (-1.91, 1.91) & (-1.91, 1.91)\\
\cline{2-7}
 & $2~\sigma$ & (-4.48, 4.66) & (-2.09, 2.09) & (-2.09, 2.09) & (-2.09, 2.09) & (-2.09, 2.09)\\
\hline
\end{tabular}
\caption{The expected $1\sigma$ and $2\sigma$ bounds at HL-LHC for the parameters $\delta_{hhh},\mathrm{Re}y_L^{tT},\mathrm{Im}y_L^{tT},\mathrm{Re}y_R^{tT},\mathrm{Im}y_R^{tT}$ under the benchmark point $m_T=$ 400 GeV and $s_L=0.2$. Here we adopt two different methods: (1) turn on one parameter at a time, namely the individual method; (2) turn on all the five parameters, namely the marginalized method.}\label{tab:mT400:bound}
\end{center}
\end{table}

As mentioned above, there are four interesting parameters $\mathrm{Re}(y_L^{tT}),\mathrm{Re}(y_R^{tT}),\mathrm{Im}(y_L^{tT}),\mathrm{Im}(y_R^{tT})$. Then we can plot the reached two-dimensional parameter space by setting two of them to be zero or imposing two conditions. Here we choose six scenarios: \textcircled{1} $y_{L,R}^{tT}$ are both real (it is similar to the both imaginary number case); \textcircled{2} $y_R^{tT}$ is real and $y_L^{tT}$ is imaginary (it is similar to the real $y_L^{tT}$ and imaginary $y_R^{tT}$ case); \textcircled{3} $y_R^{tT}=0$ (similar to the $y_L^{tT}=0$ case); \textcircled{4} $y_L^{tT}=y_R^{tT}$; \textcircled{5} $y_L^{tT}=-y_R^{tT}$; \textcircled{6} $y_L^{tT}=(y_R^{tT})^*$.

\begin{figure}[!h]
\begin{center}
\includegraphics[scale=0.28]{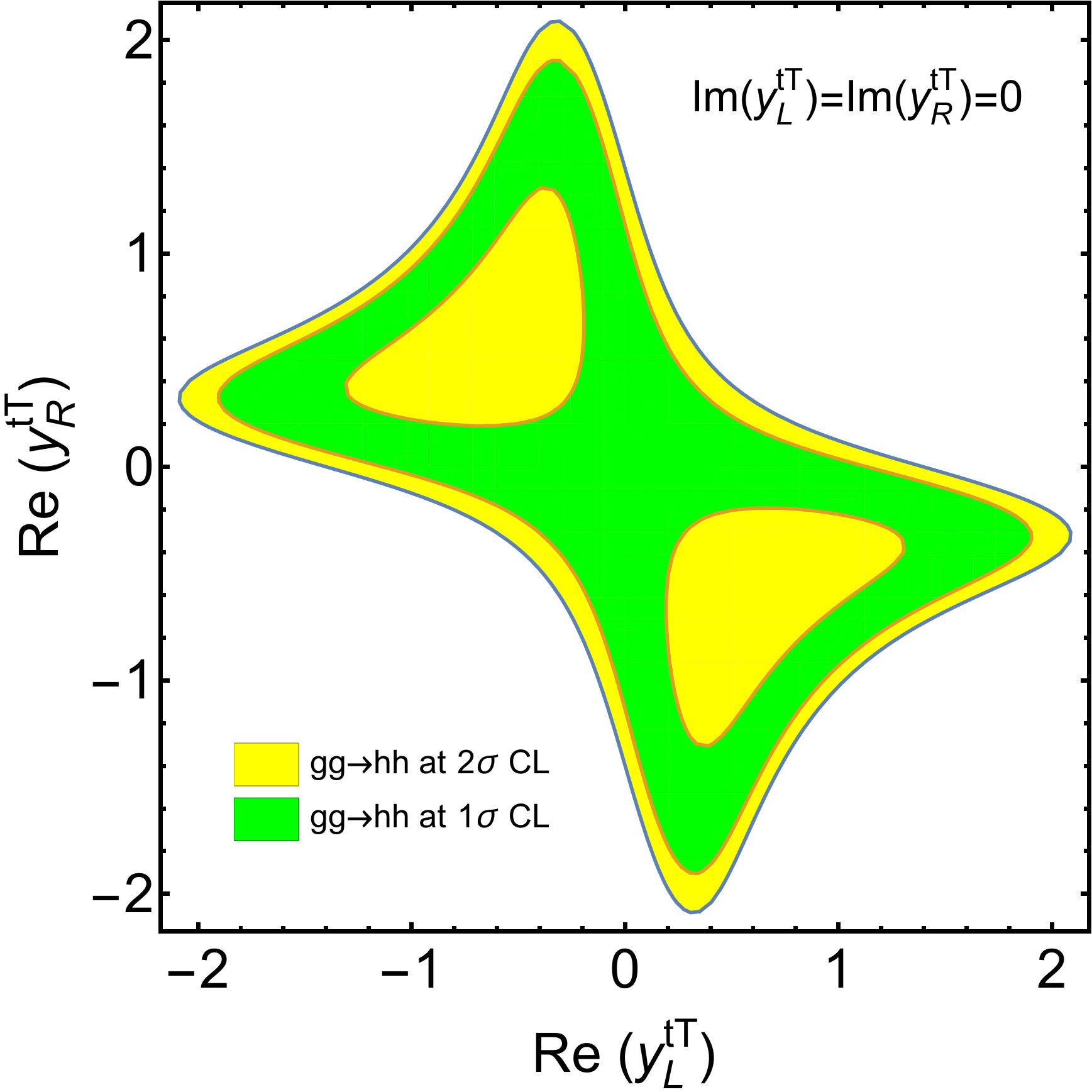}
\includegraphics[scale=0.28]{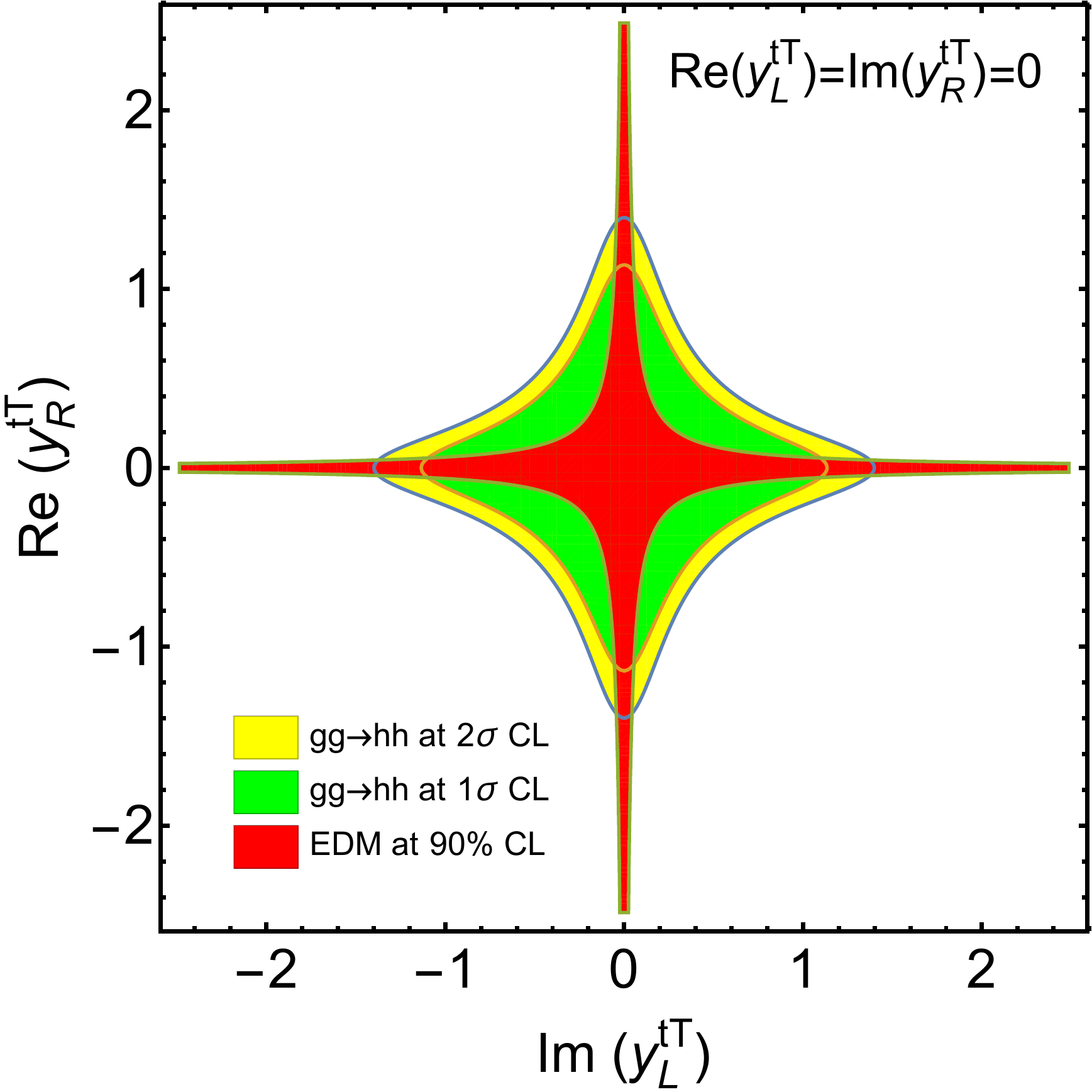}
\includegraphics[scale=0.28]{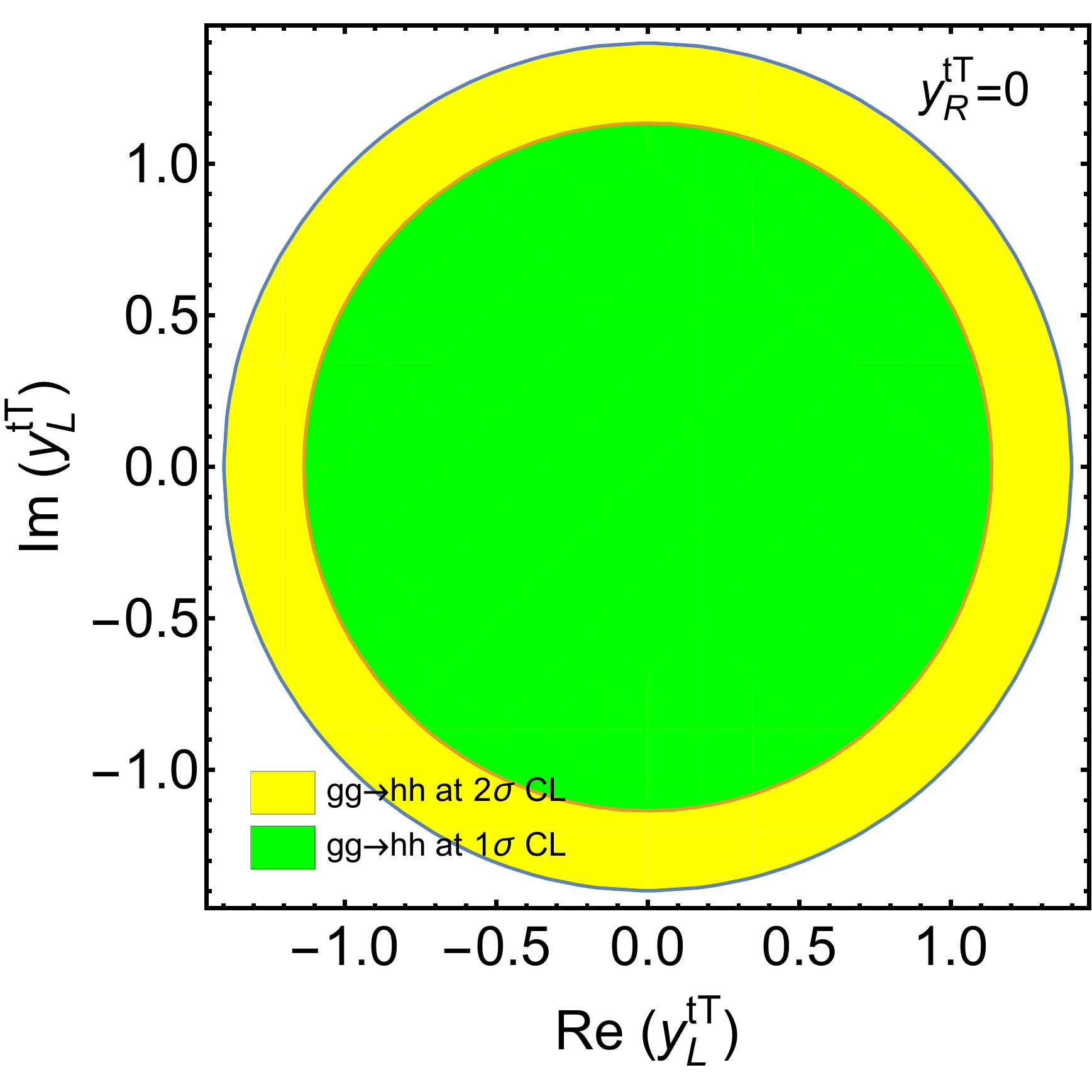}\\
\includegraphics[scale=0.28]{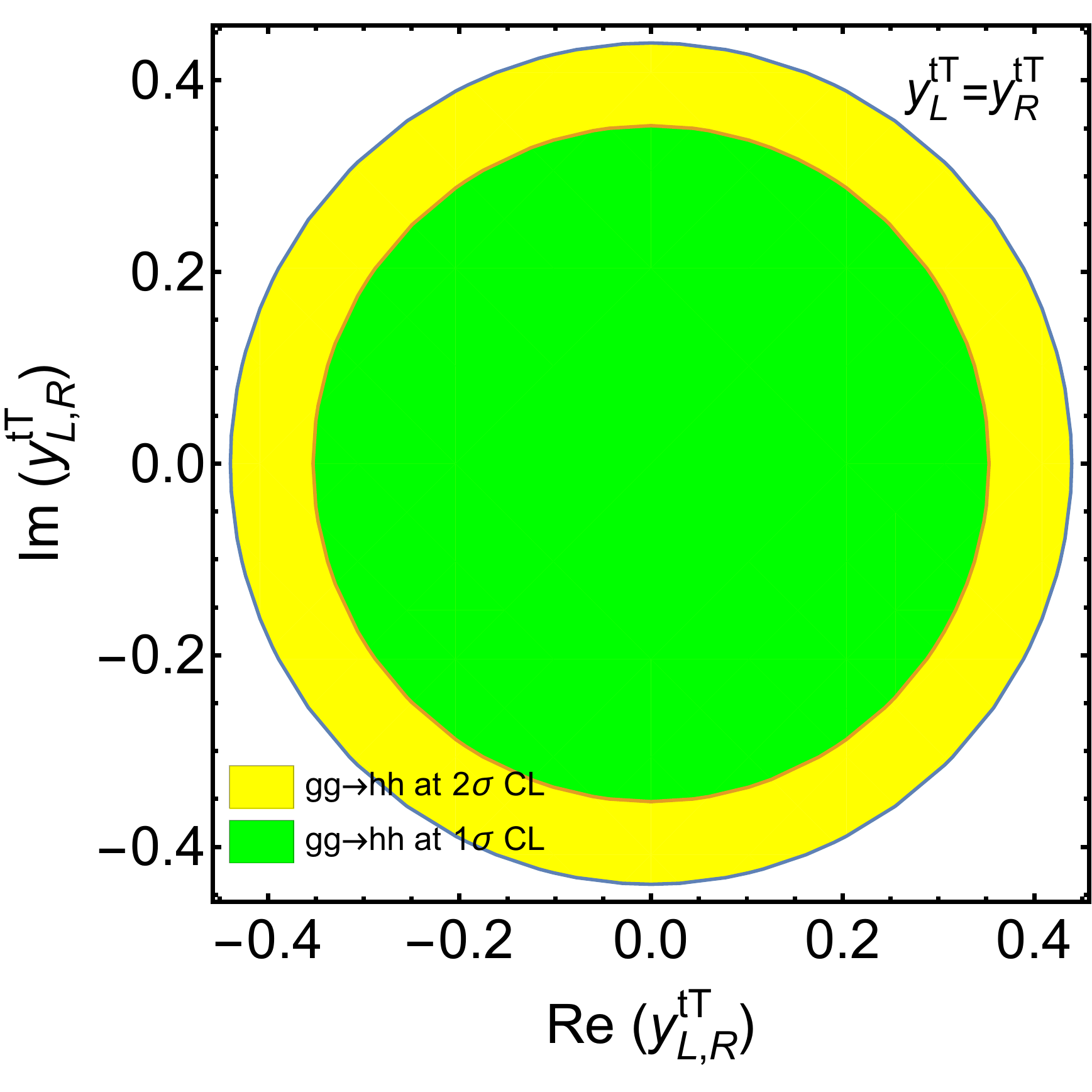}
\includegraphics[scale=0.28]{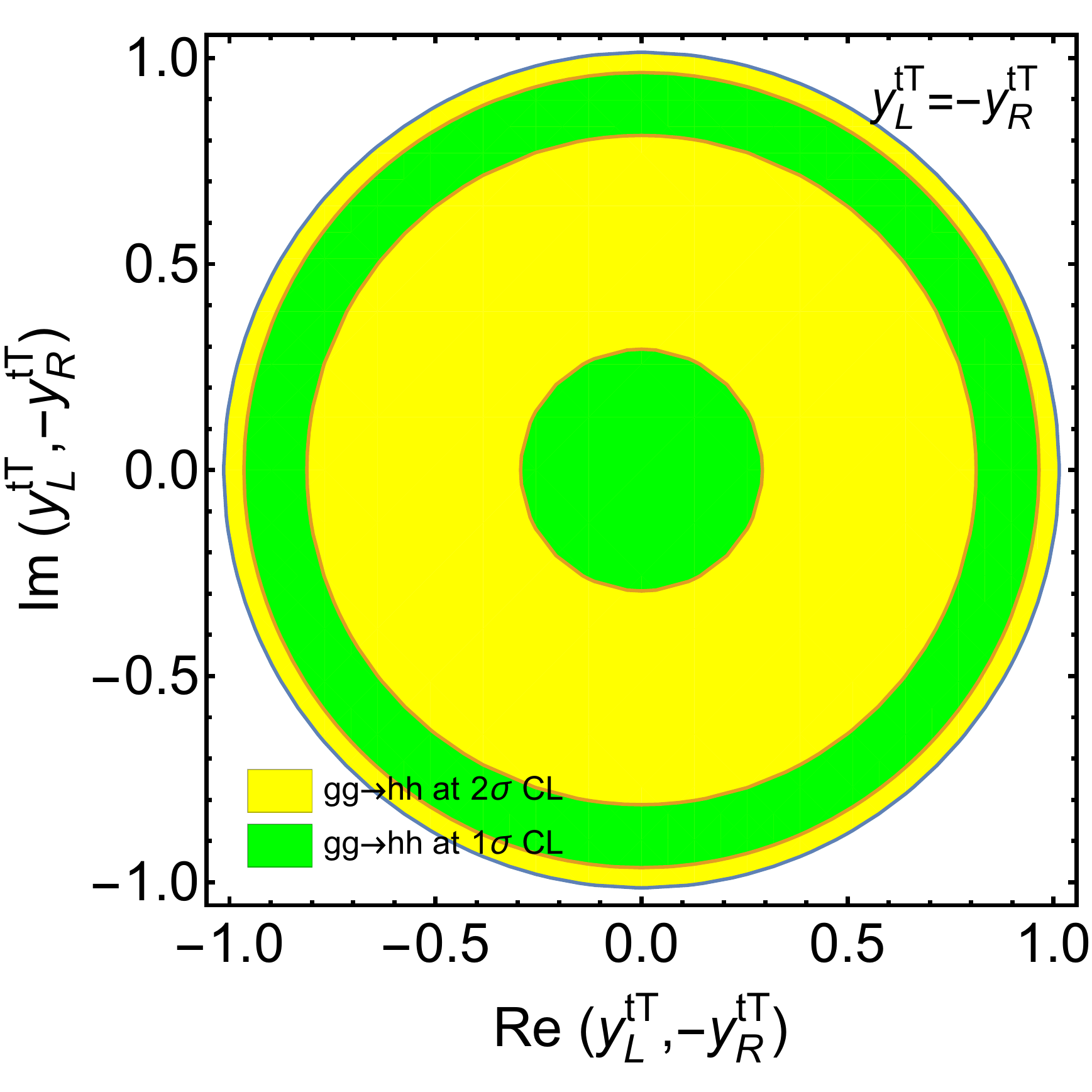}
\includegraphics[scale=0.28]{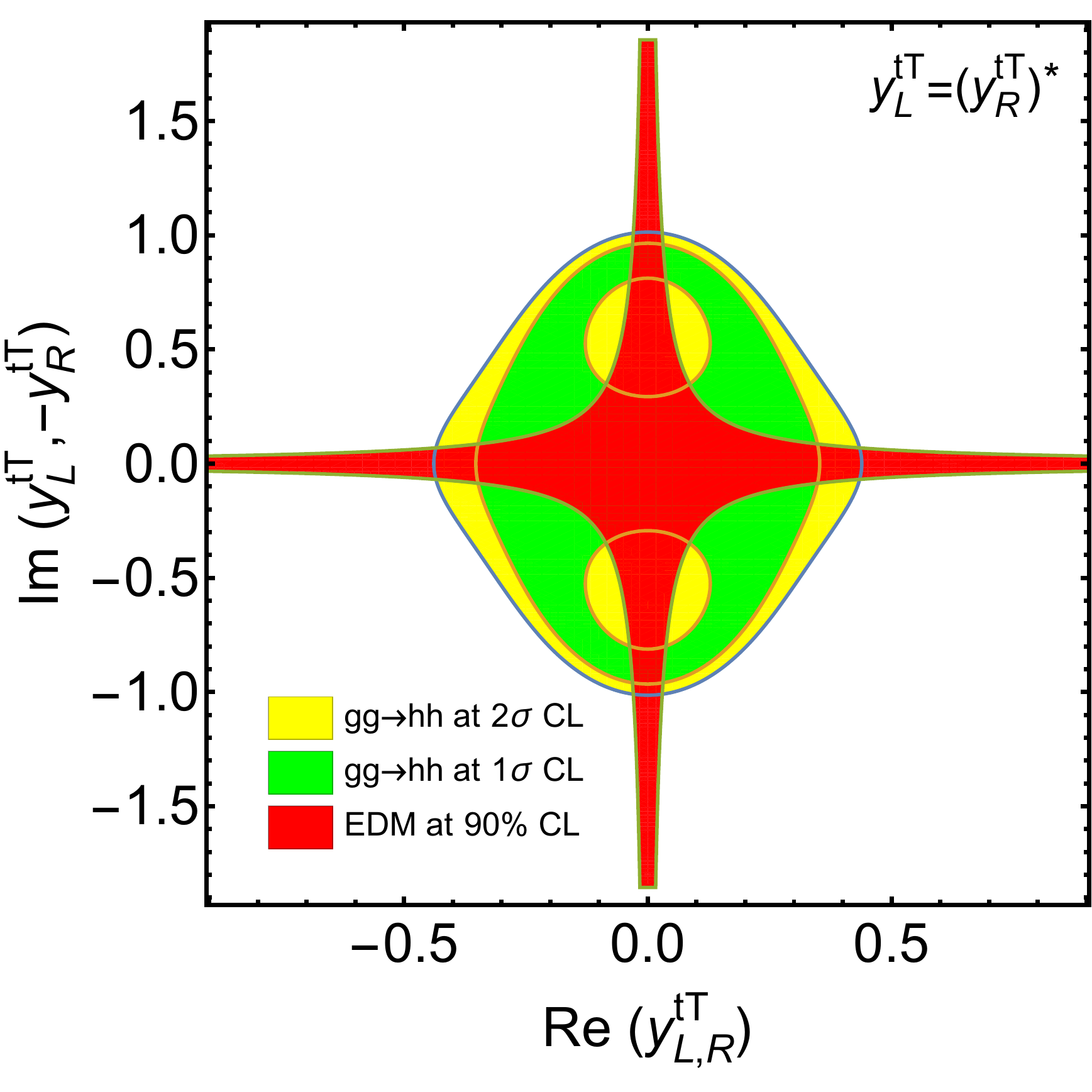}
\caption{The reach regions of $y_L^{tT},y_R^{tT}$ at HL-LHC with $\delta_{hhh}=0$ for the case of $m_T$ = 400 GeV and $s_L=0.2$. In the above plots, we take $\mathrm{Im}(y_L^{tT})=\mathrm{Im}(y_R^{tT})=0$ (upper left), $\mathrm{Re}(y_L^{tT})=\mathrm{Im}(y_R^{tT})=0$ (upper central), $y_R^{tT}=0$ (upper right), $y_L^{tT}=y_R^{tT}$ (lower left), $y_L^{tT}=-y_R^{tT}$ (lower central), and $y_L^{tT}=(y_R^{tT})^*$ (lower right) respectively. We also take into account the top quark EDM constraint for the scenarios $\mathrm{Re}(y_L^{tT})=\mathrm{Im}(y_R^{tT})=0$ (upper central) and $y_L^{tT}=(y_R^{tT})^*$ (lower right), where the reach regions of $y_L^{tT},y_R^{tT}$ are shown in red at $90\%$ CL.}\label{fig:ytT:case1:dhhh0}
\end{center}
\end{figure}

\begin{figure}[!h]
\begin{center}
\includegraphics[scale=0.28]{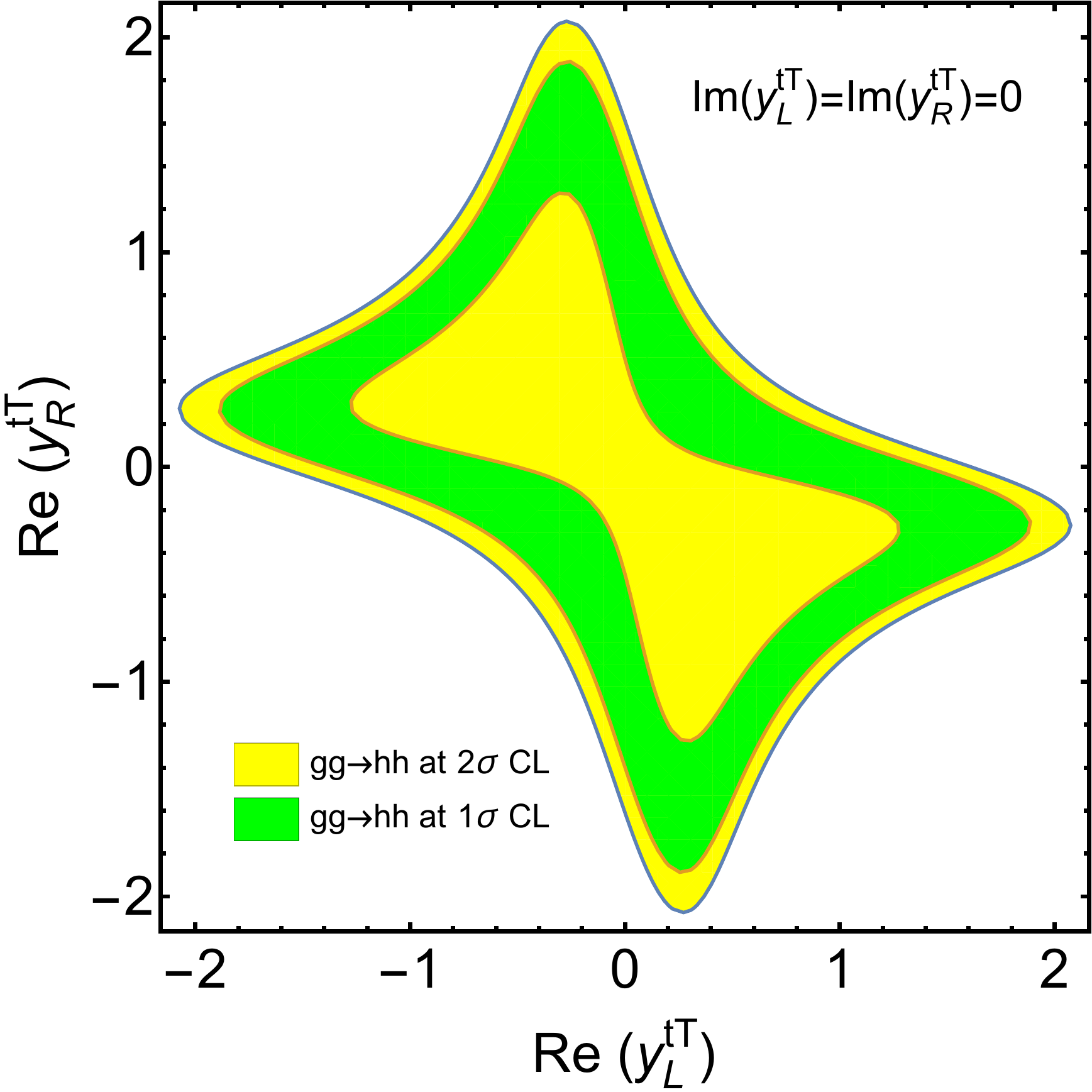}
\includegraphics[scale=0.28]{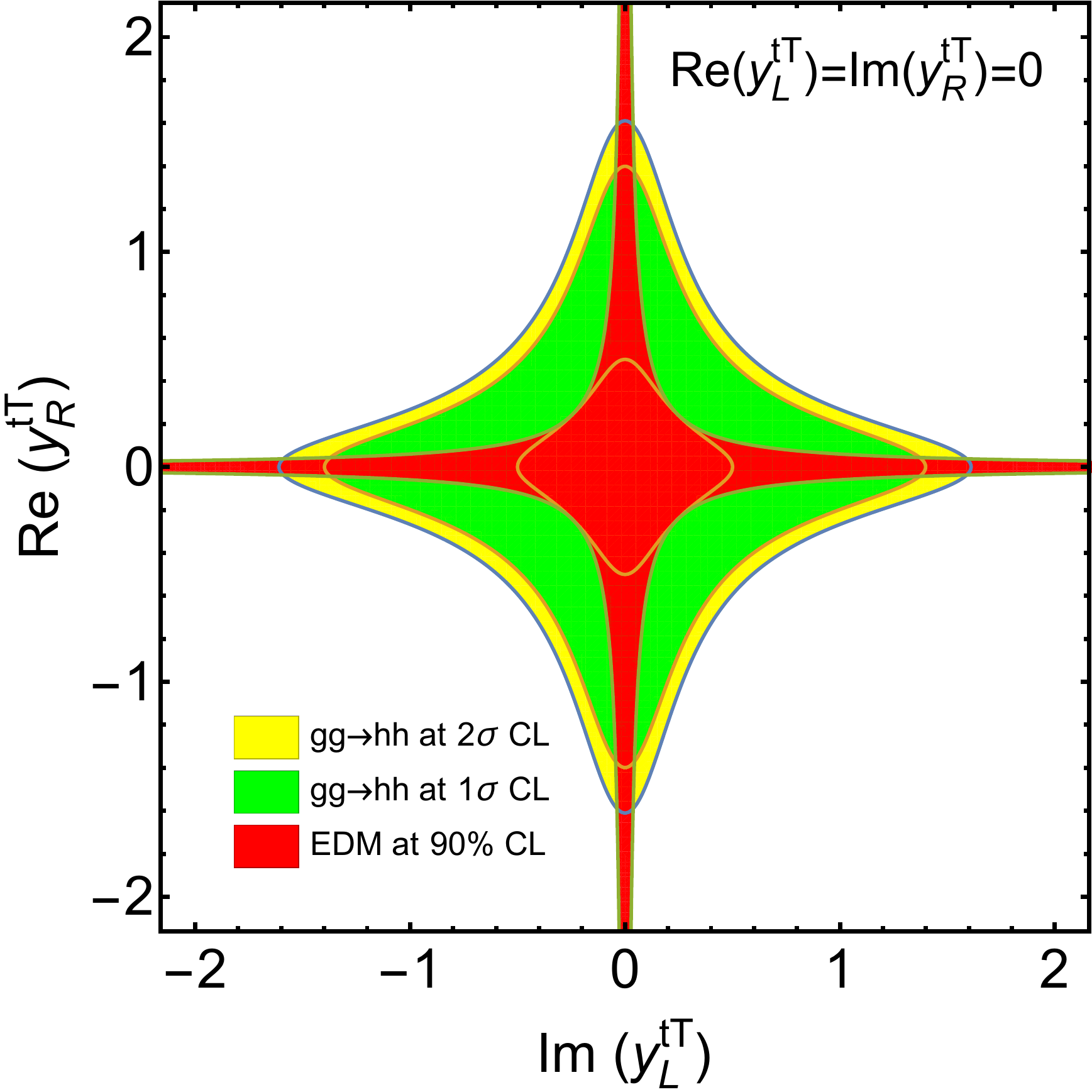}
\includegraphics[scale=0.28]{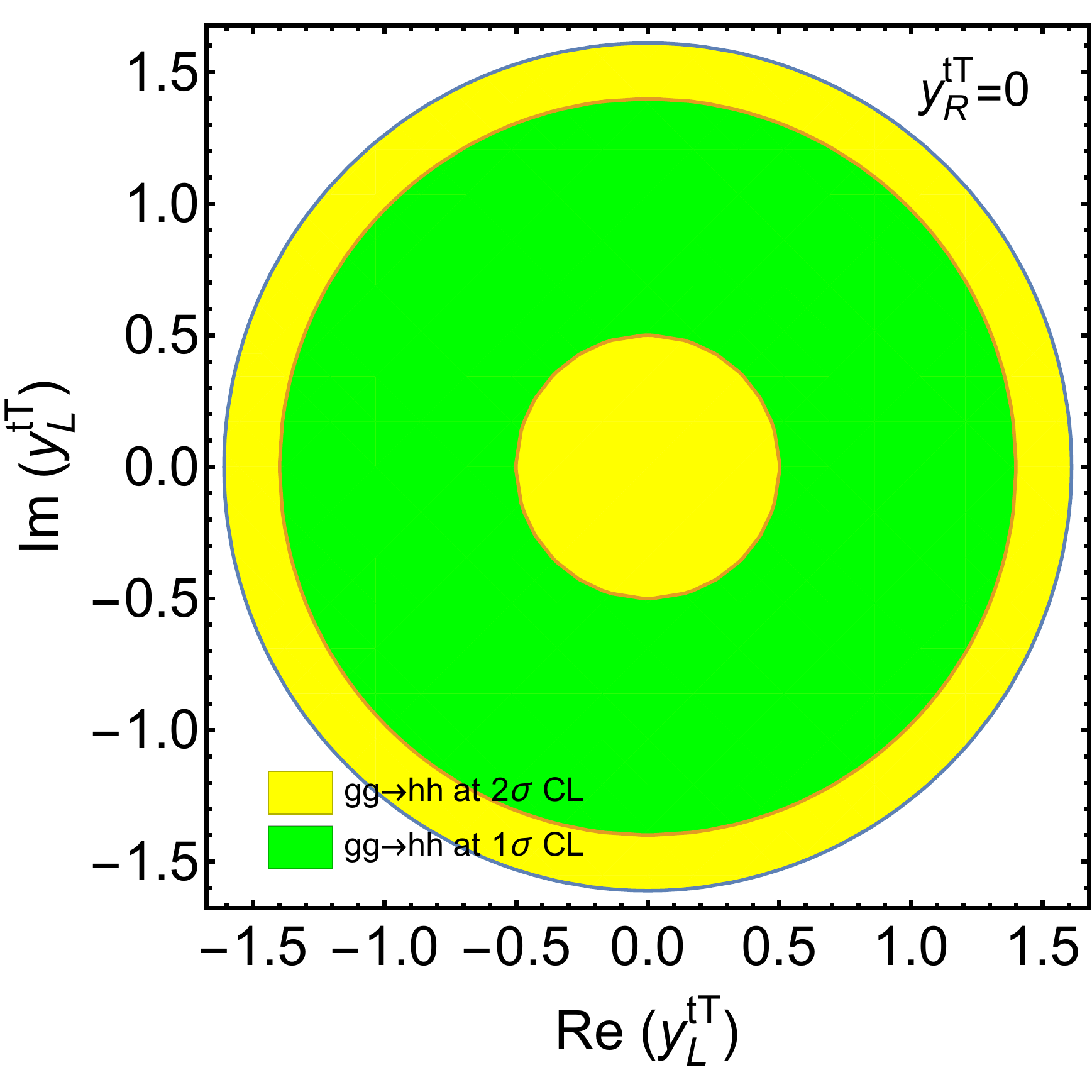}\\
\includegraphics[scale=0.28]{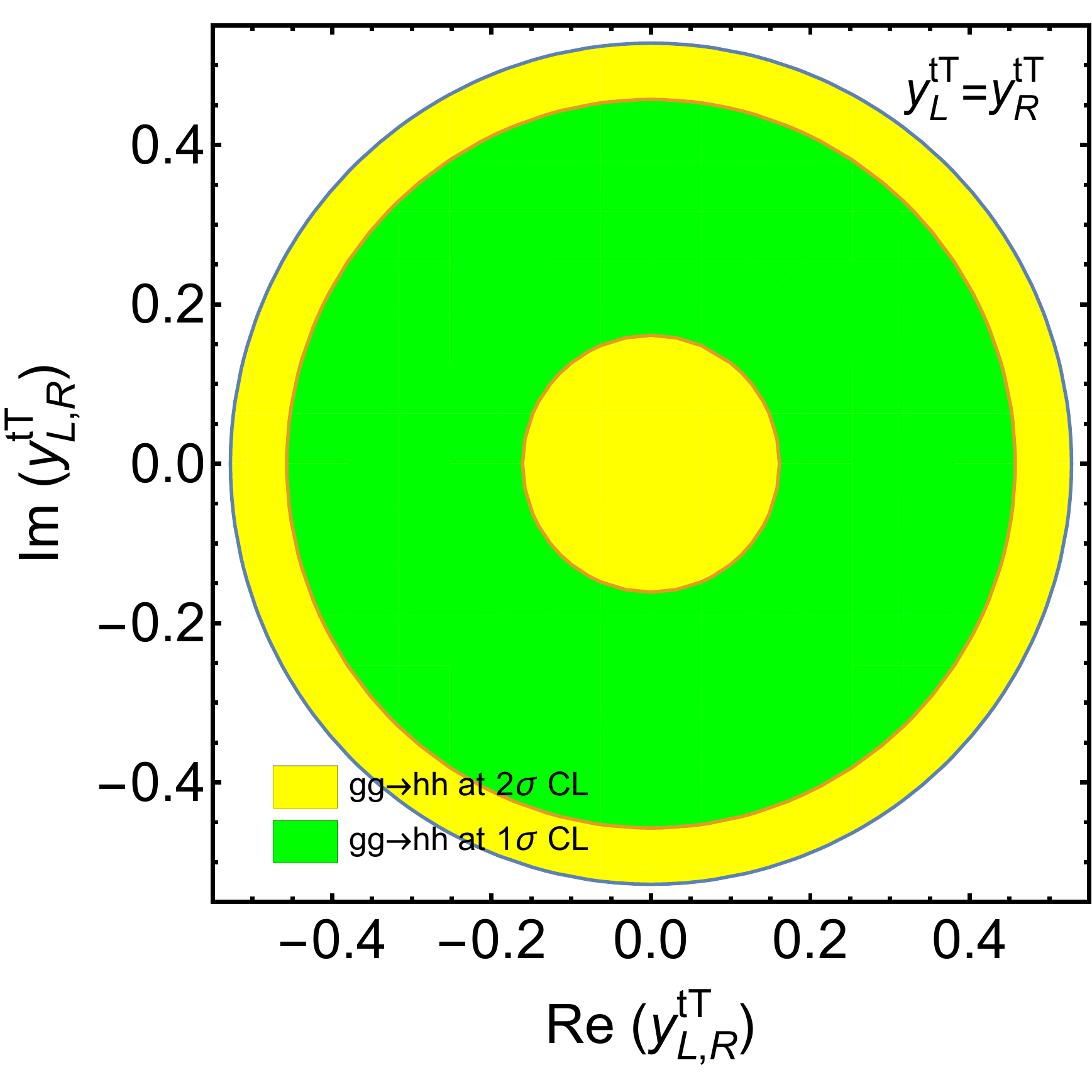}
\includegraphics[scale=0.28]{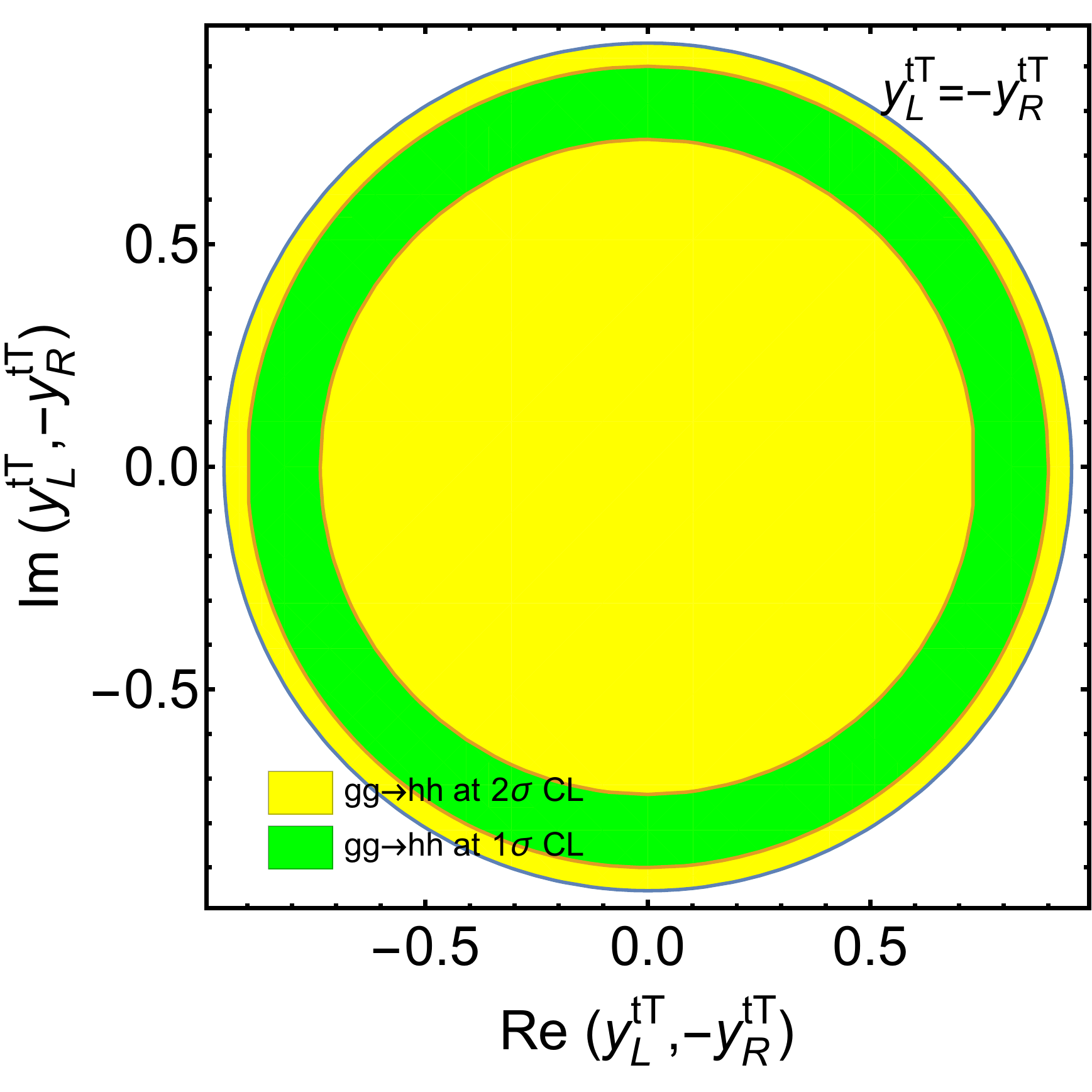}
\includegraphics[scale=0.28]{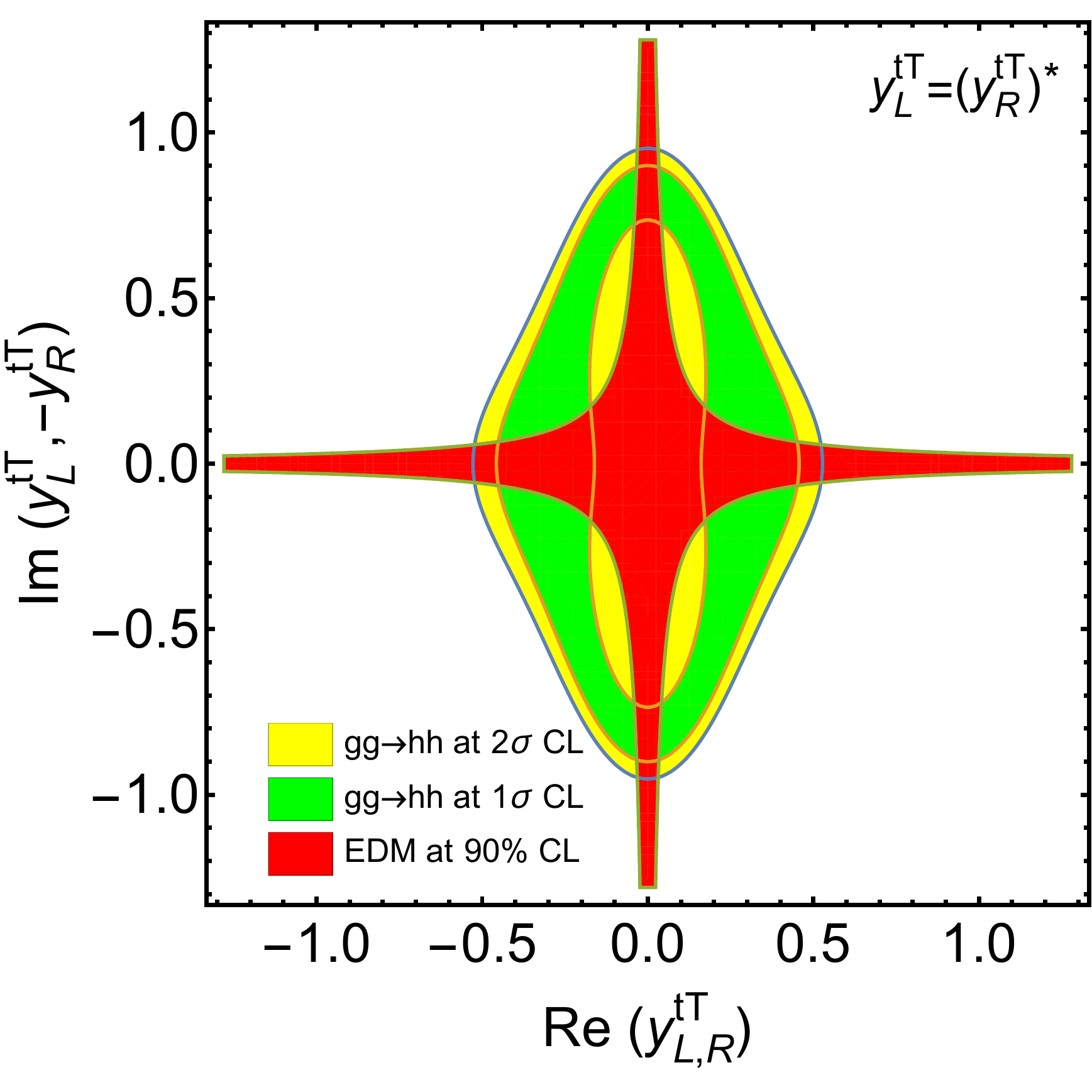}
\caption{The reach regions of $y_L^{tT},y_R^{tT}$ at HL-LHC with $\delta_{hhh}=0.5$ for the case of $m_T$ = 400 GeV and $s_L=0.2$. In the above plots, we take $\mathrm{Im}(y_L^{tT})=\mathrm{Im}(y_R^{tT})=0$ (upper left), $\mathrm{Re}(y_L^{tT})=\mathrm{Im}(y_R^{tT})=0$ (upper central), $y_R^{tT}=0$ (upper right), $y_L^{tT}=y_R^{tT}$ (lower left), $y_L^{tT}=-y_R^{tT}$ (lower central), and $y_L^{tT}=(y_R^{tT})^*$ (lower right) respectively. We also take into account the top quark EDM constraint for the scenarios $\mathrm{Re}(y_L^{tT})=\mathrm{Im}(y_R^{tT})=0$ (upper central) and $y_L^{tT}=(y_R^{tT})^*$ (lower right), where the reach regions of $y_L^{tT},y_R^{tT}$ are shown in red at $90\%$ CL.}\label{fig:ytT:case1:dhhh0d5}
\end{center}
\end{figure}

In Fig.~\ref{fig:ytT:case1:dhhh0} and Fig.~\ref{fig:ytT:case1:dhhh0d5}, we show the plots with $\delta_{hhh}=0$ and $\delta_{hhh}=0.5$, respectively. From these plots, we find that $\mathrm{Re}(y_{L,R}^{tT})$ and $\mathrm{Im}(y_{L,R}^{tT})$ are constrained to be in the range $(-2,~2)$ roughly at $2~\sigma$ CL. In some of these scenarios, the $2~\sigma$ interval can be tight as  $(-0.4,~0.4)$. The reach regions of $1~\sigma$ and $2~\sigma$ are quite different. The value of $\delta_{hhh}$ has significant effects on the extraction of $y_{L,R}^{tT}$. For the first (upper left) plot, the first and third quadrants are more constrained. This can be understood from the Eq.~\eqref{eqn:xs:ff}, because there is constructive interference between the positive $[y_L^{tT}(y_R^{tT})^*+y_R^{tT}(y_L^{tT})^*]$ term and other box diagram induced terms. While it is destructive interference if $[y_L^{tT}(y_R^{tT})^*+y_R^{tT}(y_L^{tT})^*]$ is negative. The last five plots are symmetric with respect to the horizontal and vertical axes. For the second (upper central) and sixth (lower right) plots, they receive the contributions from the $[y_L^{tT}(y_R^{tT})^*-y_R^{tT}(y_L^{tT})^*]$ term. Because $|f_C|^2$ in Eq.~\eqref{eqn:xs:partonic} is always positive, the constraints are stronger. For the fourth (lower left) plot, positive $[y_L^{tT}(y_R^{tT})^*+y_R^{tT}(y_L^{tT})^*]$ induces the constructive interference, thus the bounds are also stronger. For the third (upper right) plot, both $[y_L^{tT}(y_R^{tT})^*+y_R^{tT}(y_L^{tT})^*]$ and $[y_L^{tT}(y_R^{tT})^*-y_R^{tT}(y_L^{tT})^*]$ vanish, thus it is less constrained compared to other plots. Besides, there can be more cancellation between the triangle and box diagrams for larger $\delta_{hhh}$. Thus the constraints are usually looser than the zero $\delta_{hhh}$ ones.

In fact, we find that the di-Higgs production at HL-LHC can give stronger constraints than those from perturbative unitarity and $h\rightarrow\gamma Z$ decay. When we take into account the top quark EDM bound, the two scenarios $\mathrm{Re}(y_L^{tT})=\mathrm{Im}(y_R^{tT})=0$ and $y_L^{tT}=(y_R^{tT})^*$ can also be constrained. For the other scenarios $y_R^{tT}=0,y_L^{tT}=\pm y_R^{tT},\mathrm{Im}y_L^{tT}=\mathrm{Im}y_R^{tT}=0$, they are insensitive to the top quark EDM. Because we have the relation $d_t^{EDM}\sim y_R^{tT}(y_L^{tT})^*-y_L^{tT}(y_R^{tT})^*=2i(\mathrm{Re}y_L^{tT}\mathrm{Im}y_R^{tT}-\mathrm{Re}y_R^{tT}\mathrm{Im}y_L^{tT})$. For the scenarios $\mathrm{Re}(y_L^{tT})=\mathrm{Im}(y_R^{tT})=0$ and $y_L^{tT}=(y_R^{tT})^*$, we compare the bounds from di-Higgs production and top quark EDM for $\delta_{hhh}=0$ (Fig.~\ref{fig:ytT:case1:dhhh0}) and $\delta_{hhh}=0.5$ (Fig.~\ref{fig:ytT:case1:dhhh0d5}), respectively. From these plots, we can find that the off-axis regions can be strongly bounded by the top EDM, while it will lose the constraining power in the near axis regions.

\subsection{The benchmark point $m_T=$ 800 GeV and $s_L=0.1$}
For the case of $m_T=$ 800 GeV and $s_L=0.1$, the numerical results of $\mu_{hh}$ are evaluated as
\begin{align}
&\mu_{hh}=1-0.04939-0.7814~\delta_{hhh}+0.281~\delta_{hhh}^2+(0.1279-0.04494~\delta_{hhh})(|y_L^{tT}|^2+|y_R^{tT}|^2)\nonumber\\
&+(1.087-0.5731~\delta_{hhh})[y_L^{tT}(y_R^{tT})^*+y_R^{tT}(y_L^{tT})^*]+0.01943(|y_L^{tT}|^2+|y_R^{tT}|^2)^2+0.378[y_L^{tT}(y_R^{tT})^*+y_R^{tT}(y_L^{tT})^*]^2\nonumber\\
&+0.0711(|y_L^{tT}|^2+|y_R^{tT}|^2)[y_L^{tT}(y_R^{tT})^*+y_R^{tT}(y_L^{tT})^*]-0.9907[y_L^{tT}(y_R^{tT})^*-y_R^{tT}(y_L^{tT})^*]^2.
\end{align}
In this case, the present di-Higgs production experiments give the constraints $\delta_{hhh}\in(-3.42, 6.20)$ and $\mathrm{Re}y_L^{tT},\mathrm{Im}y_L^{tT},\mathrm{Re}y_R^{tT},\mathrm{Im}y_R^{tT}\in(-3.81, 3.81)$ at $95\%$ CL by setting one parameter at a time. In Tab.~\ref{tab:mT800:bound}, we give the expected constraints on the parameters $\delta_{hhh},\mathrm{Re}y_L^{tT},\mathrm{Im}y_L^{tT},\mathrm{Re}y_R^{tT},\mathrm{Im}y_R^{tT}$ at HL-LHC.
\begin{table}[!h]
\begin{center}
\begin{tabular}{c|c|c|c|c|c|c}
\hline
\multicolumn{2}{c|}{\diagbox[width=3.2cm,trim=lr]{method}{parameters}} &$\delta_{hhh}$ & $\mathrm{Re}y_L^{tT}$ & $\mathrm{Im}y_L^{tT}$ & $\mathrm{Re}y_R^{tT}$ & $\mathrm{Im}y_R^{tT}$\\
\hline
\multirow{2}{*}{individual}& $1~\sigma$ & (-0.499, 0.541)$\cup$(2.24, 3.28) & (-1.61, 1.61) & (-1.61, 1.61) & (-1.61, 1.61) & (-1.61, 1.61)\\
\cline{2-7}
 & $2~\sigma$ & (-0.852, 3.63) & (-2.04, 2.04) & (-2.04, 2.04) & (-2.04, 2.04) & (-2.04, 2.04)\\
\hline
\multirow{2}{*}{marginalized}& $1~\sigma$ & (-4.12, 3.89) & (-2.82, 2.82) & (-2.82, 2.82) & (-2.82, 2.82) & (-2.82, 2.82)\\
\cline{2-7}
 & $2~\sigma$ & (-4.80, 4.51) & (-3.05, 3.05) & (-3.05, 3.05) & (-3.05, 3.05) & (-3.05, 3.05)\\
\hline
\end{tabular}
\caption{The expected $1\sigma$ and $2\sigma$ bounds at HL-LHC for the parameters $\delta_{hhh},\mathrm{Re}y_L^{tT},\mathrm{Im}y_L^{tT},\mathrm{Re}y_R^{tT},\mathrm{Im}y_R^{tT}$ under the benchmark point $m_T=$ 800 GeV and $s_L=0.1$. Here we adopt two different methods: (1) turn on one parameter at a time, namely the individual method; (2) turn on all the five parameters, namely the marginalized method.}\label{tab:mT800:bound}
\end{center}
\end{table}

\begin{figure}[!h]
\begin{center}
\includegraphics[scale=0.28]{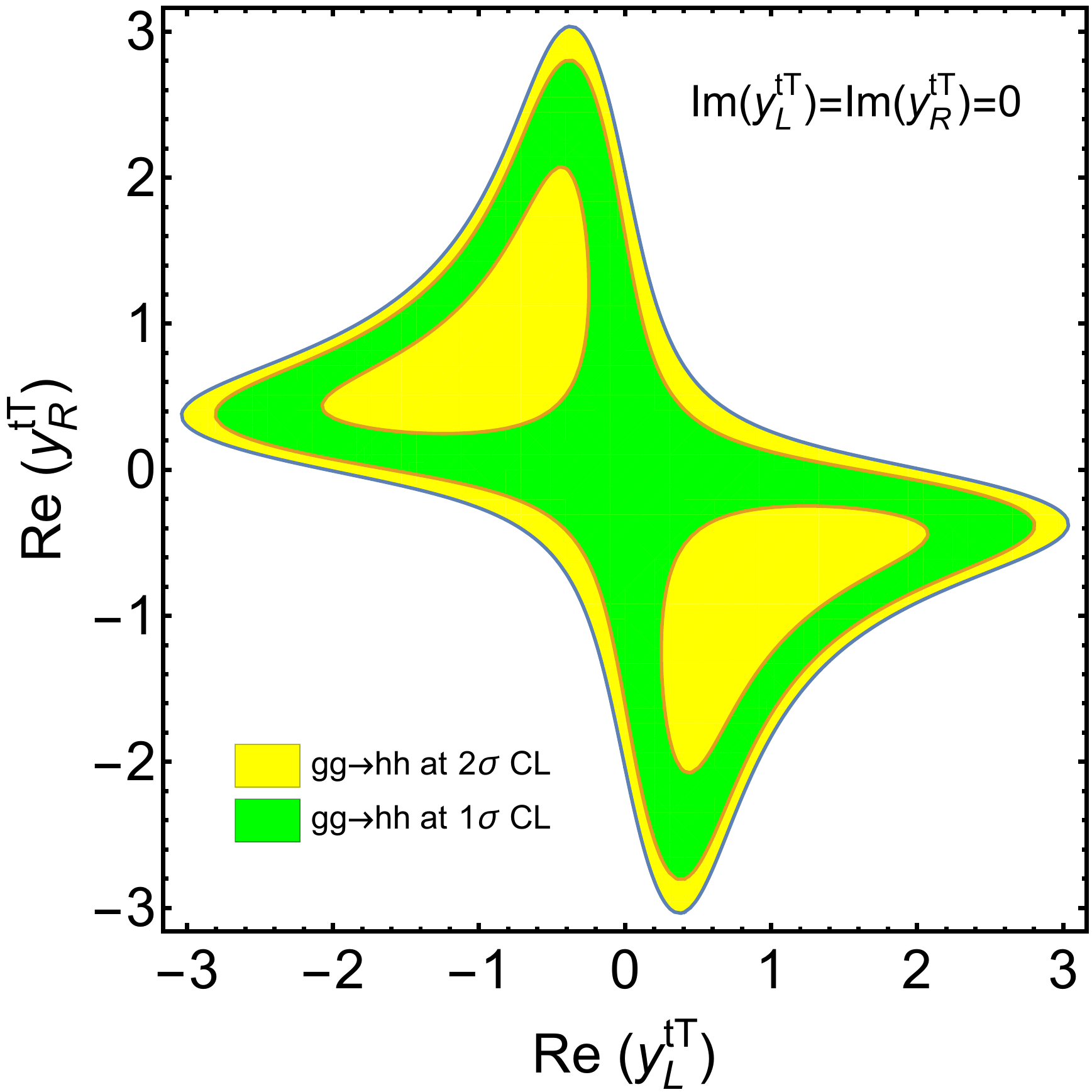}
\includegraphics[scale=0.28]{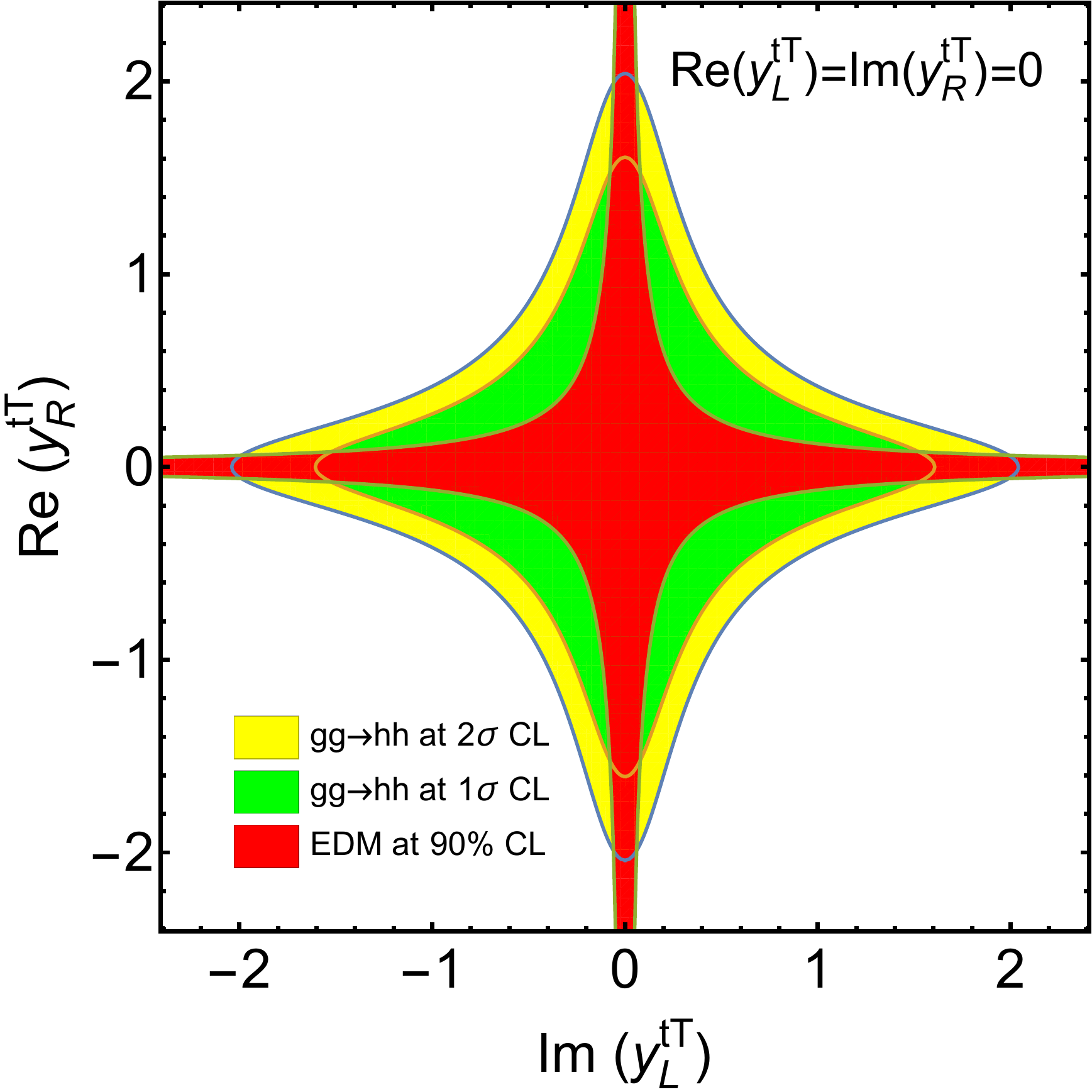}
\includegraphics[scale=0.28]{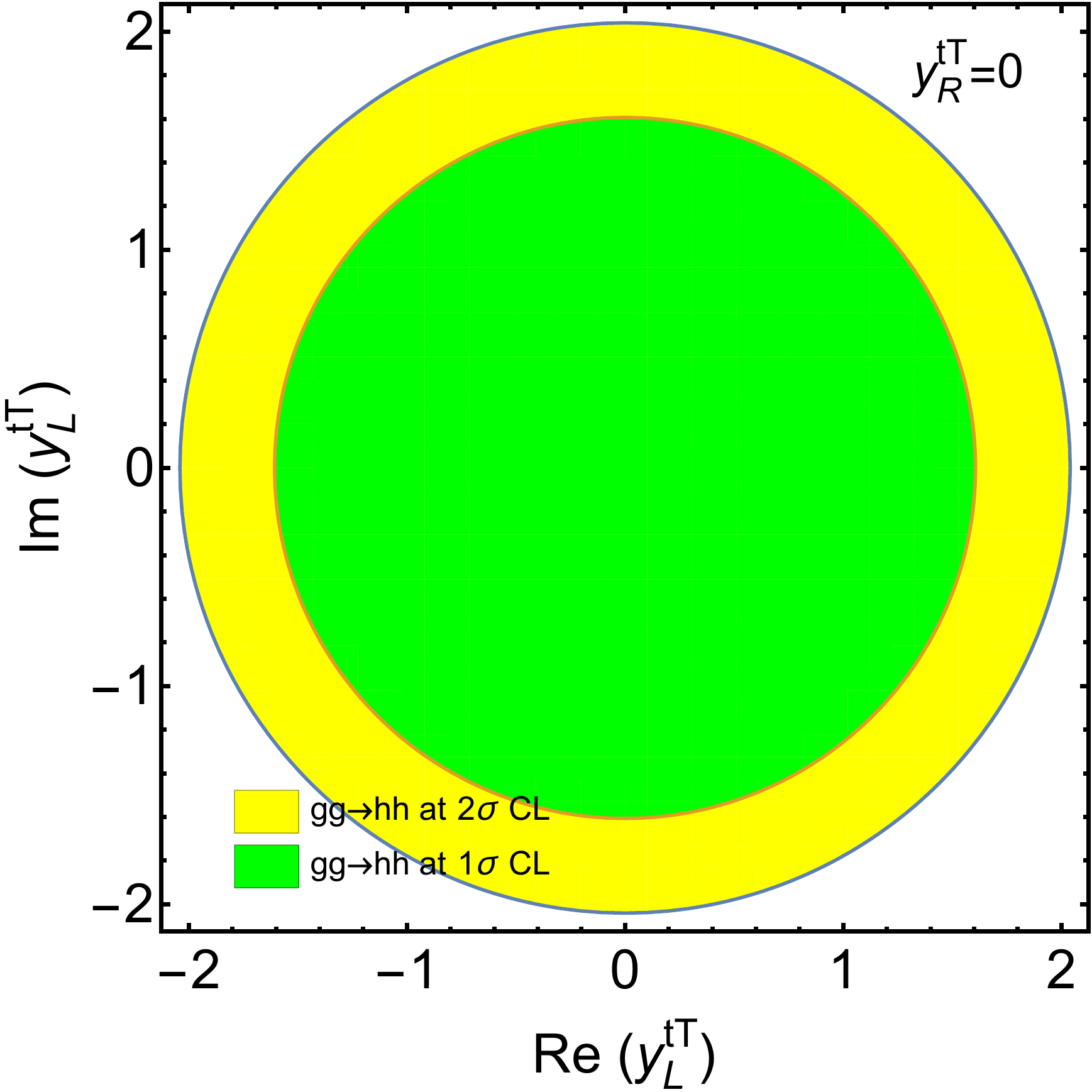}\\
\includegraphics[scale=0.28]{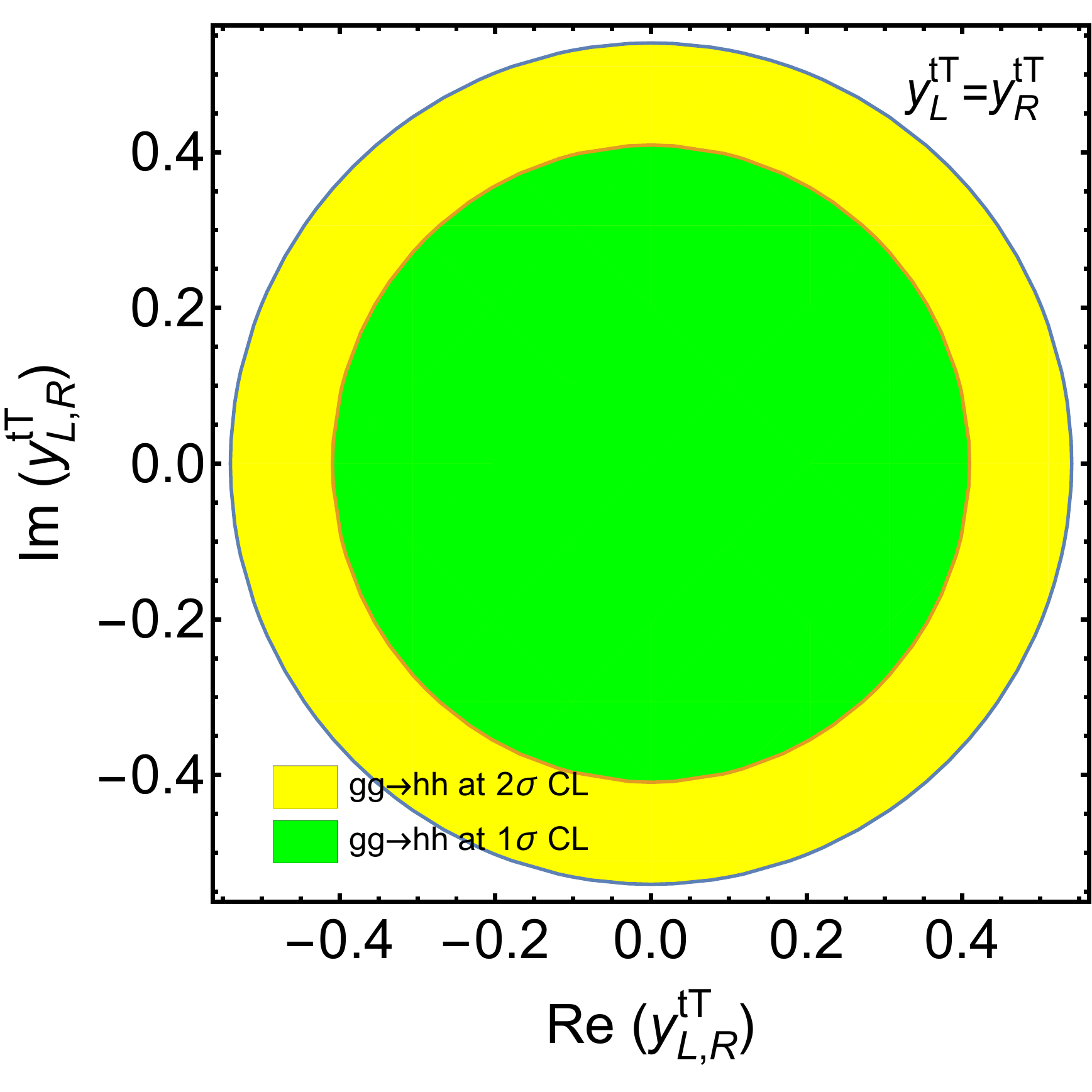}
\includegraphics[scale=0.28]{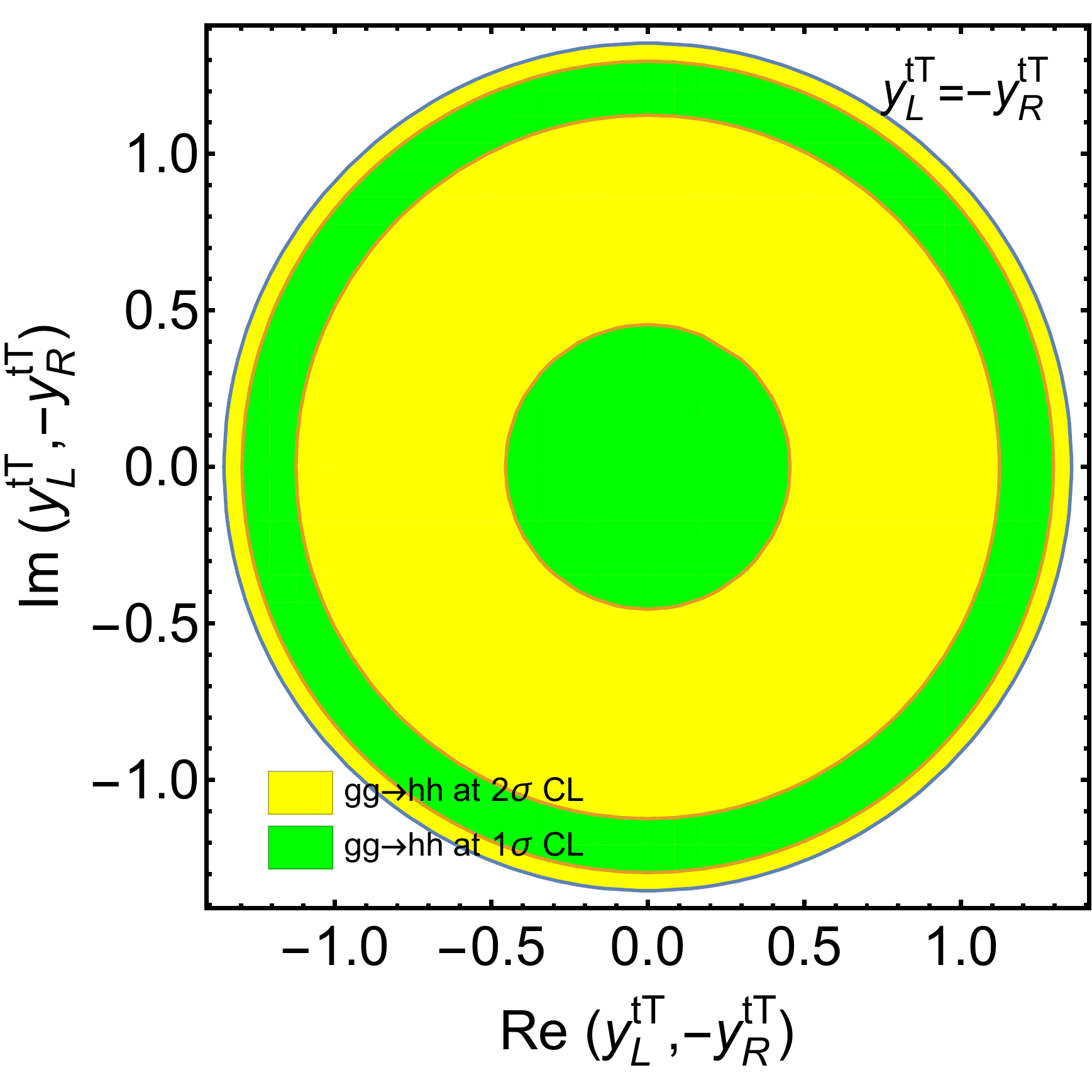}
\includegraphics[scale=0.28]{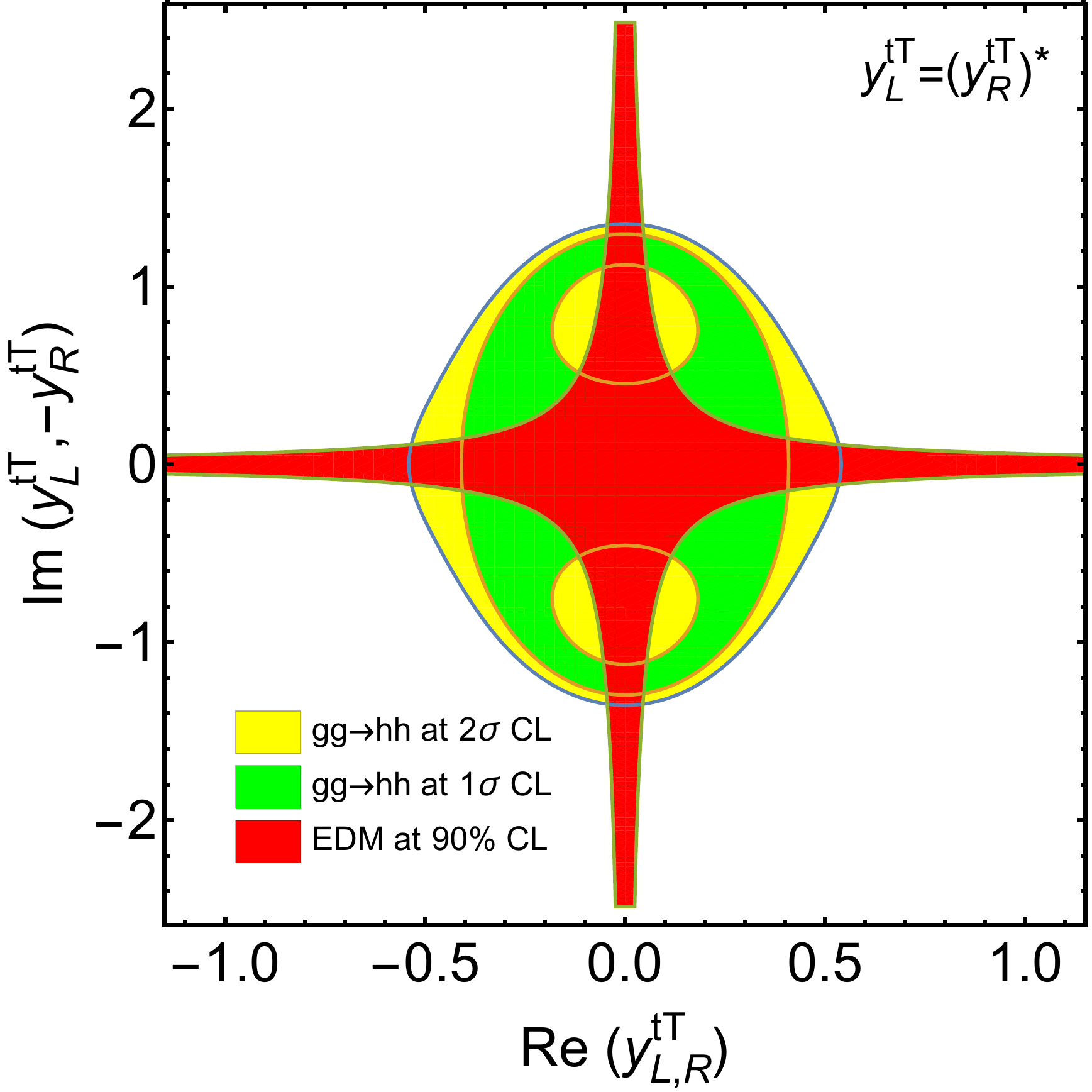}
\caption{The reach regions of $y_L^{tT},y_R^{tT}$ at HL-LHC with $\delta_{hhh}=0$ for the case of $m_T$ = 800 GeV and $s_L=0.1$. In the above plots, we take $\mathrm{Im}(y_L^{tT})=\mathrm{Im}(y_R^{tT})=0$ (upper left), $\mathrm{Re}(y_L^{tT})=\mathrm{Im}(y_R^{tT})=0$ (upper central), $y_R^{tT}=0$ (upper right), $y_L^{tT}=y_R^{tT}$ (lower left), $y_L^{tT}=-y_R^{tT}$ (lower central), and $y_L^{tT}=(y_R^{tT})^*$ (lower right) respectively. We also take into account the top quark EDM constraint for the scenarios $\mathrm{Re}(y_L^{tT})=\mathrm{Im}(y_R^{tT})=0$ (upper central) and $y_L^{tT}=(y_R^{tT})^*$ (lower right), where the reach regions of $y_L^{tT},y_R^{tT}$ are shown in red at $90\%$ CL.}\label{fig:ytT:case2:dhhh0}
\end{center}
\end{figure}

\begin{figure}[!h]
\begin{center}
\includegraphics[scale=0.28]{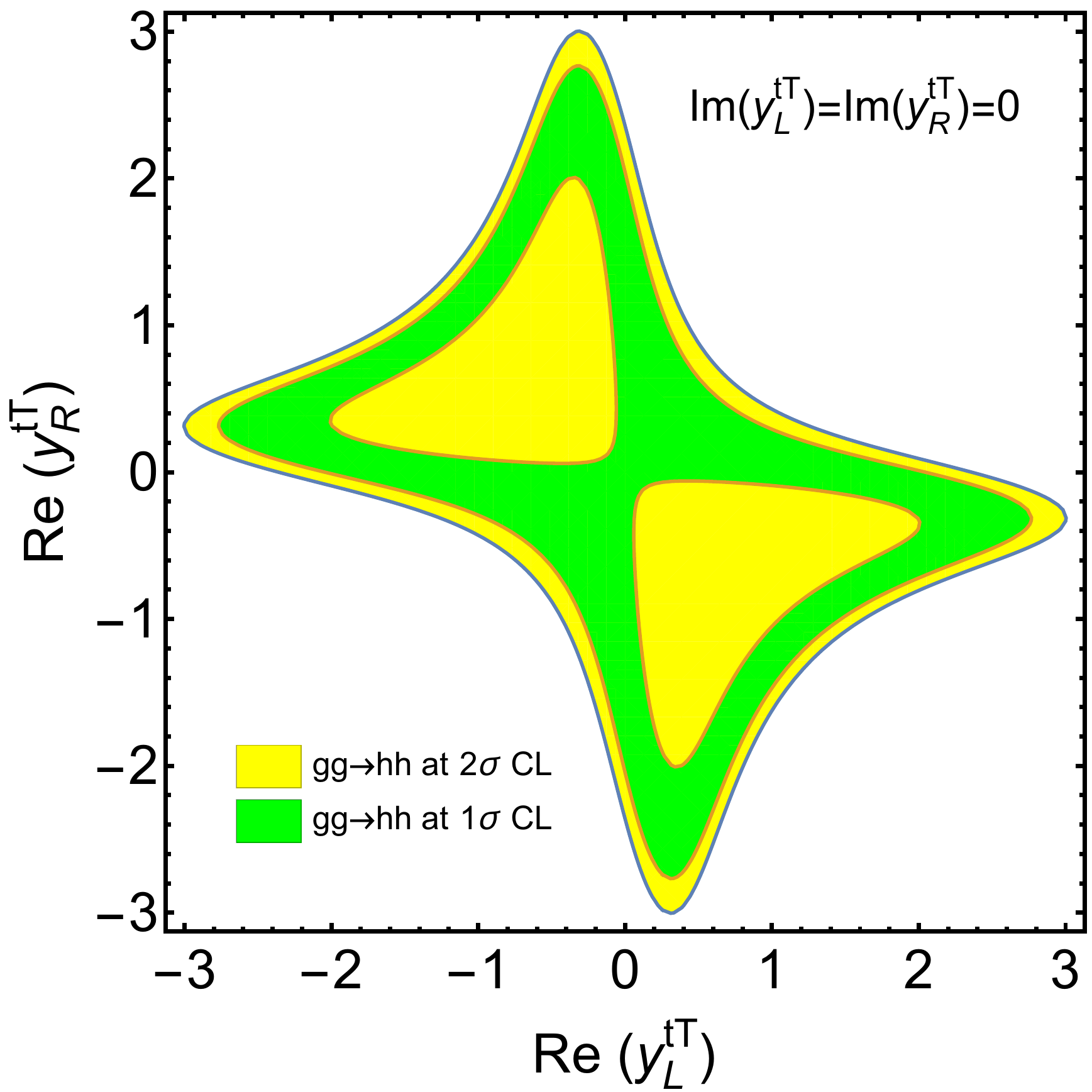}
\includegraphics[scale=0.28]{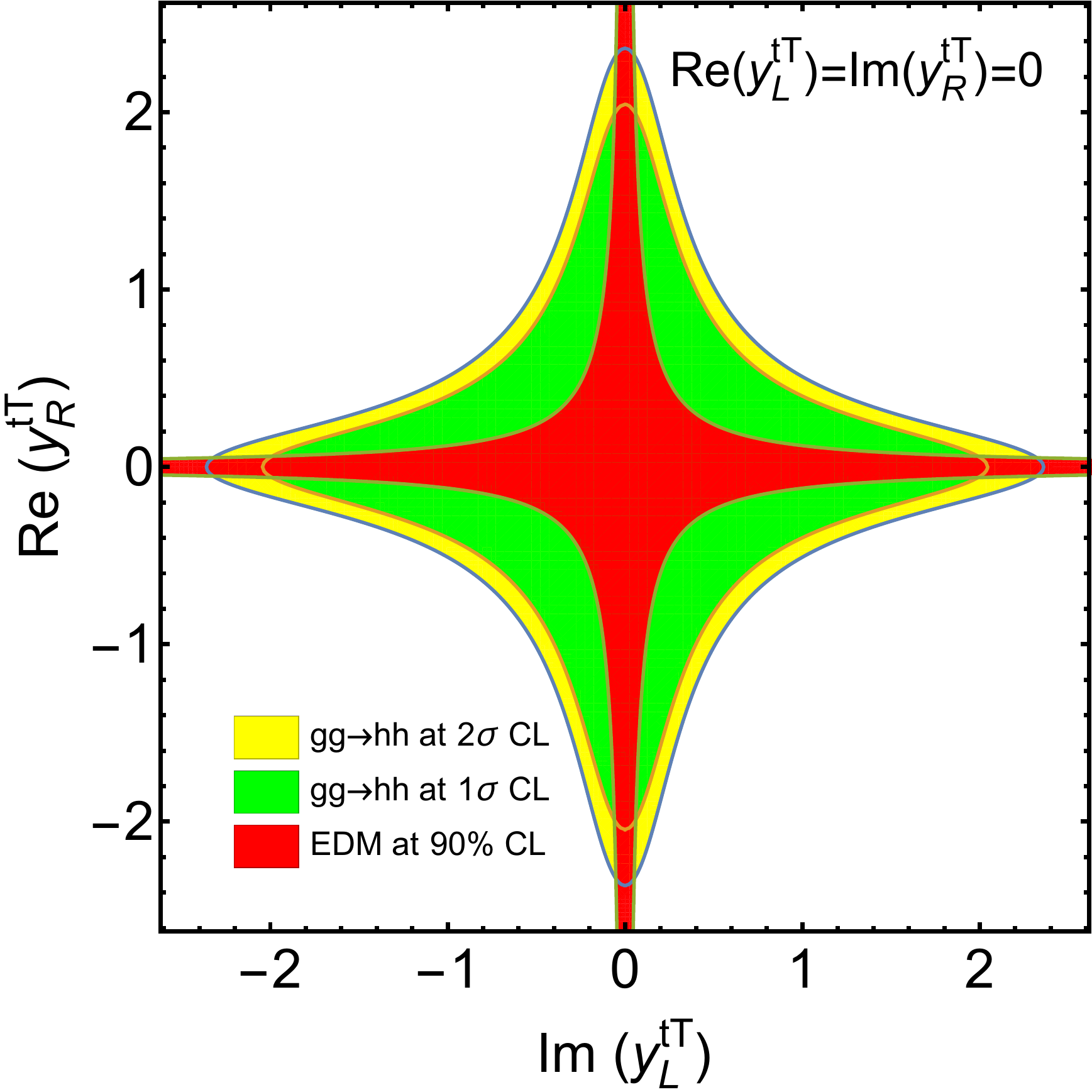}
\includegraphics[scale=0.28]{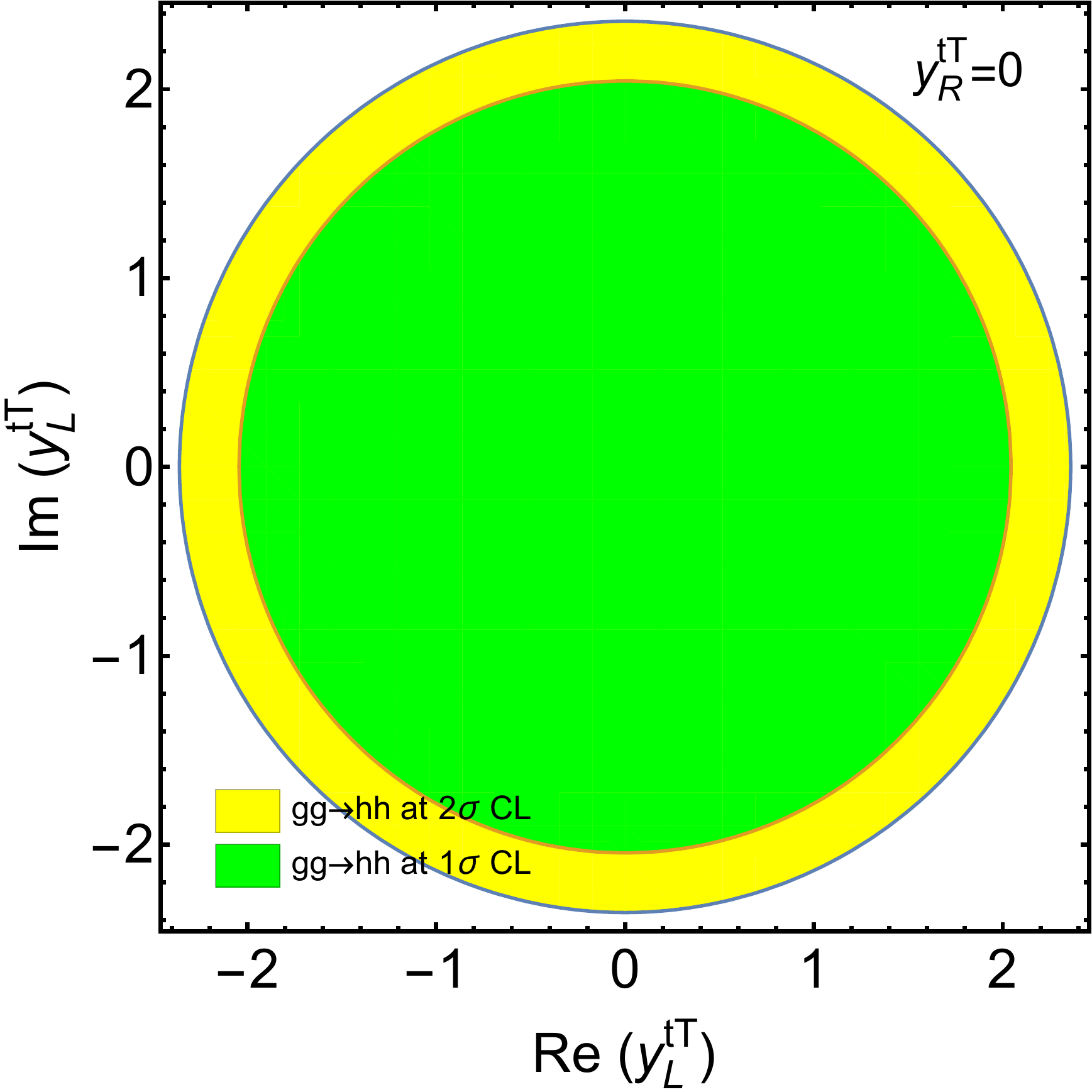}\\
\includegraphics[scale=0.28]{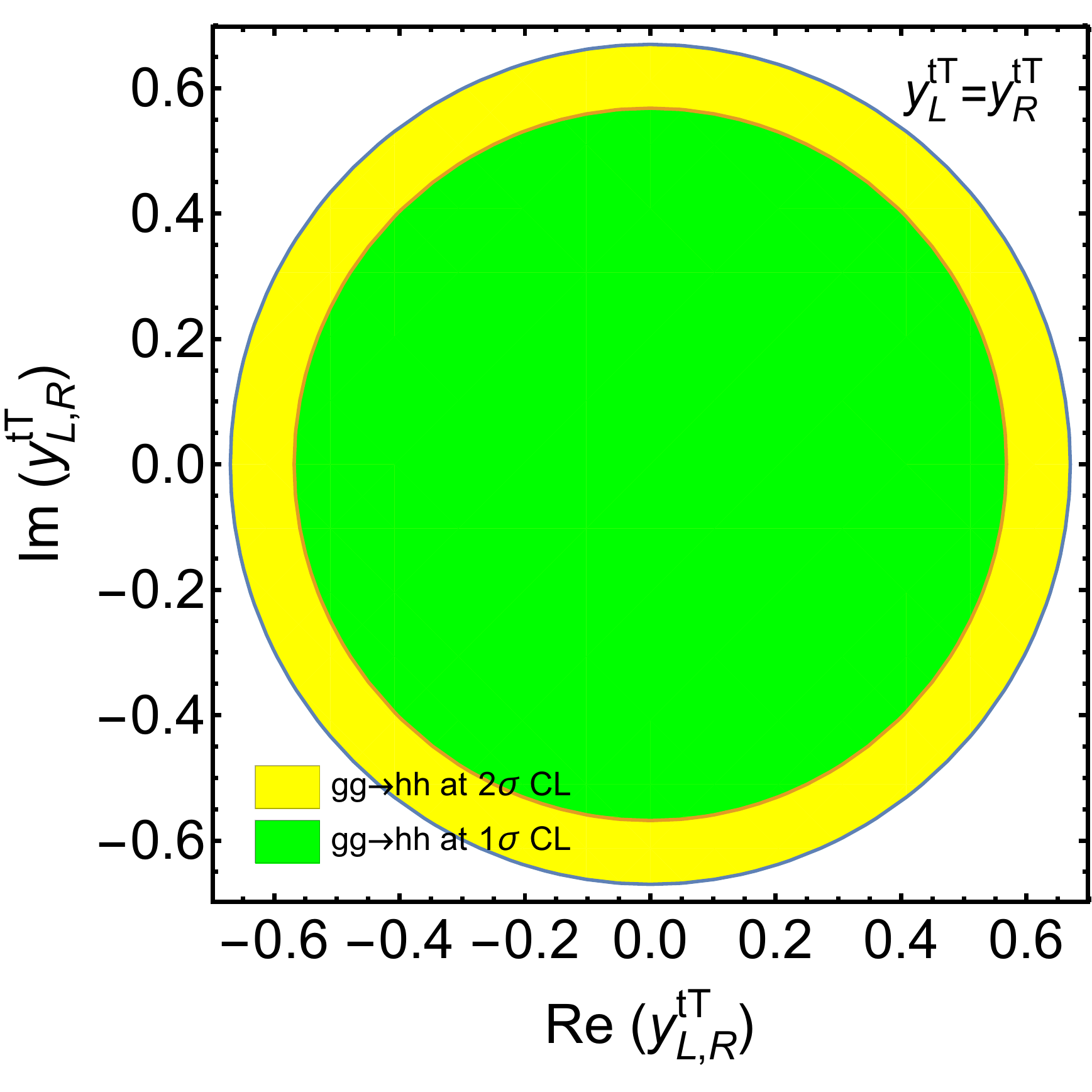}
\includegraphics[scale=0.28]{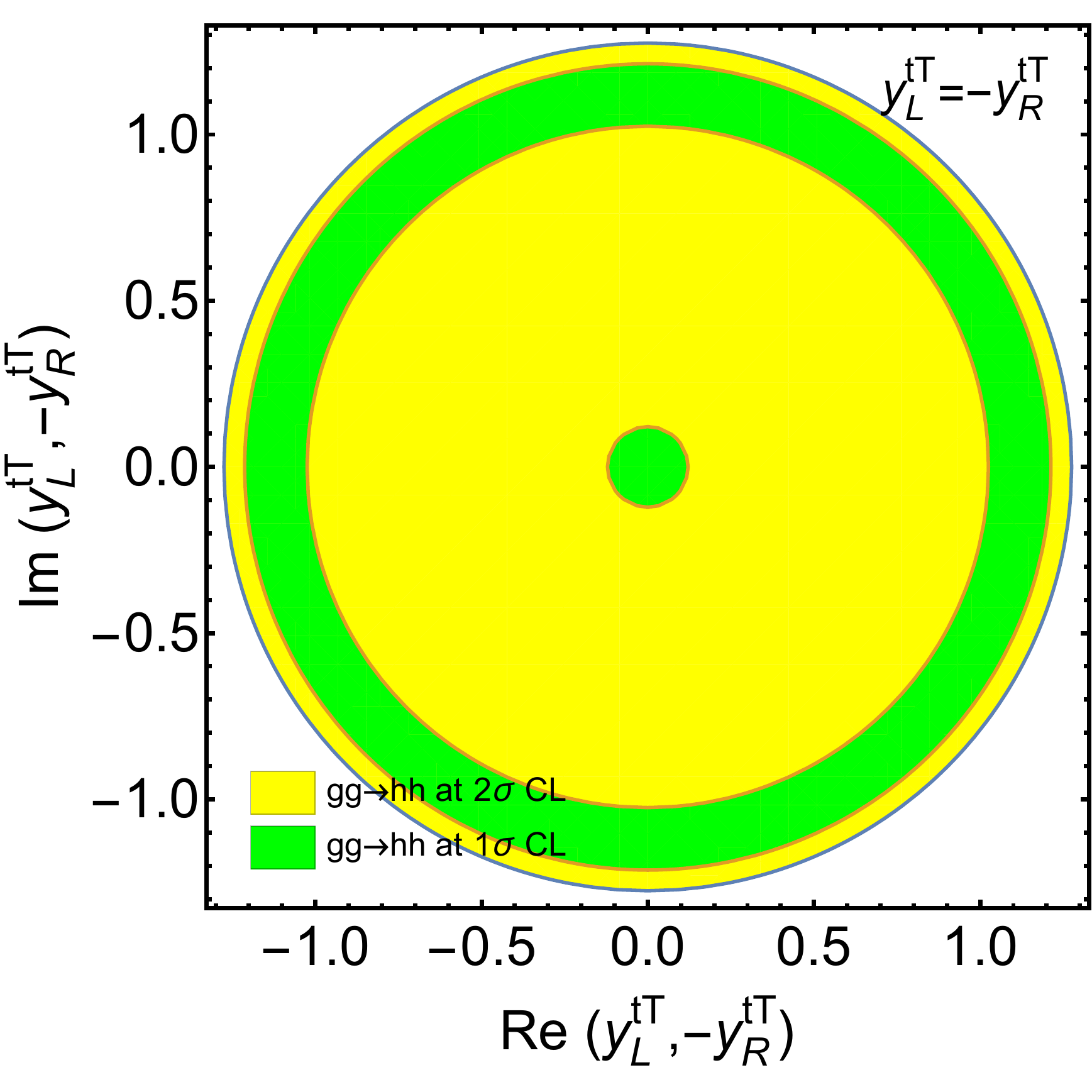}
\includegraphics[scale=0.28]{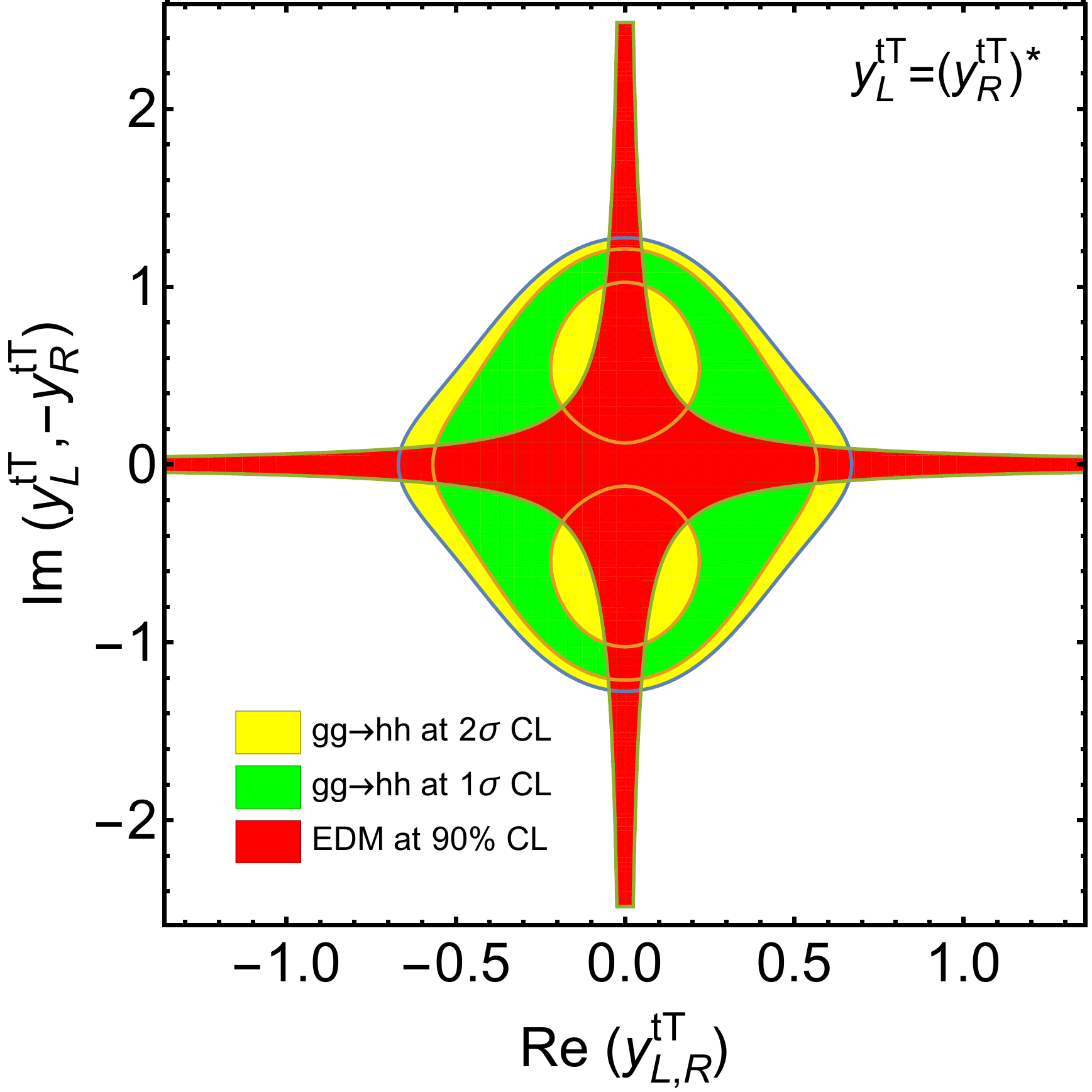}
\caption{The reach regions of $y_L^{tT},y_R^{tT}$ at HL-LHC with $\delta_{hhh}=0.5$ for the case of $m_T$ = 800 GeV and $s_L=0.1$. In the above plots, we take $\mathrm{Im}(y_L^{tT})=\mathrm{Im}(y_R^{tT})=0$ (upper left), $\mathrm{Re}(y_L^{tT})=\mathrm{Im}(y_R^{tT})=0$ (upper central), $y_R^{tT}=0$ (upper right), $y_L^{tT}=y_R^{tT}$ (lower left), $y_L^{tT}=-y_R^{tT}$ (lower central), and $y_L^{tT}=(y_R^{tT})^*$ (lower right) respectively. We also take into account the top quark EDM constraint for the scenarios $\mathrm{Re}(y_L^{tT})=\mathrm{Im}(y_R^{tT})=0$ (upper central) and $y_L^{tT}=(y_R^{tT})^*$ (lower right), where the reach regions of $y_L^{tT},y_R^{tT}$ are shown in red at $90\%$ CL.}\label{fig:ytT:case2:dhhh0d5}
\end{center}
\end{figure}

We will also plot the reached two-dimensional parameter space by setting two of them to be zero or imposing two conditions. In Fig.~\ref{fig:ytT:case2:dhhh0} and Fig.~\ref{fig:ytT:case2:dhhh0d5}, similar plots are presented for the six scenarios with $\delta_{hhh}=0$ and $\delta_{hhh}=0.5$, respectively. From these plots, we find that $\mathrm{Re}(y_{L,R}^{tT})$ and $\mathrm{Im}(y_{L,R}^{tT})$ are constrained to be in the range $(-3,~3)$ roughly at $2~\sigma$ CL. In some of these scenarios, the $2~\sigma$ interval can be tight as  $(-0.5,~0.5)$. The reach regions are similar to those in the $m_T=$ 400 GeV and $s_L=0.2$ case. For the scenarios $\mathrm{Re}(y_L^{tT})=\mathrm{Im}(y_R^{tT})=0$ and $y_L^{tT}=(y_R^{tT})^*$, we also compare the bounds from di-Higgs production and top quark EDM for $\delta_{hhh}=0$ (Fig.~\ref{fig:ytT:case2:dhhh0}) and $\delta_{hhh}=0.5$ (Fig.~\ref{fig:ytT:case2:dhhh0d5}), respectively. When $m_T$ becomes larger, $y_{L,R}^{tT}$ are constrained more loosely. When $s_L$ becomes very small, the pure top quark contributions are SM-like and the pure $T$ quark contributions are highly suppressed. Thus, the main deviation of $\mu_{hh}$ is from the FCNY interactions.

By the way, the FCNY coupling may be probed through other processes too. For example, we can probe the FCNY coupling $Tth$ through direct production processes $pp\rightarrow T\bar{t}h,T\bar{t},ThW,Thj$. But they suffer from low event rate, the detailed analyses in these channels are beyond the scope of this work.
\subsection{Comments on the doublet and triplet vector-like quarks}
We have assumed $T_{L,R}$ to be singlets throughout this work, while they can be components of the doublet or triplet VLQs. There are two doublet and two triplet VLQs containing the $T$ quark: $(X,T)_{L,R},(T,B)_{L,R},(X,T,B)_{L,R},(T,B,Y)_{L,R}$. Here $X,B,Y$ carry $\frac{4}{3},-\frac{1}{3},-\frac{4}{3}$ electric charges, respectively. For the doublet $(X,T)_{L,R}$, the Higgs particle only interact with the $T_{L,R}$. For the doublet $(T,B)_{L,R}$ and triplets $(X,T,B)_{L,R},(T,B,Y)_{L,R}$, the $B$ quark can mix with the SM bottom quark. Thus, the Higgs particle will interact with both the $T$ and $B$ quarks.
Let us denote left (right) up-type and down-type quark mixing angles as $\theta_L^t$ ($\theta_R^t$) and $\theta_L^b$ ($\theta_R^b$). They can be related with each other \cite{Aguilar-Saavedra:2013qpa}.
\begin{itemize}
\item For the triplet $(X,T,B)_{L,R}$, we have the relations $\tan\theta_R^t=\frac{m_t}{m_T}\tan\theta_L^t$ and $\tan\theta_R^b=\frac{m_b}{m_B}\tan\theta_L^b$. $\theta_L^t$ and $\theta_L^b$ can be related through the identity $\sin2\theta_L^b=\sqrt{2}\frac{m_T^2-m_t^2}{m_B^2-m_b^2}\sin2\theta_L^t$. Thus, there is only one independent mixing angle $\theta_L^t$.
\item For the triplet $(T,B,Y)_{L,R}$, we have the relations $\tan\theta_R^t=\frac{m_t}{m_T}\tan\theta_L^t$ and $\tan\theta_R^b=\frac{m_b}{m_B}\tan\theta_L^b$. $\theta_L^t$ and $\theta_L^b$ can be related through the identity $\sin2\theta_L^b=\frac{1}{\sqrt{2}}\frac{m_T^2-m_t^2}{m_B^2-m_b^2}\sin2\theta_L^t$. Thus, there is only one independent mixing angle $\theta_L^t$.
\item For the doublet $(X,T)_{L,R}$, we have the relation $\tan\theta_L^t=\frac{m_t}{m_T}\tan\theta_R^t$. Thus, there is only one independent mixing angle $\theta_R^t$.
\item For the doublet $(T,B)_{L,R}$, we have the relations $\tan\theta_L^t=\frac{m_t}{m_T}\tan\theta_R^t$ and $\tan\theta_L^b=\frac{m_b}{m_B}\tan\theta_R^b$. Thus, there are two independent mixing angles $\theta_R^t$ and $\theta_R^b$.
\end{itemize}

For the doublet $(X,T)_{L,R}$, the constraints on FCNY couplings from di-Higgs production are similar to those in the singlet $T_{L,R}$ case. Compared to the singlet $T_{L,R}$, there are extra $BBh,Bbh$ type Yukawa interactions for the doublet $(T,B)_{L,R}$ and triplets $(X,T,B)_{L,R},(T,B,Y)_{L,R}$. Thus, it is expected that the constraints on FCNY interactions are looser.
\section{Summary and conclusions}\label{sec:summary}
Top partners are well motivated in many new physics models and FCNY interactions can appear between top quark and the new heavy quark. To unveil the nature of flavor structure and EWSB, it is important to probe the FCNY interactions. However, it is challenging to constrain the $Tth$ coupling at both current and future experiments directly.

In this paper, we have introduced a simplified model and summarized the main constraints from theoretical and experimental viewpoints first. Then we calculate the amplitude of di-Higgs production. After choosing $m_T=400~\mathrm{GeV},s_L=0.2$ and $m_T=800~\mathrm{GeV},s_L=0.1$ as benchmark points, we evaluate the numerical cross sections. It is found that the present constraints from di-Higgs production have already surpassed the unitarity bound because of the $(y_{L,R}^{tT})^4$ behavior in di-Higgs production cross section. For the case of $m_T=400~\mathrm{GeV}$ and $s_L=0.2$, $\mathrm{Re}y_{L,R}^{tT}$ and $\mathrm{Im}y_{L,R}^{tT}$ are expected to be bounded in the range $(-2, 2)$, and even $(-0.4, 0.4)$ in some scenarios at HL-LHC with $2\sigma$ CL roughly. For the case of $m_T=800~\mathrm{GeV}$ and $s_L=0.1$, $\mathrm{Re}y_{L,R}^{tT}$ and $\mathrm{Im}y_{L,R}^{tT}$ are expected to be bounded in the range $(-3, 3)$, and even $(-0.5, 0.5)$ in some scenarios at HL-LHC with $2\sigma$ CL roughly. The value of $\delta_{hhh}$ can have significant effects on the constraints of $y_{L,R}^{tT}$. Simply speaking, larger $\delta_{hhh}$ leads to looser constraints on $y_{L,R}^{tT}$, because there can be more cancellation between the triangle and box diagrams. Finally, we find that the top quark EDM can give stronger bounds of $y_{L,R}^{tT}$ in the off-axis regions for some scenarios.
\section*{Acknowledgements}
We would like to thank Gang Li, Zhao Li, Ying-nan Mao, and Hao Zhang for helpful discussions. We also thank Olivier Mattelaer for MadGraph program discussions through the \href{https://answers.launchpad.net/mg5amcnlo}{launchpad} platform.
\bibliographystyle{unsrt}
\bibliography{gg2hh}
\section*{Appendix}
\begin{appendices}

\section{Asymptotic behaviors of the loop functions}\label{app:A}
\subsection{The shorthand notations of $C_0$ and $D_0$ functions}
The definitions of $C_0$ and $D_0$ function related with pure top quark loops are given as
\begin{align}
&C_0^t(\hat{s})\equiv C_0(0,0,\hat{s},m_t^2,m_t^2,m_t^2),\nonumber\\
&C_0^t(m_h^2)\equiv C_0(m_h^2,m_h^2,\hat{s},m_t^2,m_t^2,m_t^2),\nonumber\\
&C_0^t(\hat{t})\equiv C_0(0,m_h^2,\hat{t},m_t^2,m_t^2,m_t^2),\nonumber\\
&C_0^t(\hat{u})\equiv C_0(0,m_h^2,\hat{u},m_t^2,m_t^2,m_t^2),\nonumber\\
&D_0^t(\hat{t},\hat{s})\equiv D_0(m_h^2,0,0,m_h^2,\hat{t},\hat{s},m_t^2,m_t^2,m_t^2,m_t^2),\nonumber\\
&D_0^t(\hat{u},\hat{s})\equiv D_0(m_h^2,0,0,m_h^2,\hat{u},\hat{s},m_t^2,m_t^2,m_t^2,m_t^2),\nonumber\\
&D_0^t(\hat{t},\hat{u})\equiv D_0(m_h^2,0,m_h^2,0,\hat{t},\hat{u},m_t^2,m_t^2,m_t^2,m_t^2).
\end{align}
The definitions of $C_0$ and $D_0$ function related with pure $T$ quark loops are given as
\begin{align}
&C_0^T(\hat{s})\equiv C_0(0,0,\hat{s},m_T^2,m_T^2,m_T^2),\nonumber\\
&C_0^T(m_h^2)\equiv C_0(m_h^2,m_h^2,\hat{s},m_T^2,m_T^2,m_T^2),\nonumber\\
&C_0^T(\hat{t})\equiv C_0(0,m_h^2,\hat{t},m_T^2,m_T^2,m_T^2),\nonumber\\
&C_0^T(\hat{u})\equiv C_0(0,m_h^2,\hat{u},m_T^2,m_T^2,m_T^2),\nonumber\\
&D_0^T(\hat{t},\hat{s})\equiv D_0(m_h^2,0,0,m_h^2,\hat{t},\hat{s},m_T^2,m_T^2,m_T^2,m_T^2),\nonumber\\
&D_0^T(\hat{u},\hat{s})\equiv D_0(m_h^2,0,0,m_h^2,\hat{u},\hat{s},m_T^2,m_T^2,m_T^2,m_T^2),\nonumber\\
&D_0^T(\hat{t},\hat{u})\equiv D_0(m_h^2,0,m_h^2,0,\hat{t},\hat{u},m_T^2,m_T^2,m_T^2,m_T^2).
\end{align}
The definitions of $C_0$ and $D_0$ function related with mixed $t$ and $T$ quark loops are given as
\begin{align}
&C_0^{tT}(m_h^2)\equiv C_0(m_h^2,m_h^2,\hat{s},m_t^2,m_T^2,m_t^2),\nonumber\\
&C_0^{tT}(\hat{t})\equiv C_0(0,m_h^2,\hat{t},m_t^2,m_t^2,m_T^2),\nonumber\\
&C_0^{tT}(\hat{u})\equiv C_0(0,m_h^2,\hat{u},m_t^2,m_t^2,m_T^2),\nonumber\\
&D_0^{tT}(\hat{t},\hat{s})\equiv D_0(m_h^2,0,0,m_h^2,\hat{t},\hat{s},m_T^2,m_t^2,m_t^2,m_t^2),\nonumber\\
&D_0^{tT}(\hat{u},\hat{s})\equiv D_0(m_h^2,0,0,m_h^2,\hat{u},\hat{s},m_T^2,m_t^2,m_t^2,m_t^2),\nonumber\\
&D_0^{tT}(\hat{t},\hat{u})\equiv D_0(m_h^2,0,m_h^2,0,\hat{t},\hat{u},m_t^2,m_T^2,m_T^2,m_t^2),
\end{align}
and
\begin{align}
&C_0^{Tt}(m_h^2)\equiv C_0(m_h^2,m_h^2,\hat{s},m_T^2,m_t^2,m_T^2),\nonumber\\
&C_0^{Tt}(\hat{t})\equiv C_0(0,m_h^2,\hat{t},m_T^2,m_T^2,m_t^2),\nonumber\\
&C_0^{Tt}(\hat{u})\equiv C_0(0,m_h^2,\hat{u},m_T^2,m_T^2,m_t^2),\nonumber\\
&D_0^{Tt}(\hat{t},\hat{s})\equiv D_0(m_h^2,0,0,m_h^2,\hat{t},\hat{s},m_t^2,m_T^2,m_T^2,m_T^2),\nonumber\\
&D_0^{Tt}(\hat{u},\hat{s})\equiv D_0(m_h^2,0,0,m_h^2,\hat{u},\hat{s},m_t^2,m_T^2,m_T^2,m_T^2),\nonumber\\
&D_0^{Tt}(\hat{t},\hat{u})\equiv D_0(m_h^2,0,m_h^2,0,\hat{t},\hat{u},m_T^2,m_t^2,m_t^2,m_T^2).
\end{align}
As a matter of fact, we have the relation $D_0^{tT}(\hat{t},\hat{u})=D_0^{Tt}(\hat{t},\hat{u})$.
\subsection{Heavy quark expansion of $C_0$ function}
$C_0$ function is defined as
\begin{align}
&C_0(k_1^2,k_{12}^2,k_2^2,m_0^2,m_1^2,m_2^2)\nonumber\\
\equiv&\frac{(2\pi\mu)^{4-D}}{i\pi^2}\int d^Dq\frac{1}{(q^2-m_0^2)[(q+k_1)^2-m_1^2][(q+k_2)^2-m_2^2]}\nonumber\\
=&-\int_0^1\int_0^1\int_0^1dxdydz\frac{\delta(x+y+z-1)}{xm_0^2+ym_1^2+zm_2^2-xyk_1^2-xzk_2^2-yzk_{12}^2},
\end{align}
where $k_{12}\equiv k_1-k_2$ and $D$ is the dimension of space time. When the three internal masses are all equal, $C_0$ function can be expanded as \cite{Shao:2013bz}
\begin{align}
&C_0(k_1^2,k_{12}^2,k_2^2,m_t^2,m_t^2,m_t^2)\nonumber\\
=&-\int_0^1\int_0^1\int_0^1dxdydz\frac{\delta(x+y+z-1)}{m_t^2-xyk_1^2-xzk_2^2-yzk_{12}^2}\nonumber\\
=&-\frac{1}{2m_t^2}-\frac{k_1^2+k_2^2+k_{12}^2}{24m_t^4}-\frac{k_1^4+k_2^4+k_{12}^4+k_1^2k_2^2+k_1^2k_{12}^2+k_2^2k_{12}^2}{180m_t^6}+\mathcal{O}(\frac{k^6}{m_t^8}).
\end{align}
Especially, we have the following results:
\begin{align}
&C_0^t(\hat{s})\approx-\frac{1}{2m_t^2}(1+\frac{\hat{s}}{12m_t^2}+\frac{\hat{s}^2}{90m_t^4}),\nonumber\\
&C_0^t(m_h^2)\approx-\frac{1}{2m_t^2}(1+\frac{2m_h^2+\hat{s}}{12m_t^2}+\frac{3m_h^4+2m_h^2\hat{s}+\hat{s}^2}{90m_t^4}),\nonumber\\
&C_0^t(\hat{t})\approx-\frac{1}{2m_t^2}(1+\frac{m_h^2+\hat{t}}{12m_t^2}+\frac{m_h^4+m_h^2\hat{t}+\hat{t}^2}{90m_t^4}).
\end{align}
The expansion of $C_0^t(\hat{u})$ functions can be obtained when replacing the $\hat{t}$ in $C_0^t(\hat{t})$ by $\hat{u}$. The expansion of $C_0^T$ functions can be obtained when replacing the $m_t$ in $C_0^t$ by $m_T$. 

When the first two internal masses are equal, $C_0$ function can be expanded as
\begin{align}
&C_0(k_1^2,k_{12}^2,k_2^2,m_t^2,m_t^2,m_T^2)\nonumber\\
=&-\int_0^1\int_0^1\int_0^1dxdydz\frac{\delta(x+y+z-1)}{(x+y)m_t^2+zm_T^2-xyk_1^2-xzk_2^2-yzk_{12}^2}\nonumber\\
=&-\int_0^1\int_0^1\int_0^1dxdydz\frac{\delta(x+y+z-1)}{(x+y)m_t^2+zm_T^2}[1+\frac{xyk_1^2+xzk_2^2+yzk_{12}^2}{(x+y)m_t^2+zm_T^2}+\frac{(xyk_1^2+xzk_2^2+yzk_{12}^2)^2}{((x+y)m_t^2+zm_T^2)^2}]+\mathcal{O}(\frac{k^6}{m_{t,T}^8})\nonumber\\
\approx&\frac{1+\log r_{tT}^2-r_{tT}^2}{m_T^2(1-r_{tT}^2)^2}-\frac{2+6r_{tT}^2\log r_{tT}^2+3r_{tT}^2-6r_{tT}^4+r_{tT}^6}{12m_T^4r_{tT}^2(1-r_{tT}^2)^4}k_1^2\nonumber\\
&+\frac{5+2(1+2r_{tT}^2)\log r_{tT}^2-4r_{tT}^2-r_{tT}^4}{4m_T^4(1-r_{tT}^2)^4}(k_2^2+k_{12}^2)\nonumber\\
&-\frac{3-30r_{tT}^2-20r_{tT}^4(1+3\log r_{tT}^2)+60r_{tT}^6-15r_{tT}^8+2r_{tT}^{10}}{180m_T^6r_{tT}^4(1-r_{tT}^2)^6}k_1^4\nonumber\\
&+\frac{10+9r_{tT}^2+3(1+6r_{tT}^2+3r_{tT}^4)\log r_{tT}^2-18r_{tT}^4-r_{tT}^6}{9m_T^6(1-r_{tT}^2)^6}(k_{12}^4+k_2^2k_{12}^2+k_2^4)\nonumber\\
&-\frac{3+44r_{tT}^2+12r_{tT}^2(2+3r_{tT}^2)\log r_{tT}^2-36r_{tT}^4-12r_{tT}^6+r_{tT}^8}{36m_T^6r_{tT}^2(1-r_{tT}^2)^6}k_1^2(k_2^2+k_{12}^2).
\end{align}
When the first and third internal masses are equal, $C_0$ function can be correlated with the first two mass equal cases through the following relations
\begin{align}
&C_0(k_1^2,k_{12}^2,k_2^2,m_t^2,m_T^2,m_t^2)=C_0(k_2^2,k_{12}^2,k_1^2,m_t^2,m_t^2,m_T^2),\nonumber\\
&C_0(k_1^2,k_{12}^2,k_2^2,m_T^2,m_t^2,m_T^2)=C_0(k_2^2,k_{12}^2,k_1^2,m_T^2,m_T^2,m_t^2).
\end{align}
Especially, we have the following results
\begin{align}
&C_0^{tT}(\hat{t})\approx\frac{1}{m_T^2}\cdot\frac{1+\log r_{tT}^2-r_{tT}^2}{(1-r_{tT}^2)^2}+\frac{m_h^2+\hat{t}}{m_T^4}\cdot\frac{5+2(1+2r_{tT}^2)\log r_{tT}^2-4r_{tT}^2-r_{tT}^4}{4(1-r_{tT}^2)^4}\nonumber\\
&+\frac{m_h^4+m_h^2\hat{t}+\hat{t}^2}{m_T^6}\cdot\frac{10+3(1+6r_{tT}^2+3r_{tT}^4)\log r_{tT}^2+9r_{tT}^2-18r_{tT}^4-r_{tT}^6}{9(1-r_{tT}^2)^6},\nonumber\\
&C_0^{tT}(m_h^2)\approx\frac{1}{m_T^2}\cdot\frac{1+\log r_{tT}^2-r_{tT}^2}{(1-r_{tT}^2)^2}-\frac{\hat{s}}{m_T^4}\cdot\frac{2+6r_{tT}^2\log r_{tT}^2+3r_{tT}^2-6r_{tT}^4+r_{tT}^6}{12r_{tT}^2(1-r_{tT}^2)^4}\nonumber\\
	&+\frac{m_h^2}{m_T^4}\cdot\frac{5+2(1+2r_{tT}^2)\log r_{tT}^2-4r_{tT}^2-r_{tT}^4}{2(1-r_{tT}^2)^4}+\frac{m_h^4}{m_T^6}\cdot\frac{10+3(1+6r_{tT}^2+3r_{tT}^4)\log r_{tT}^2+9r_{tT}^2-18r_{tT}^4-r_{tT}^6}{3(1-r_{tT}^2)^6}\nonumber\\
&-\frac{m_h^2\hat{s}}{m_T^6}\cdot\frac{3+12r_{tT}^2(2+3r_{tT}^2)\log r_{tT}^2+44r_{tT}^2-36r_{tT}^4-12r_{tT}^6+r_{tT}^8}{18r_{tT}^2(1-r_{tT}^2)^6}\nonumber\\
&-\frac{\hat{s}^2}{m_T^6}\cdot\frac{3-30r_{tT}^2-60r_{tT}^4\log r_{tT}^2-20r_{tT}^4+60r_{tT}^6-15r_{tT}^8+2r_{tT}^{10}}{180r_{tT}^4(1-r_{tT}^2)^6}.
\end{align}
Keeping the terms up to $\mathcal{O}(\frac{1}{m_T^4})$ and considering the $\log r_{tT}^2$ enhanced terms, they can be simplified as
\begin{align}
&C_0^{tT}(\hat{t})\approx\frac{1}{m_T^2}\big[1+\log r_{tT}^2+r_{tT}^2(1+2\log r_{tT}^2)+\frac{m_h^2+\hat{t}}{4m_T^2}(5+2\log r_{tT}^2)\big],\nonumber\\
&C_0^{tT}(m_h^2)\approx\frac{1}{m_T^2}\big[1+\log r_{tT}^2+r_{tT}^2(1+2\log r_{tT}^2)-\frac{\hat{s}}{12m_T^2}(\frac{2}{r_{tT}^2}+11+6\log r_{tT}^2)+\frac{m_h^2}{2m_T^2}(5+2\log r_{tT}^2)\big].
\end{align}
Similarly, we can get the following results
\begin{align}
&C_0^{Tt}(\hat{t})\approx-\frac{1}{m_T^2}\cdot\frac{1+r_{tT}^2\log r_{tT}^2-r_{tT}^2}{(1-r_{tT}^2)^2}-\frac{m_h^2+\hat{t}}{m_T^4}\cdot\frac{1+2r_{tT}^2(2+r_{tT}^2)\log r_{tT}^2+4r_{tT}^2-5r_{tT}^4}{4(1-r_{tT}^2)^4}\nonumber\\
&-\frac{m_h^4+m_h^2\hat{t}+\hat{t}^2}{m_T^6}\cdot\frac{1+3r_{tT}^2(3+6r_{tT}^2+r_{tT}^4)\log r_{tT}^2+18r_{tT}^2-9r_{tT}^4-10r_{tT}^6}{9(1-r_{tT}^2)^6},\nonumber\\
&C_0^{Tt}(m_h^2)\approx-\frac{1}{m_T^2}\cdot\frac{1+r_{tT}^2\log r_{tT}^2-r_{tT}^2}{(1-r_{tT}^2)^2}-\frac{\hat{s}}{m_T^4}\cdot\frac{1-6r_{tT}^2-6r_{tT}^4\log r_{tT}^2+3r_{tT}^4+2r_{tT}^6}{12(1-r_{tT}^2)^4}\nonumber\\
	&-\frac{m_h^2}{2m_T^4}\cdot\frac{1+2r_{tT}^2(2+r_{tT}^2)\log r_{tT}^2+4r_{tT}^2-5r_{tT}^4}{(1-r_{tT}^2)^4}-\frac{m_h^4}{m_T^6}\cdot\frac{1+3r_{tT}^2(3+6r_{tT}^2+r_{tT}^4)\log r_{tT}^2+18r_{tT}^2-9r_{tT}^4-10r_{tT}^6}{3(1-r_{tT}^2)^6}\nonumber\\
&-\frac{m_h^2\hat{s}}{m_T^6}\cdot\frac{1-12r_{tT}^2-12r_{tT}^4(3+2r_{tT}^2)\log r_{tT}^2-36r_{tT}^4+44r_{tT}^6+3r_{tT}^8}{18(1-r_{tT}^2)^6}\nonumber\\
&-\frac{\hat{s}^2}{m_T^6}\cdot\frac{2-15r_{tT}^2+60r_{tT}^4+60r_{tT}^6\log r_{tT}^2-20r_{tT}^6-30r_{tT}^8+3r_{tT}^{10}}{180(1-r_{tT}^2)^6}.
\end{align}
Keeping the terms up to $\mathcal{O}(\frac{1}{m_T^4})$ and considering the $\log r_{tT}^2$ enhanced terms, they can be simplified as
\begin{align}
&C_0^{Tt}(\hat{t})\approx-\frac{1}{m_T^2}\big[1+r_{tT}^2(1+\log r_{tT}^2)+\frac{m_h^2+\hat{t}}{4m_T^2}\big],\nonumber\\
&C_0^{Tt}(m_h^2)\approx-\frac{1}{m_T^2}\big[1+r_{tT}^2(1+\log r_{tT}^2)+\frac{\hat{s}}{12m_T^2}+\frac{m_h^2}{2m_T^2}\big].
\end{align}

\subsection{Heavy quark expansion of $D_0$ function}
$D_0$ function is defined as:
\begin{align}
&D_0(k_1^2,k_{12}^2,k_{23}^2,k_3^2,k_2^2,k_{13}^2,m_0^2,m_1^2,m_2^2,m_3^2)\nonumber\\
\equiv&\frac{(2\pi\mu)^{4-D}}{i\pi^2}\int d^Dq\frac{1}{(q^2-m_0^2)[(q+k_1)^2-m_1^2][(q+k_2)^2-m_2^2][(q+k_3)^2-m_3^2]}\nonumber\\
=&\int_0^1\int_0^1\int_0^1\int_0^1dxdydzdw\frac{\delta(x+y+z+w-1)}{[xm_0^2+ym_1^2+zm_2^2+wm_3^2-xyk_1^2-xzk_2^2-xwk_3^2-yzk_{12}^2-ywk_{13}^2-zwk_{23}^2]^2},
\end{align}
where we have $k_{12}\equiv k_1-k_2$, $k_{23}\equiv k_2-k_3$, and $k_{13}\equiv k_1-k_3$. When the four internal masses are all equal, $D_0$ function can be expanded as \cite{Shao:2013bz}
\begin{align}
&D_0(k_1^2,k_{12}^2,k_{23}^2,k_3^2,k_2^2,k_{13}^2,m_t^2,m_t^2,m_t^2,m_t^2)\nonumber\\
=&\int_0^1\int_0^1\int_0^1\int_0^1dxdydzdw\frac{\delta(x+y+z+w-1)}{[m_t^2-xyk_1^2-xzk_2^2-xwk_3^2-yzk_{12}^2-ywk_{13}^2-zwk_{23}^2]^2}\nonumber\\
=&\frac{1}{6m_t^4}\Big[1+\frac{k_1^2+k_{12}^2+k_{23}^2+k_3^2+k_2^2+k_{13}^2}{10m_t^2}+\frac{1}{140m_t^4}\Big(2(k_1^4+k_2^4+k_3^4+k_1^2k_2^2+k_1^2k_3^2+k_2^2k_3^2)\nonumber\\
	&+2(k_{12}^4+k_{13}^4+k_{23}^4+k_{12}^2k_{13}^2+k_{12}^2k_{23}^2+k_{13}^2k_{23}^2)+2k_1^2(k_{12}^2+k_{13}^2)+2k_2^2(k_{12}^2+k_{23}^2)+2k_3^2(k_{13}^2+k_{23}^2)\nonumber\\
	&+(k_1^2k_{23}^2+k_2^2k_{13}^2+k_3^2k_{12}^2)\Big)+\mathcal{O}(\frac{k^6}{m_t^6})\Big].
\end{align}
Especially, we have the following results
\begin{align}
&D_0^t(\hat{t},\hat{s})\approx\frac{1}{6m_t^4}[1+\frac{2m_h^2+\hat{s}+\hat{t}}{10m_t^2}+\frac{6m_h^4+4m_h^2(\hat{s}+\hat{t})+2\hat{s}^2+2\hat{t}^2+\hat{s}\hat{t}}{140m_t^4}],\nonumber\\
&D_0^t(\hat{t},\hat{u})\approx\frac{1}{6m_t^4}[1+\frac{2m_h^2+\hat{t}+\hat{u}}{10m_t^2}+\frac{5m_h^4+4m_h^2(\hat{t}+\hat{u})+2\hat{t}^2+2\hat{u}^2+\hat{t}\hat{u}}{140m_t^4}].
\end{align}
The expansion of $D_0^t(\hat{u},\hat{s})$ can be obtained when replacing the $\hat{t}$ in $D_0^t(\hat{t},\hat{s})$ by $\hat{u}$. The expansion of $D_0^T$ functions can be obtained when replacing the $m_t$ in $D_0^t$ by $m_T$.

When three internal masses are equal, $D_0$ function can be expanded as
\begin{align}
&D_0(k_1^2,k_{12}^2,k_{23}^2,k_3^2,k_2^2,k_{13}^2,m_T^2,m_t^2,m_t^2,m_t^2)\nonumber\\
=&\int_0^1\int_0^1\int_0^1\int_0^1dxdydzdw\frac{\delta(x+y+z+w-1)}{[xm_T^2+(y+z+w)m_t^2-xyk_1^2-xzk_2^2-xwk_3^2-yzk_{12}^2-ywk_{13}^2-zwk_{23}^2]^2}\nonumber\\
=&\int_0^1\!\int_0^1\!\int_0^1\!\int_0^1\!dxdydzdw\frac{\delta(x+y+z+w-1)}{[xm_T^2+(y+z+w)m_t^2]^2}[1\!+\!\frac{2(xyk_1^2+xzk_2^2+xwk_3^2+yzk_{12}^2+ywk_{13}^2+zwk_{23}^2)}{xm_T^2+(y+z+w)m_t^2}]\!+\!\mathcal{O}(\frac{k^4}{m_{t,T}^8})\nonumber\\
\approx&\frac{1}{m_T^4}\cdot\frac{1+2r_{tT}^2\log r_{tT}^2-r_{tT}^4}{2r_{tT}^2(1-r_{tT}^2)^3}+\frac{(k_1^2+k_2^2+k_3^2)}{m_T^6}\cdot\frac{1+6r_{tT}^2(1+r_{tT}^2)\log r_{tT}^2+9r_{tT}^2-9r_{tT}^4-r_{tT}^6}{6r_{tT}^2(1-r_{tT}^2)^5}\nonumber\\
&+\frac{(k_{12}^2+k_{23}^2+k_{13}^2)}{m_T^6}\cdot\frac{1-8r_{tT}^2-12r_{tT}^4\log r_{tT}^2+8r_{tT}^6-r_{tT}^8}{24r_{tT}^4(1-r_{tT}^2)^5}.
\end{align}
We only expand it up to $\mathcal{O}(\frac{k^2}{m_{t,T}^6})$, because the general results will be quite lengthy. For the integral $D_0^{tT}(\hat{t},\hat{s})$, we obtain the expression up to $\mathcal{O}(\frac{k^4}{m_{t,T}^8})$
\begin{align}
&D_0^{tT}(\hat{t},\hat{s})=\int_0^1\int_0^1\int_0^1\int_0^1dxdydzdw\frac{\delta(x+y+z+w-1)}{[xm_T^2+(y+z+w)m_t^2-x(y+w)m_h^2-xzt-yws]^2}\nonumber\\
=&\int_0^1\int_0^1\int_0^1\int_0^1dxdydzdw\frac{\delta(x+y+z+w-1)}{[xm_T^2+(y+z+w)m_t^2]^2}\cdot\nonumber\\
	&\left(1+\frac{2[x(y+w)m_h^2+xzt+yws]}{xm_T^2+(y+z+w)m_t^2}+\frac{3[x(y+w)m_h^2+xzt+yws]^2}{[xm_T^2+(y+z+w)m_t^2]^2}\right)+\mathcal{O}(\frac{1}{m_{t,T}^{10}})\nonumber\\
\approx&\frac{1}{m_T^4}\cdot\frac{1+2r_{tT}^2\log r_{tT}^2-r_{tT}^4}{2r_{tT}^2(1-r_{tT}^2)^3}+\frac{(2m_h^2+\hat{t})}{m_T^6}\cdot\frac{1+6r_{tT}^2(1+r_{tT}^2)\log r_{tT}^2+9r_{tT}^2-9r_{tT}^4-r_{tT}^6}{6r_{tT}^2(1-r_{tT}^2)^5}\nonumber\\
&+\frac{\hat{s}}{m_T^6}\cdot\frac{1-8r_{tT}^2-12r_{tT}^4\log r_{tT}^2+8r_{tT}^6-r_{tT}^8}{24r_{tT}^4(1-r_{tT}^2)^5} +\frac{\hat{s}^2}{m_T^8}\cdot\frac{1-9r_{tT}^2+45r_{tT}^4-45r_{tT}^8+9r_{tT}^{10}-r_{tT}^{12}+60r_{tT}^6\log r_{tT}^2}{180r_{tT}^6(1-r_{tT}^2)^7}\nonumber\\
&+\frac{\hat{s}(\hat{t}+4m_h^2)}{m_T^8}\cdot\frac{1-15r_{tT}^2-80r_{tT}^4+80r_{tT}^6+15r_{tT}^8-r_{tT}^{10}-60r_{tT}^4(1+r_{tT}^2)\log r_{tT}^2}{120r_{tT}^4(1-r_{tT}^2)^7}\nonumber\\
	&+\frac{\hat{t}^2+2m_h^2\hat{t}+3m_h^4}{m_T^8}\cdot\frac{1+28r_{tT}^2-28r_{tT}^6-r_{tT}^8+12r_{tT}^2(1+3r_{tT}^2+r_{tT}^4)\log r_{tT}^2}{12r_{tT}^2(1-r_{tT}^2)^7}.
\end{align}
Keeping the terms up to $\mathcal{O}(\frac{1}{m_T^6})$ and considering the $\log r_{tT}^2$ enhanced terms, they can be simplified as
\begin{align}
&D_0^{tT}(\hat{t},\hat{s})\approx\frac{1}{2m_T^4}\big[\frac{1}{r_{tT}^2}+3+2\log r_{tT}^2+r_{tT}^2(5+6\log r_{tT}^2)\big]+\frac{(2m_h^2+\hat{t})}{6m_T^6}(\frac{1}{r_{tT}^2}+6\log r_{tT}^2+14)\nonumber\\
	&+\frac{\hat{s}}{24m_T^6}(\frac{1}{r_{tT}^4}-\frac{3}{r_{tT}^2}-12\log r_{tT}^2-25).
\end{align}
Similarly, we can get the following results
\begin{align}
&D_0^{Tt}(\hat{t},\hat{s})\approx\frac{1}{m_T^4}\cdot\frac{1+2r_{tT}^2\log r_{tT}^2-r_{tT}^4}{2(1-r_{tT}^2)^3}+\frac{(2m_h^2+\hat{t})}{m_T^6}\cdot
\frac{1+9r_{tT}^2+6r_{tT}^2(1+r_{tT}^2)\log r_{tT}^2-9r_{tT}^4-r_{tT}^6}{6(1-r_{tT}^2)^5}\nonumber\\
&+\frac{\hat{s}}{m_T^6}\cdot\frac{1-8r_{tT}^2-12r_{tT}^4\log r_{tT}^2+8r_{tT}^6-r_{tT}^8}{24(1-r_{tT}^2)^5} +\frac{\hat{s}^2}{m_T^8}\cdot\frac{1-9r_{tT}^2+45r_{tT}^4-45r_{tT}^8+9r_{tT}^{10}-r_{tT}^{12}+60r_{tT}^6\log r_{tT}^2}{180(1-r_{tT}^2)^7}\nonumber\\
&+\frac{\hat{s}(\hat{t}+4m_h^2)}{m_T^8}\cdot\frac{1-15r_{tT}^2-80r_{tT}^4+80r_{tT}^6+15r_{tT}^8-r_{tT}^{10}-60r_{tT}^4(1+r_{tT}^2)\log r_{tT}^2}{120(1-r_{tT}^2)^7}\nonumber\\
	&+\frac{\hat{t}^2+2m_h^2\hat{t}+3m_h^4}{m_T^8}\cdot\frac{1+28r_{tT}^2-28r_{tT}^6-r_{tT}^8+12r_{tT}^2(1+3r_{tT}^2+r_{tT}^4)\log r_{tT}^2}{12(1-r_{tT}^2)^7}.
\end{align}
Keeping the terms up to $\mathcal{O}(\frac{1}{m_T^6})$ and considering the $\log r_{tT}^2$ enhanced terms, they can be simplified as
\begin{align}
&D_0^{Tt}(\hat{t},\hat{s})\approx\frac{1}{2m_T^4}(1+3r_{tT}^2+2r_{tT}^2\log r_{tT}^2)+\frac{2m_h^2+\hat{t}}{6m_T^6}+\frac{\hat{s}}{24m_T^6}.
\end{align}
When two internal masses are equal individually, we have the following relations 
\begin{align}
&D_0(k_1^2,k_{12}^2,k_{23}^2,k_3^2,k_2^2,k_{13}^2,m_t^2,m_T^2,m_T^2,m_t^2)=D_0(k_1^2,k_3^2,k_{23}^2,k_{12}^2,k_{13}^2,k_2^2,m_T^2,m_t^2,m_t^2,m_T^2)\nonumber\\
&=D_0(k_{23}^2,k_3^2,k_1^2,k_{12}^2,k_2^2,k_{13}^2,m_T^2,m_t^2,m_t^2,m_T^2).
\end{align}
$D_0$ function can be expanded as
\begin{align}
&D_0(k_1^2,k_{12}^2,k_{23}^2,k_3^2,k_2^2,k_{13}^2,m_t^2,m_T^2,m_T^2,m_t^2)\nonumber\\
=&\int_0^1\int_0^1\int_0^1\int_0^1dxdydzdw\frac{\delta(x+y+z+w-1)}{[(y+z)m_T^2+(x+w)m_t^2-xyk_1^2-xzk_2^2-xwk_3^2-yzk_{12}^2-ywk_{13}^2-zwk_{23}^2]^2}\nonumber\\
=&\int_0^1\!\int_0^1\!\int_0^1\!\int_0^1\!dxdydzdw\frac{\delta(x+y+z+w-1)}{[(y+z)m_T^2+(x+w)m_t^2]^2}[1\!+\!\frac{2(xyk_1^2+xzk_2^2+xwk_3^2+yzk_{12}^2+ywk_{13}^2+zwk_{23}^2)}{(y+z)m_T^2+(x+w)m_t^2}]\!+\!\mathcal{O}(\frac{k^4}{m_{t,T}^8})\nonumber\\
\approx&-\frac{1}{m_T^4}\cdot\frac{2+(1+r_{tT}^2)\log r_{tT}^2-2r_{tT}^2}{(1-r_{tT}^2)^3}-\frac{k_1^2+k_2^2+k_{13}^2+k_{23}^2}{m_T^6}\cdot\frac{3+(1+4r_{tT}^2+r_{tT}^4)\log r_{tT}^2-3r_{tT}^4}{2(1-r_{tT}^2)^5}\nonumber\\
&+\frac{k_3^2}{m_T^6}\cdot\frac{1+6r_{tT}^2(1+r_{tT}^2)\log r_{tT}^2+9r_{tT}^2-9r_{tT}^4-r_{tT}^6}{6r_{tT}^2(1-r_{tT}^2)^5}+\frac{k_{12}^2}{m_T^6}\cdot\frac{1+6r_{tT}^2(1+r_{tT}^2)\log r_{tT}^2+9r_{tT}^2-9r_{tT}^4-r_{tT}^6}{6(1-r_{tT}^2)^5}.
\end{align}
We only expand it up to $\mathcal{O}(\frac{k^2}{m_{t,T}^6})$, because the general results will be quite lengthy. For the integral $D_0^{tT}(\hat{t},\hat{u})$, we obtain the expression up to $\mathcal{O}(\frac{k^4}{m_{t,T}^8})$
\begin{align}
&D_0^{tT}(\hat{t},\hat{u})=\int_0^1\int_0^1\int_0^1\int_0^1dxdydzdw\frac{\delta(x+y+z+w-1)}{[(x+w)m_t^2+(y+z)m_T^2-(xy+zw)m_h^2-xzt-ywu]^2}\nonumber\\
=&\int_0^1\int_0^1\int_0^1\int_0^1dxdydzdw\frac{\delta(x+y+z+w-1)}{[(x+w)m_t^2+(y+z)m_T^2]^2}\cdot\nonumber\\
	&\left(1+\frac{2[(xy+zw)m_h^2+xzt+ywu]}{(x+w)m_t^2+(y+z)m_T^2}+\frac{3[(xy+zw)m_h^2+xzt+ywu]^2}{[(x+w)m_t^2+(y+z)m_T^2]^2}\right)+\mathcal{O}(\frac{1}{m_{t,T}^{10}})\nonumber\\
\approx&-\frac{1}{m_T^4}\cdot\frac{2+(1+r_{tT}^2)\log r_{tT}^2-2r_{tT}^2}{(1-r_{tT}^2)^3}-\frac{2m_h^2+\hat{t}+\hat{u}}{m_T^6}\cdot\frac{3+(1+4r_{tT}^2+r_{tT}^4)\log r_{tT}^2-3r_{tT}^4}{2(1-r_{tT}^2)^5}\nonumber\\
-&\frac{5m_h^4+4m_h^2(\hat{t}+\hat{u})+2\hat{t}^2+\hat{t}\hat{u}+2\hat{u}^2}{m_T^8}\cdot\frac{11+27r_{tT}^2-27r_{tT}^4-11r_{tT}^6+3(1+9r_{tT}^2+9r_{tT}^4+r_{tT}^6)\log r_{tT}^2}{18(1-r_{tT}^2)^7}.
\end{align}
Keeping the terms up to $\mathcal{O}(\frac{1}{m_T^4})$ and considering the $\log r_{tT}^2$ enhanced terms, they can be simplified as
\begin{align}
&D_0^{tT}(\hat{t},\hat{u})\approx-\frac{1}{m_T^4}\big[2+\log r_{tT}^2+4r_{tT}^2(1+\log r_{tT}^2)\big]-\frac{4m_h^2-\hat{s}}{2m_T^6}(3+\log r_{tT}^2).
\end{align}
In the above calculations, the $t$ and $T$ quark mixed $C_0$ and $D_0$ integrals will agree with the pure top quark integrals in the limit of $m_t=m_T$ (or $r_{tT}\rightarrow1$). Besides, these expansion results have been checked by the LoopTools numerically \cite{Hahn:1998yk}.
\end{appendices}
\clearpage
\newpage
\end{document}